\documentclass[twocolumn]{aastex631}

\usepackage{savesym}
\savesymbol{tablenum}
\usepackage{siunitx}
\restoresymbol{SIX}{tablenum}
\usepackage[version=4]{mhchem}
\usepackage{graphicx,epsf}
\usepackage{subfigure}
\usepackage{graphicx}	
\usepackage{amsmath}	
\usepackage{amssymb}	
\usepackage{newtxtext,newtxmath}
\usepackage{siunitx}
\usepackage{footnote}

\newcommand{\Msun}{{\rm M}_\odot}
\newcommand{\Rsun}{{\rm R}_\odot}

\newcommand{\kms}{\textrm{km}\,\textrm{s}^{-1}}

\newcommand{\mdot}{M$_{\odot}$~yr$^{-1}$}

\DeclareRobustCommand{\ion}[2]{\relax\ifmmode\ifx\testbx\f@series{\mathbf{#1\,\mathsc{#2}}}\else{\mathrm{#1\,\mathsc{#2}}}\fi\else\textup{#1\,{\mdseries\textsc{#2}}}\fi}

\newcommand{\code}[1]{\texttt{#1}}
\def\heracles{{\code{HERACLES}}}
\def\cmfgen{{\code{CMFGEN}}}

\usepackage[scale=0.94]{tgbonum}
\usepackage[mode=text]{siunitx}

\makeatletter
\DeclareTextCompositeCommand{\r}{OT1}{A}{%
  \leavevmode\vbox{%
    \offinterlineskip
    \ialign{\hfil##\hfil\cr\char23\cr\noalign{\kern-1.15ex}A\cr}%
  }%
}
\makeatother

\shorttitle{Final Moments III}
\shortauthors{Jacobson-Gal\'an et al.}

\begin{document}

\title{Final Moments III: Explosion Properties and Progenitor Constraints of CSM-Interacting Type II Supernovae}

\correspondingauthor{Wynn Jacobson-Gal\'{a}n (he, him, his)}
\email{wynnjg@caltech.edu}

\author[0000-0002-3934-2644]{W.~V.~Jacobson-Gal\'{a}n}
\altaffiliation{NASA Hubble Fellow}
\affil{Department of Astronomy and Astrophysics, California Institute of Technology, Pasadena, CA 91125, USA}
\affil{Department of Astronomy, University of California, Berkeley, CA 94720-3411, USA}

\author[0000-0003-0599-8407]{L.~Dessart}
\affil{Institut d’Astrophysique de Paris, CNRS-Sorbonne Université, 98 bis boulevard Arago, F-75014 Paris, France}

\author[0000-0002-5680-4660]{K.~W.~Davis}
\affil{Department of Astronomy and Astrophysics, University of California, Santa Cruz, CA 95064, USA}

\author[0000-0002-4924-444X]{K.~A.~Bostroem}
\affil{Steward Observatory, University of Arizona, 933 North Cherry Avenue, Tucson, AZ 85721-0065, USA}
\altaffiliation{LSST-DA Catalyst Fellow}

\author[0000-0002-5740-7747]{C.~D.~Kilpatrick}
\affil{Center for Interdisciplinary Exploration and Research in Astrophysics (CIERA), Northwestern University, Evanston, IL 60202, USA}
\affiliation{Department of Physics and Astronomy, Northwestern University, Evanston, IL 60208, USA}

\author[0000-0003-4768-7586]{R.~Margutti}
\affil{Department of Astronomy, University of California, Berkeley, CA 94720-3411, USA}
\affil{Department of Physics, University of California, Berkeley, CA 94720-7300, USA}

\author[0000-0003-3460-0103]{A.~V.~Filippenko}
\affil{Department of Astronomy, University of California, Berkeley, CA 94720-3411, USA}

\author[0000-0002-2445-5275]{R.~J.~Foley}
\affiliation{Department of Astronomy and Astrophysics, University of California, Santa Cruz, CA 95064, USA}

\author[0000-0002-7706-5668]{R.~Chornock}
\affil{Department of Astronomy, University of California, Berkeley, CA 94720-3411, USA}

\author[0000-0003-0794-5982]{G.~Terreran}
\affil{Las Cumbres Observatory, 6740 Cortona Dr. Suite 102, Goleta, CA, 93117}

\author[0000-0002-1125-9187]{D.~Hiramatsu}
\affil{Center for Astrophysics \textbar{} Harvard \& Smithsonian, 60 Garden Street, Cambridge, MA 02138-1516, USA}
\affil{The NSF AI Institute for Artificial Intelligence and Fundamental Interactions, USA}

\author[0000-0001-9570-0584]{M.~Newsome}
\affil{Las Cumbres Observatory, 6740 Cortona Dr. Suite 102, Goleta, CA, 93117}
\affil{Department of Physics, University of California, Santa Barbara, Santa Barbara, CA, USA, 93111}

\author[0000-0003-0209-9246]{E.~Padilla~Gonzalez}
\affil{Las Cumbres Observatory, 6740 Cortona Dr. Suite 102, Goleta, CA, 93117}
\affil{Department of Physics, University of California, Santa Barbara, Santa Barbara, CA, USA, 93111}

\author[0000-0002-7472-1279]{C.~Pellegrino}
\affil{Department of Astronomy, University of Virginia, Charlottesville, VA 22904, USA}

\author[0000-0003-4253-656X]{D.~A.~Howell}
\affil{Las Cumbres Observatory, 6740 Cortona Dr. Suite 102, Goleta, CA, 93117}
\affil{Department of Physics, University of California, Santa Barbara, Santa Barbara, CA, USA, 93111}


\author[0000-0003-0227-3451]{J.~P.~Anderson}
\affil{European Southern Observatory, Alonso de C\'ordova 3107, Casilla 19, Santiago, Chile }
\affil{Millennium Institute of Astrophysics MAS, Nuncio Monsenor Sotero Sanz 100, Off. 104, Providencia, Santiago, Chile}

\author[0000-0002-4269-7999]{C.~R.~Angus}
\affil{DARK, Niels Bohr Institute, University of Copenhagen, Jagtvej 128, 2200 Copenhagen, Denmark}
\affil{Astrophysics Research Centre, School of Mathematics and Physics, Queen’s University Belfast, Belfast BT7 1NN, UK}

\author[0000-0002-4449-9152]{K.~Auchettl}
\affil{Department of Astronomy and Astrophysics, University of California, Santa Cruz, CA 95064, USA}
\affil{School of Physics, The University of Melbourne, VIC 3010, Australia}

\author[0000-0001-5955-2502]{T.~G.~Brink}
\affil{Department of Astronomy, University of California, Berkeley, CA 94720-3411, USA}

\author[0000-0003-4553-4033]{R.~Cartier}
\affil{Centro de Astronom\'ia (CITEVA), Universidad de Antofagasta, Avenida Angamos 601, Antofagasta, Chile}

\author[0000-0003-4263-2228]{D.~A.~Coulter}
\affil{Space Telescope Science Institute, Baltimore, MD 21218, USA}

\author[0000-0001-5486-2747]{T.~de~Boer}
\affil{Institute for Astronomy, University of Hawaii, 2680 Woodlawn Drive, Honolulu, HI 96822, USA}

\author[0000-0001-7081-0082]{M.~R.~Drout}
\affil{David A. Dunlap Department of Astronomy and Astrophysics, University of Toronto, 50 St. George Street, Toronto, Ontario, M5S 3H4, Canada}

\author[0000-0003-1714-7415]{N.~Earl}
\affil{Department of Astronomy, University of Illinois at Urbana-Champaign, 1002 W. Green St., Urbana, IL 61801, USA}

\author[0000-0001-7251-8368]{K.~Ertini}
\affil{Facultad de Ciencias Astron\'omicas y Geofísicas, Universidad Nacional de La Plata, Paseo del Bosque S/N, B1900FWA, La Plata, Argentina}
\affil{Instituto de Astrofisica de La Plata (IALP), CCT-CONICET-UNLP, Paseo del Bosque S/N, B1900FWA, La Plata, Argentina}

\author[0000-0003-4914-5625]{J.~R.~Farah}
\affil{Las Cumbres Observatory, 6740 Cortona Dr. Suite 102, Goleta, CA, 93117}
\affil{Department of Physics, University of California, Santa Barbara, Santa Barbara, CA, USA, 93111}

\author[0000-0002-6886-269X]{D.~Farias}
\affiliation{DARK, Niels Bohr Institute, University of Copenhagen, Jagtvej 128, 2200 Copenhagen, Denmark}

\author[0000-0002-8526-3963]{C.~Gall}
\affil{DARK, Niels Bohr Institute, University of Copenhagen, Jagtvej 128, 2200 Copenhagen, Denmark}

\author[0000-0003-1015-5367]{H.~Gao}
\affil{Institute for Astronomy, University of Hawaii, 2680 Woodlawn Drive, Honolulu, HI 96822, USA}

\author{M.~A.~Gerlach}
\affil{Department of Astrophysics, Pontificia Universidad Catolica de Chile, Santiago, Chile}

\author{F.~Guo}
\affil{Department of Physics, Tsinghua University, Shuangqing Road, Beijing, China}

\author[0000-0003-4287-4577]{A.~Haynie}
\affiliation{The Observatories of the Carnegie Institute for Science, 813 Santa Barbara St., Pasadena, CA 91101, USA}
\affiliation{Department of Physics \& Astronomy, University of Southern California, Los Angeles, CA 90089, USA}

\author[0000-0002-0832-2974]{G.~Hosseinzadeh}
\affil{Department of Astronomy \& Astrophysics, University of California, San Diego, 9500 Gilman Drive, MC 0424, La Jolla, CA 92093-0424, USA}

\author[0000-0003-2405-2967]{A.~L.~Ibik}
\affiliation{David A. Dunlap Department of Astronomy and Astrophysics, University of Toronto, 50 St. George Street, Toronto, Ontario, M5S 3H4, Canada}

\author[0000-0001-8738-6011]{S.~W.~Jha}
\affil{Department of Physics and Astronomy, Rutgers, the State University of New Jersey, 136 Frelinghuysen Road, Piscataway, NJ 08854, USA}

\author[0000-0002-6230-0151]{D.~O.~Jones}
\affiliation{Institute for Astronomy, University of Hawai'i, 640 N. A'ohoku Pl., Hilo, HI 96720, USA}

\author[0000-0001-5710-8395]{D.~Langeroodi}
\affiliation{DARK, Niels Bohr Institute, University of Copenhagen, Jagtvej 128, 2200 Copenhagen, Denmark}

\author[0000-0002-2249-0595]{N.~LeBaron}
\affil{Department of Astronomy, University of California, Berkeley, CA 94720-3411, USA}

\author[0000-0002-7965-2815]{E.~A.~Magnier}
\affil{Institute for Astronomy, University of Hawaii, 2680 Woodlawn Drive, Honolulu, HI 96822, USA}

\author[0000-0001-6806-0673]{A.~L.~Piro}
\affil{The Observatories of the Carnegie Institute for Science, 813 Santa Barbara St., Pasadena, CA 91101, USA}

\author[0000-0002-6248-398X]{S.~I.~Raimundo}
\affil{DARK, Niels Bohr Institute, University of Copenhagen, Jagtvej 128, 2200 Copenhagen, Denmark}
\affil{Department of Physics and Astronomy, University of Southampton, Highfield, Southampton SO17 1BJ, UK}

\author[0000-0002-4410-5387]{A.~Rest}
\affil{Department of Physics and Astronomy, The Johns Hopkins University, Baltimore, MD 21218, USA}
\affil{Space Telescope Science Institute, Baltimore, MD 21218, USA}

\author[0000-0002-3825-0553]{S.~Rest}
\affil{Department of Physics and Astronomy, The Johns Hopkins University, Baltimore, MD 21218, USA}

\author[0000-0003-0427-8387]{R.~Michael~Rich}
\affil{Department Physics and Astronomy, University of California, Los Angeles, Los Angeles, CA, 90095-1547}

\author[0000-0002-7559-315X]{C.~Rojas-Bravo}
\affiliation{Department of Astronomy and Astrophysics, University of California, Santa Cruz, CA 95064, USA}

\author[0000-0001-8023-4912]{H.~Sears}
\affiliation{Department of Physics and Astronomy, Rutgers, the State University of New Jersey, 136 Frelinghuysen Road, Piscataway, NJ 08854-8019, USA}

\author[0000-0002-5748-4558]{K.~Taggart}
\affil{Department of Astronomy and Astrophysics, University of California, Santa Cruz, CA 95064, USA}

\author[0000-0002-1125-9187]{V.~A.~Villar}
\affil{Center for Astrophysics \textbar{} Harvard \& Smithsonian, 60 Garden Street, Cambridge, MA 02138-1516, USA}

\author[0000-0002-1341-0952]{R.~J.~Wainscoat}
\affil{Institute for Astronomy, University of Hawaii, 2680 Woodlawn Drive, Honolulu, HI 96822, USA}

\author[0000-0002-7334-2357]{X.-F.~Wang}
\affil{Department of Physics, Tsinghua University, Shuangqing Road, Beijing, China}

\author[0000-0002-4186-6164]{A.~R.~Wasserman}
\affil{Department of Astronomy, University of Illinois at Urbana-Champaign, 1002 W. Green St., IL 61801, USA}
\affil{Center for Astrophysical Surveys, National Center for Supercomputing Applications, Urbana, IL, 61801, USA}

\author[0009-0004-4256-1209]{S.~Yan}
\affil{Department of Physics, Tsinghua University, Shuangqing Road, Beijing, China}

\author{Y.~Yang}
\affil{Department of Astronomy, University of California, Berkeley, CA 94720-3411, USA}
\affil{Department of Physics, Tsinghua University, Shuangqing Road, Beijing, China}

\author[0000-0002-8296-2590]{J.~Zhang}
\affil{Yunnan Observatories (YNAO), Chinese Academy of Sciences, Kunming 650216, China}
\affil{Key Laboratory for the Structure and Evolution of Celestial Objects, CAS, Kunming, 650216, China}

\author{W.~Zheng}
\affil{Department of Astronomy, University of California, Berkeley, CA 94720-3411, USA}

\begin{abstract}

We present analysis of the plateau and late-time phase properties of a sample of 39 Type II supernovae (SNe~II) that show narrow, transient, high-ionization emission lines (i.e., ``IIn-like'') in their early-time spectra from interaction with confined, dense circumstellar material (CSM). Originally presented by \cite{wjg24}, this sample also includes multicolor light curves and spectra extending to late-time phases of 35 SNe with no evidence for IIn-like features at $<$2~days after first light. We measure photospheric phase light-curve properties for the distance-corrected sample and find that SNe II with IIn-like features have significantly higher luminosities and decline rates at +50 days than the comparison sample, which could be connected to inflated progenitor radii, lower ejecta mass, and/or persistent CSM interaction. However, we find no statistical evidence that the measured plateau durations and $^{56}$Ni masses of SNe~II with and without IIn-like features arise from different distributions. We estimate progenitor zero-age main sequence (ZAMS) masses for all SNe with nebular spectroscopy through spectral model comparisons and find that most objects, both with and without IIn-like features, are consistent with progenitor masses $\leq12.5~\Msun$. Combining progenitor ZAMS masses with CSM densities inferred from early-time spectra suggests multiple channels for enhanced mass loss in the final years before core collapse such as a convection-driven chromosphere or binary interaction. Finally, we find spectroscopic evidence for ongoing ejecta-CSM interaction at radii $>10^{16}$~cm, consistent with substantial progenitor mass-loss rates of $\sim 10^{-4}$--$10^{-5}$~\mdot\ ($v_w < 50~\kms$) in the final centuries to millennia before explosion.

\end{abstract}


\section{Introduction} \label{sec:intro}

A fundamental goal in the study of Type II supernovae (SNe II) is to connect explosion properties to the evolution of red supergiant (RSG) stars in their final years before core collapse. After shock breakout and the rise to peak brightness, SNe~II are powered by hydrogen recombination and the cooling of a shocked, expanding RSG envelope during their light-curve ``plateau'' --- this phase being defined observationally by the plateau duration, brightness, and slope \citep{anderson14, Sanders15, valenti16, Martinez22}. Theoretically, variations in these observables map to differences in kinetic energy, progenitor radius, ejecta mass, and/or $^{56}$Ni mass \citep{Kasen06, dessart13, Hiramatsu21b}. By calibrating to numerical simulations, scaling relations between these quantities have been constructed that relate SN~II light-curve properties to regions of the progenitor/explosion parameter space despite known degeneracies \citep{kasen09, Dessart19SNII, Goldberg19}. However, SN ejecta interaction with intervening circumstellar material (CSM) can also impact the plateau phase \citep{Hillier19, Matsumoto25}.

As the ejecta turn optically thin (i.e., the nebular phase), SNe~II settle on their light-curve ``tail'' wherein the standard power source is the radioactive decay of ${}^{56}\rm{Ni} \rightarrow {}^{56}\rm{Co} \rightarrow {}^{56}$Fe \citep{arnett82}. Modeling of this post-plateau decline rate and luminosity allows for observational constraints on the total amount of $^{56}$Ni synthesized in explosive nucleosynthesis \citep{anderson19, Martinez22b}. Based on SN~II simulations, the total amount of $^{56}$Ni created after core collapse is highly dependent on the explosion energy as well as progenitor core structure, mass, and composition \citep{Sukhbold16, Burrows21, Curtis21, Laplace21, burrows24, Janka25}. However, construction of more three-dimensional (3D), end-to-end core-collapse SN simulations is necessary to understand consistencies with the observed distributions of SN~II observables, in particular $^{56}$Ni mass \citep{Vartanyan25}. In addition to the late-time light curve, nebular spectra are another powerful probe of the progenitor identity and its core mass prior to explosion \citep{Jerkstrand14, Dessart21}. Moreover, while radioactive decay is expected to be the main modulator of SN~II late-time luminosity, additional energy injection from the persistent collision of SN ejecta with distant CSM can lead to a flattening of the light curve and enhanced UV emission \citep{Weil20, Dessart23PWR}. Consequently, monitoring of CSM-interacting SNe~II at years to decades post-explosion is a direct tracer of the RSG mass-loss history in the final centuries to millenia before core collapse. 

To date, many large sample studies have attempted to constrain the SN~II progenitor parameter space using the observables described above (e.g., \citealt{valenti16, Gutierrez17a, Gutierrez17b, Martinez22}). However, such works do not use information gained about the CSM from very early-time ($<2$~days after first light) spectra as an additional prior that can be used to differentiate between objects included in SN~II samples. However, we now know that a significant fraction (possibly $>40\%$; \citealt{bruch23}) of SNe~II show narrow, high-ionization emission lines in their early-time spectra (i.e., ``IIn-like'' or ``flash'' features) from the collision of SN ejecta with confined, dense CSM created in the final years before core collapse \citep{Leonard00, fassia01, galyam14-1, Khazov16, yaron17, dessart17, bruch21, wjg24}. Consequently, SNe~II with confined CSM are likely typical in samples of SNe~II, which could have a significant impact on the correlations inferred from observables. 

Intriguingly, SNe~II with IIn-like features are best modeled by progenitor mass-loss rates of $\sim10^{-3}$--$10^{-1}$~\mdot\ (e.g., \citealt{groh14, shivvers15, Tartaglia21, terreran22, Bostroem23, Dessart23, wjg23, wjg24ggi, Zhang2023, Zhang24ggi, Shrestha24ggi, Zimmerman24}), which implies CSM densities that are orders of magnitude larger than those inferred for typical SNe~II (e.g., \citealt{Szalai19, Hosseinzadeh22, Pearson23, Shrestha24}) as well as populations of local RSGs \citep{Beasor20, Beasor22}. These enhanced CSM densities close to the progenitor star are also confirmed by early-time radio and X-ray observations of SNe~II with IIn-like features (e.g., \citealt{Berger23, Chandra24, Grefenstette23, Berger23, Panjkov23, Nayana24}). Given that these SNe evolve to resemble SNe~II both photometrically (e.g., light-curve plateau and radioactive-decay tail) and spectroscopically (e.g., Doppler-broadened P-Cygni absorption/emission), it is necessary to understand how the plateau/late-time properties of SNe~II with IIn-like features differ from SNe~II without early-time CSM interaction. Furthermore, comparison of late-time photometry and spectroscopy to model predictions enables constraints on the types of RSG progenitors associated with CSM-interacting SNe~II.


In this paper, we extend the analysis of the SN~II sample presented by \cite{wjg24} (hereafter WJG24a) to include properties of CSM-interacting SNe II during and after the light-curve plateau phase. In \S\ref{sec:obs}, we present specifics of the sample and additional late-time spectroscopy to be included in this study. \S\ref{sec:analysis} presents an analysis of the plateau phase, radioactive-decay decline, and nebular spectra of SNe~II with and without IIn-like features. Our results are discussed in \S\ref{sec:dis} and our conclusions are in \S\ref{sec:conclusion}. All phases reported in this paper are with respect to the adopted time of first light (WJG24a) and are in rest-frame days. When possible, we use redshift-independent host-galaxy distances and adopt standard $\Lambda$CDM cosmology (H$_{0}$ = 70~km~s$^{-1}$~Mpc$^{-1}$, $\Omega_M = 0.27$, $\Omega_{\Lambda} = 0.73$) if only redshift information is available for a given object. 

\section{Observations} \label{sec:obs}

This work utilizes the same sample of 74 SNe~II originally published by WJG24a. The sample includes 39 objects with IIn-like features in their early-time ($\delta t < 10$~days) spectra and 35 SNe~II with spectra obtained at $\delta t < 2$~days but showing no detectable IIn-like features (i.e., the ``comparison sample''). For objects with IIn-like features, we divide the total sample into two subsamples: ``gold sample'' events having spectra at $\delta t < 2$~days and ``silver sample'' events only having spectra at $\delta t > 2$~days. As in WJG24a, gold/silver objects are placed in three classes based on the properties of their IIn-like features: ``Class 1'' objects (plotted in blue) show narrow, high-ionization emission lines of \ion{N}{iii}, \ion{He}{ii}, and \ion{C}{iv}; ``Class 2'' objects (plotted in yellow) have no \ion{N}{iii} emission but do exhibit \ion{He}{ii} and \ion{C}{iv}; and ``Class 3'' objects (plotted in red) only show weaker, narrow \ion{He}{ii} emission superimposed on a blueshifted, Doppler-broadened \ion{He}{ii} line. Finally, when performing statistical tests, we only include SNe with distances $D > 40$~Mpc in order to obtain consistent distance/redshift distributions among  all subsamples. 

We adopt the sample multiband and pseudobolometric light curves presented by WJG24a in our analysis of the later-phase photometric evolution of all sample objects. Additional variations of the pseudobolometric light curves are generated for the $^{56}$Ni mass estimates (e.g., \S \ref{sec:ni56}). In addition to the photospheric-phase spectra presented by WJG24a, we now include nebular spectra of gold-, silver-, and comparison-sample objects (e.g., \S \ref{sec:nebspec}). Nebular spectra were obtained with the Kast spectrograph on the 3~m Shane telescope at Lick Observatory \citep{KAST} and Keck/LRIS \citep{oke95}. For all of these spectroscopic observations, standard CCD processing and spectrum extraction were accomplished with \textsc{IRAF}\footnote{\url{https://github.com/msiebert1/UCSC\_spectral\_pipeline}}. The data were extracted using the optimal algorithm of \citet{1986PASP...98..609H}. Low-order polynomial fits to calibration-lamp spectra were used to establish the wavelength scale and small adjustments derived from night-sky lines in the object frames were applied. Additional spectra were obtained with Keck/DEIMOS, Binospec on the MMT \citep{Fabricant19}, and Gemini Multi-Object Spectrographs (GMOS). These spectroscopic observations were reduced with a variety of pipelines such as {\tt Pypeit} \citep{Prochaska20}, {\tt Lpipe} \citep{Perley19}, and {\tt DRAGONS} \citep{dragons}. A log of previously unpublished observations is presented in Table \ref{tab:spec_all}. 

\begin{figure*}[t!]
\centering
\subfigure{\includegraphics[width=0.51\textwidth]{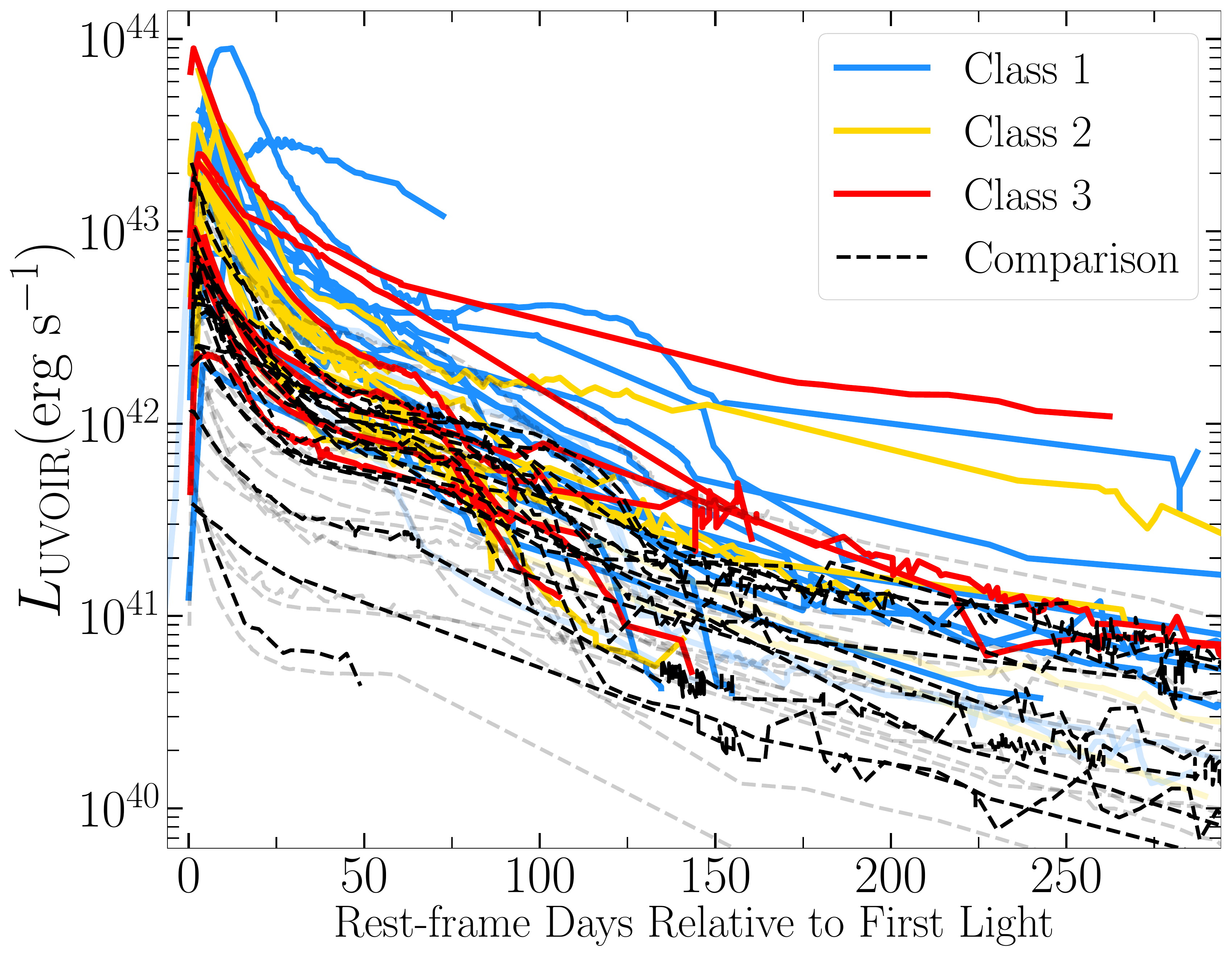}}
\subfigure{\includegraphics[width=0.485\textwidth]{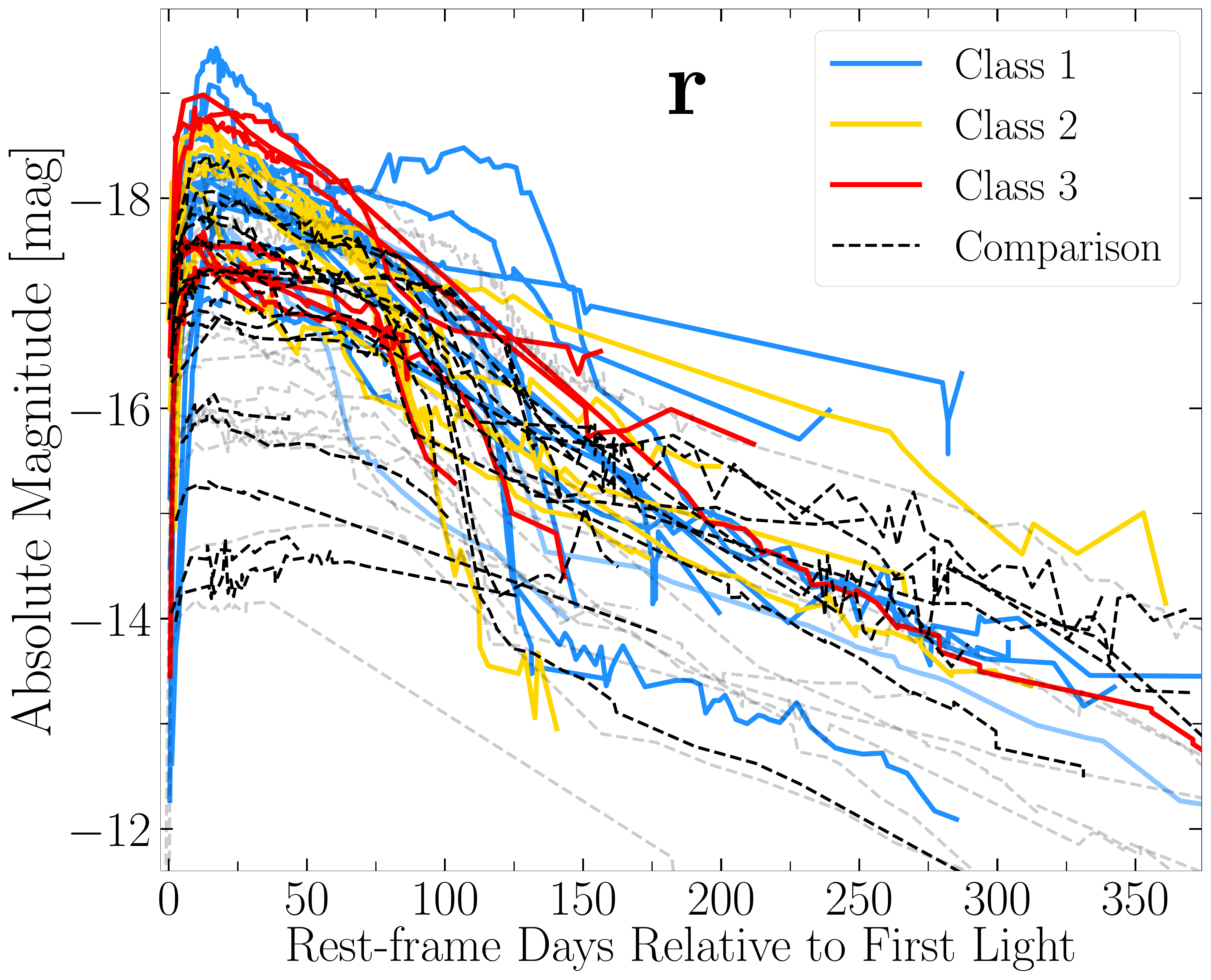}}
\caption{{\it Left:} Pseudobolometric (i.e., UVOIR) light curves of gold/silver samples (blue/yellow/red solid lines) and the comparison sample (dashed black lines). Solid black and colored points/curves represent the subsample of objects at $D>40$~Mpc. Objects at $D<40$~Mpc are plotted as lower opacity curves/points. {\it Right:} Extinction-corrected \textit{r}-band light curves of gold/silver- and comparison-sample objects. Compared to SNe~II without IIn-like features (i.e., comparison sample), objects with confirmed IIn-like signatures are more luminous during the plateau phase.     
\label{fig:Lbol_Mr}}
\end{figure*}

\section{Analysis} \label{sec:analysis}
\subsection{Photospheric Phase Light-Curve Properties} \label{sec:plateau}

\begin{figure*}[t!]
\centering
\subfigure{\includegraphics[width=0.50\textwidth]{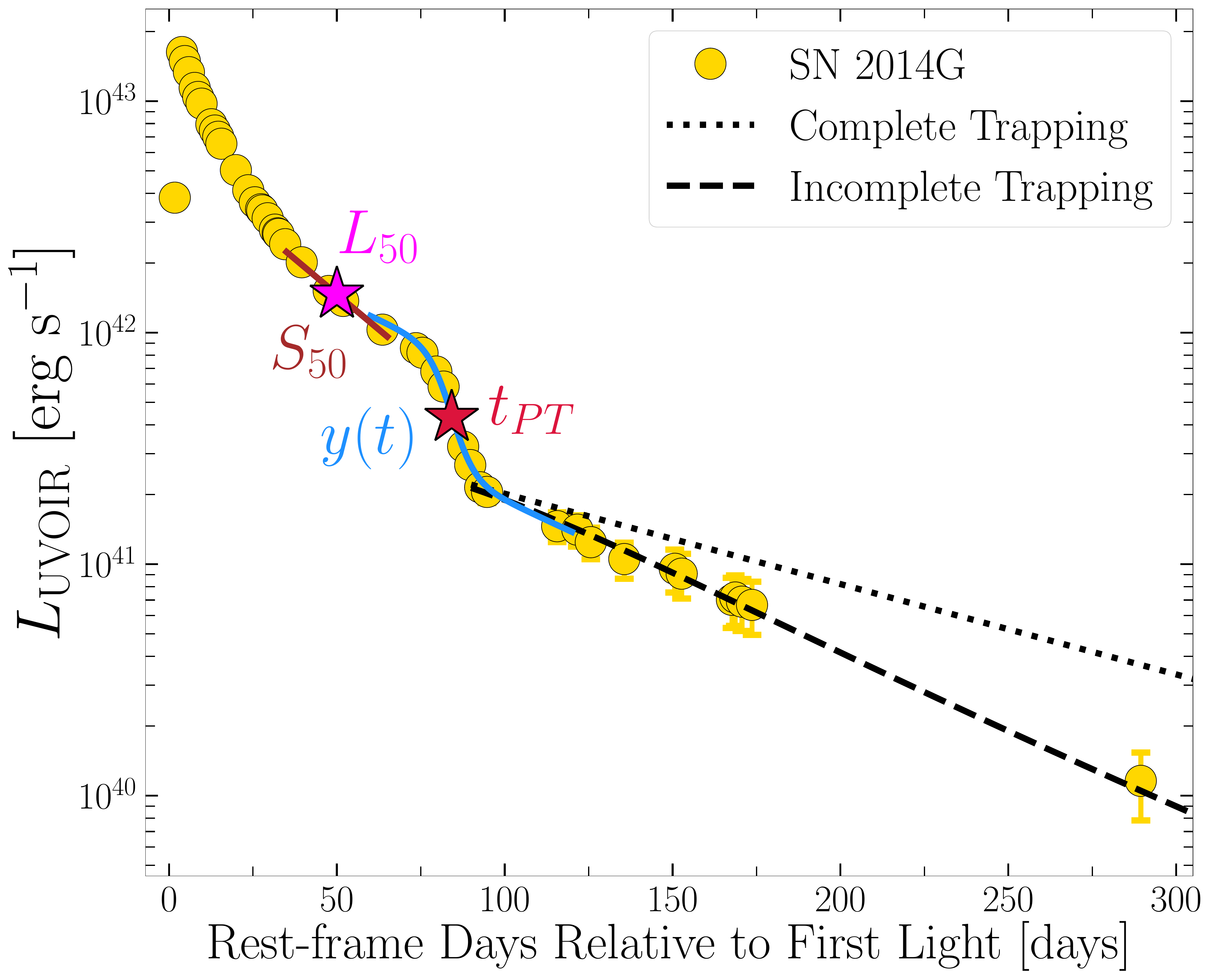}}
\subfigure{\includegraphics[width=0.495\textwidth]{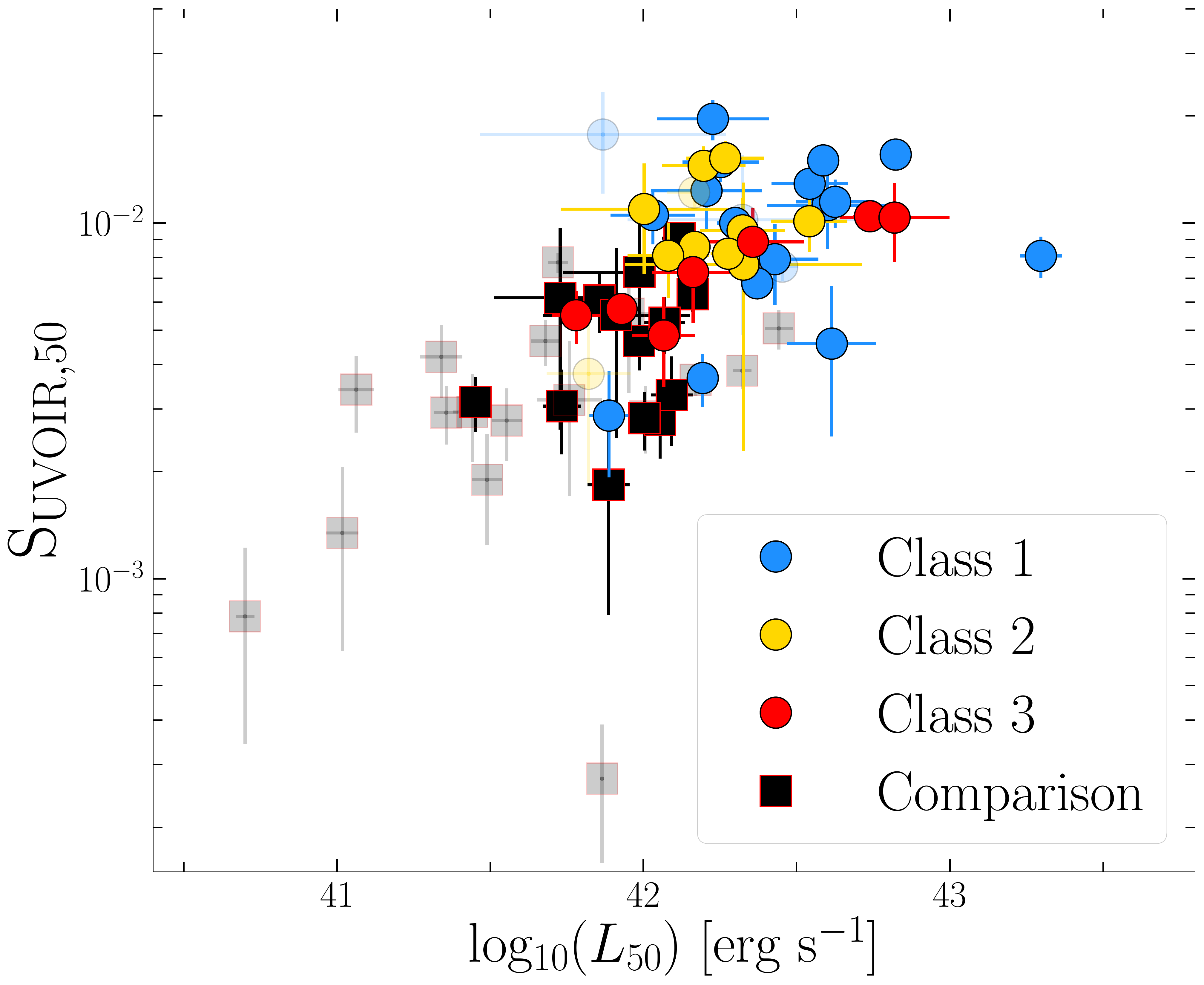}}\\
\subfigure{\includegraphics[width=0.485\textwidth]{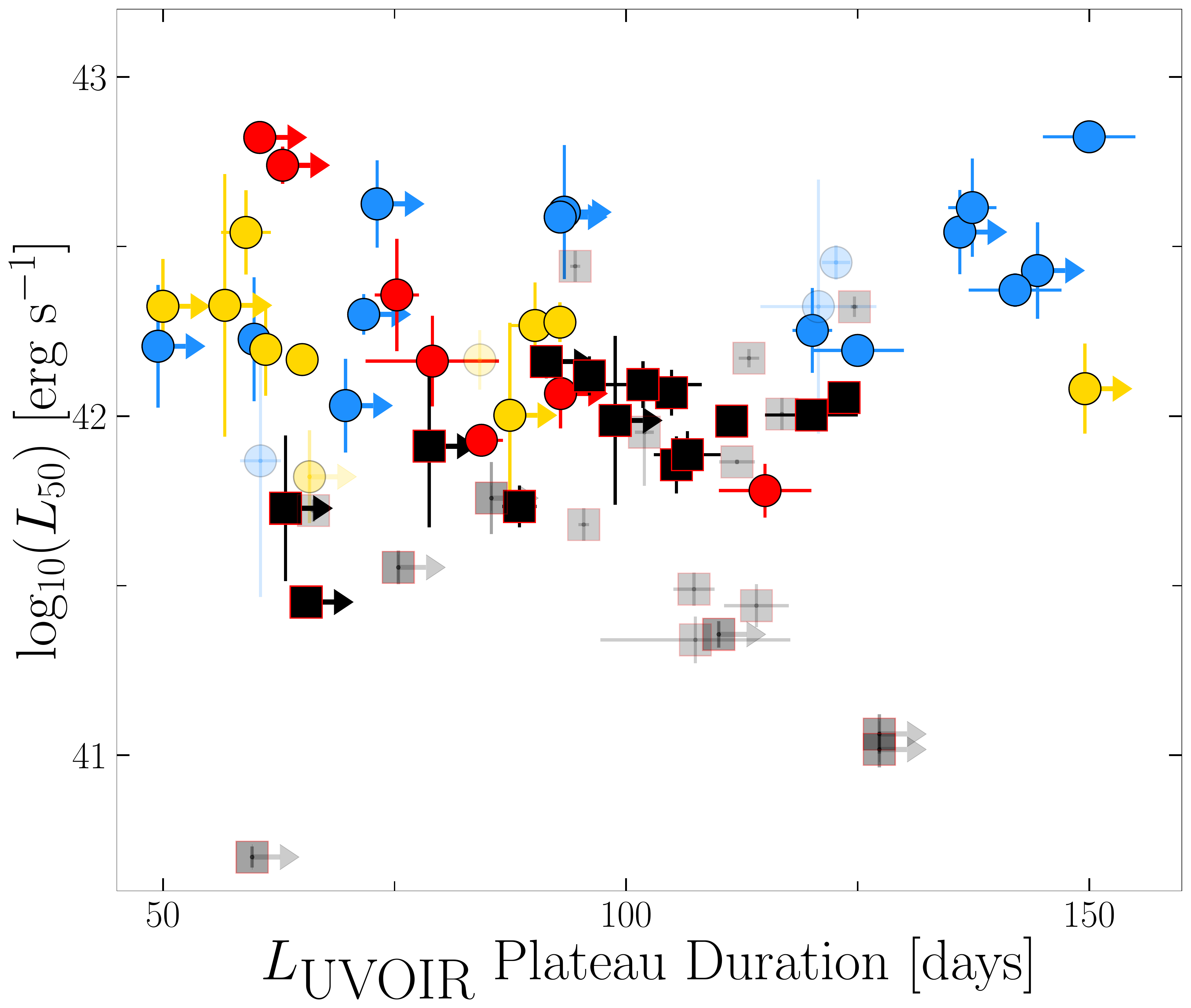}}
\subfigure{\includegraphics[width=0.50\textwidth]{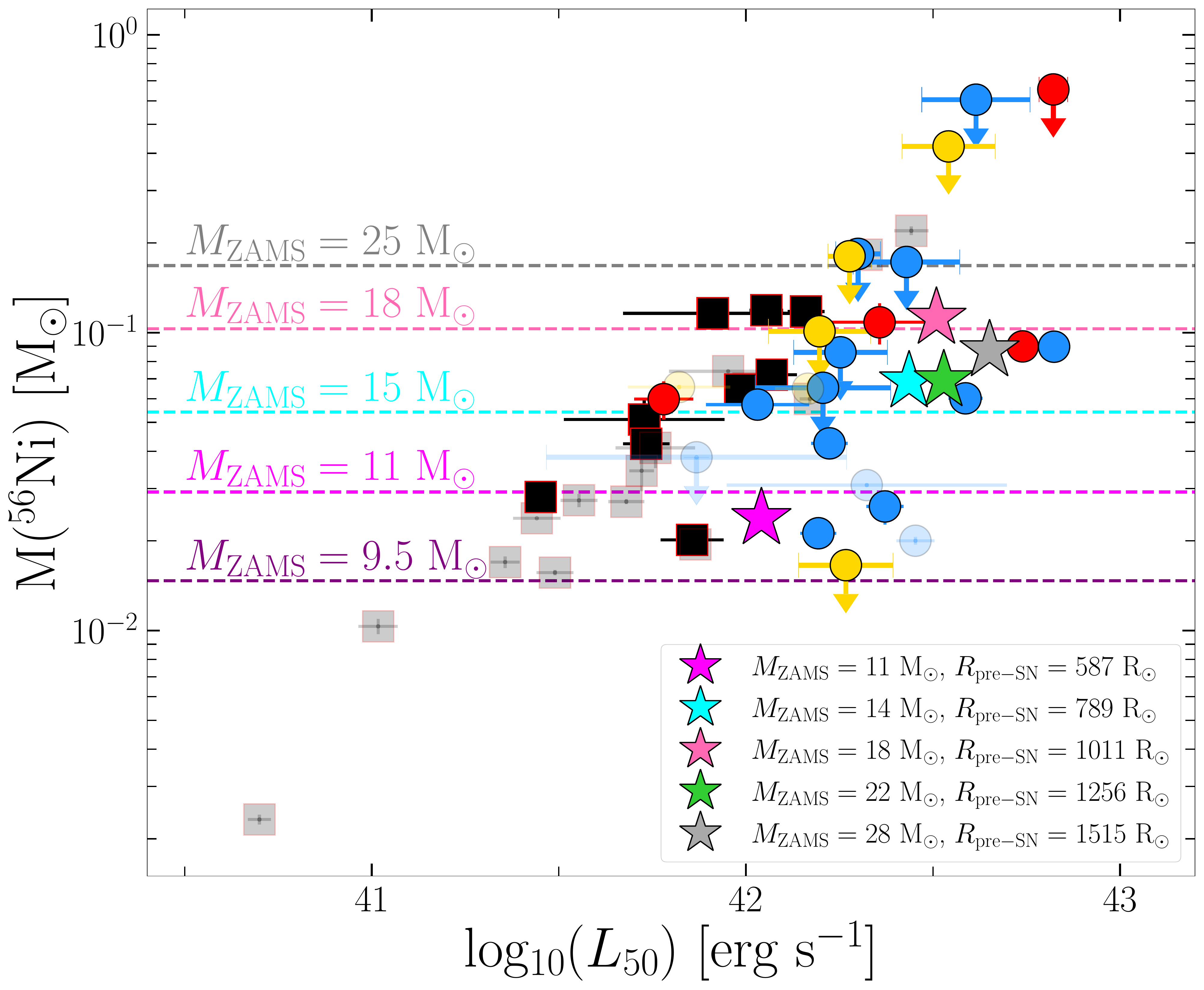}}
\caption{{\it Top left:} Pseudobolometric light curve of Class 2 object SN~2014G (yellow circles) with light-curve parameters derived for all sample objects (\S\ref{sec:plateau}) labeled for visualization: the luminosity at 50 days ($L_{50}$, magenta star), light-curve decline rate at 50 days ($S_{50}$, brown solid line), photospheric-phase duration ($t_{\rm PT}$, red star), and Fermi-Dirac function ($y(t)$, solid blue line). Additionally, radioactive decay powered models for complete and incomplete $\gamma$-ray trapping are plotted as dotted and dashed black lines. {\it Top right:} Pseudobolometric light curve luminosity and decline rate at $\delta t = 50$~days for gold/silver- (blue, yellow, red circles) and comparison-sample (black squares) objects. SNe~II with IIn-like features are more luminous and decline faster during their photospheric phase. Solid colored points/curves represent the subsample of objects at $D>40$~Mpc. {\it Bottom left:} Luminosity at 50 days ($L_{50}$) versus photospheric phase duration. {\it Bottom right:} $^{56}$Ni mass versus pseudobolometric luminosity at +50 days. Theoretical predictions for $^{56}$Ni yields from 9.5--25~$\Msun$ \citep{burrows24} progenitor models shown as dashed lines and 11--28~$\Msun$ \citep{Curtis21} progenitor models shown as stars.
\label{fig:Lbol_tp_S50_L50} }
\end{figure*}

We present the complete pseudobolometric and $r$-band light curves for gold-, silver- and control-sample objects in Figure \ref{fig:Lbol_Mr}. For a number of events in each subsample, we are able to estimate the plateau (i.e., ``photospheric phase'') luminosity, duration, and decline rate during this phase. Similar to other sample studies of SNe~II (e.g., \citealt{anderson14,valenti16}), we attempt to measure specific parameters that define the light-curve photospheric phase for all objects where the observations allow it. As shown in the upper-left panel of Figure \ref{fig:Lbol_tp_S50_L50}, we measure the pseudobolometric luminosity at $\delta t = 50$~days ($L_{50}$), the light-curve slope at $\delta t = 50$~days ($S_{50}$), and the photospheric phase duration as inferred from the fitting function $y(t)$ (e.g., \citealt{valenti16}). The $L_{50}$ and $S_{50}$ parameters are calculated through a linear function fit to the light-curve plateau, while $t_{\rm PT}$ is measured after fitting the light curve with the $y(t)$ function. We present these quantities for each class in the gold/silver samples (37 out of 39 objects) as well as the comparison sample (31 out of 35 objects) in Figure \ref{fig:Lbol_tp_S50_L50}. Notably, SNe~II with IIn-like features display steeper light curves (i.e., ``IIL-like'') than comparison-sample objects during the photospheric phase in addition to being more luminous than SNe~II without any IIn-like features. However, there is no obvious correlation present when comparing $L_{50}$ to the plateau duration;  gold/silver-sample objects show a significant diversity in plateau durations, similar to comparison-sample objects. It is worth noting that for the objects with constrained (i.e., not lower limits) plateau durations, seven SNe~II with IIn-like features have $t_{\rm PT} < 80$~days (SNe PTF10abyy, 2017ahn, 2021afkk, 2021can, 2021ont, 2020lfn, and 2021aaqn), with only one comparison-sample object (SN~2020jfo) having such a short plateau duration. Furthermore, six (three detections, three lower limits) gold-sample objects have plateau durations $>130$~days (SNe~2021aek, 2021dbg, 2021tyw, 2021zj, 2022pgf, 2016blz), with no comparison-sample objects showing such long plateau durations. We discuss the physical interpretation of these photospheric phase properties in \S\ref{sec:dis}. 

\begin{figure*}[t!]
\centering
\subfigure{\includegraphics[width=0.34\textwidth]{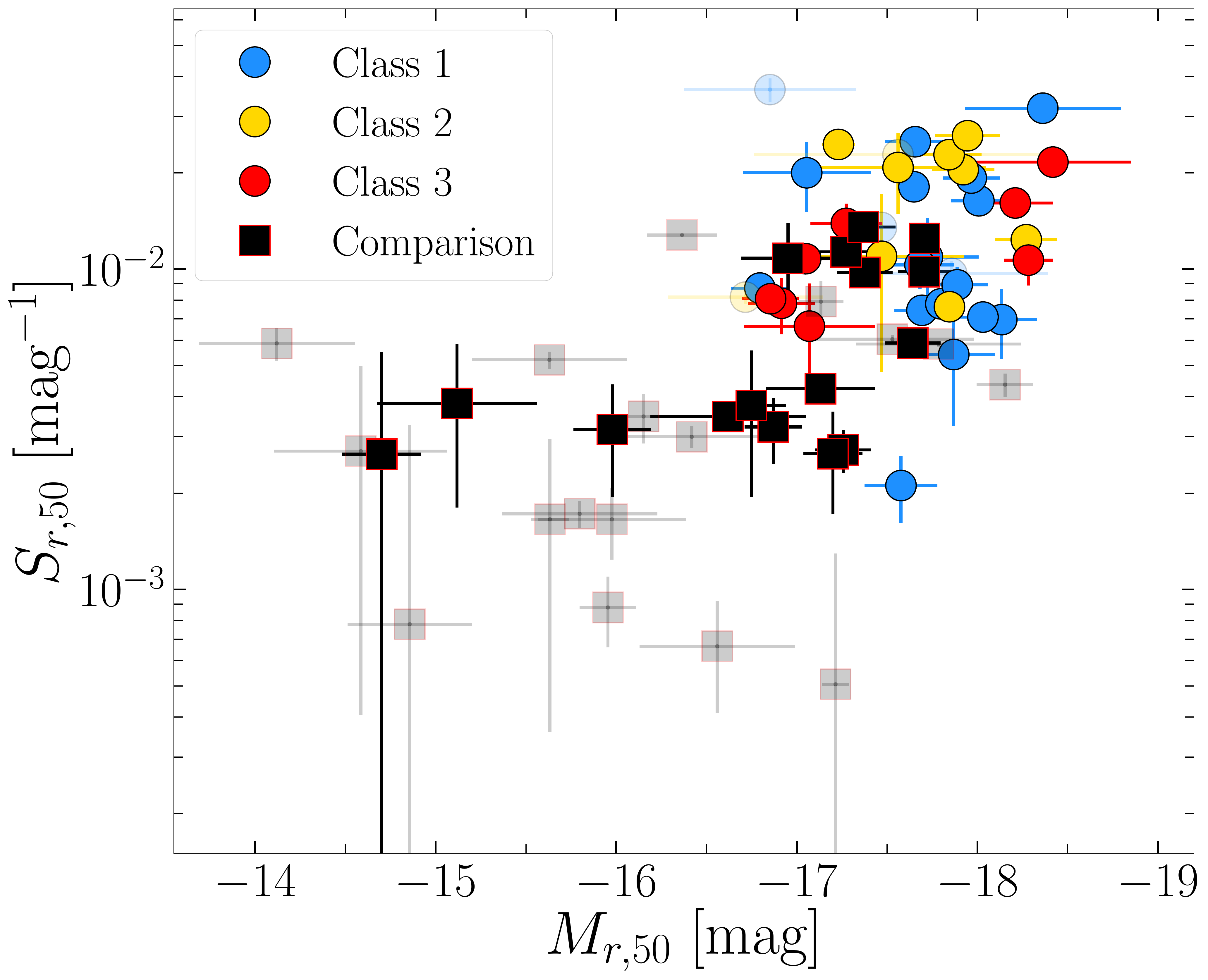}}
\subfigure{\includegraphics[width=0.32\textwidth]{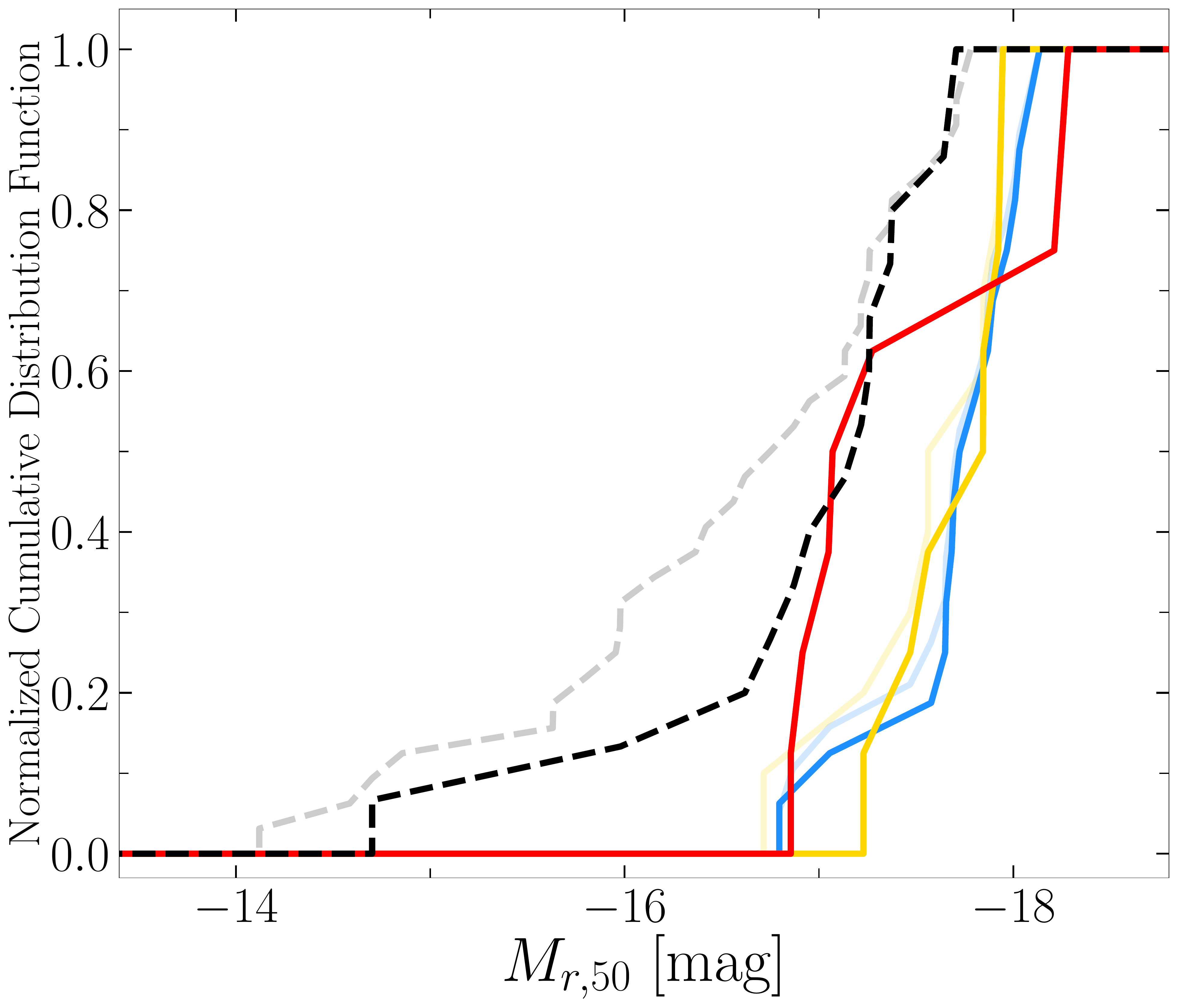}}
\subfigure{\includegraphics[width=0.325\textwidth]{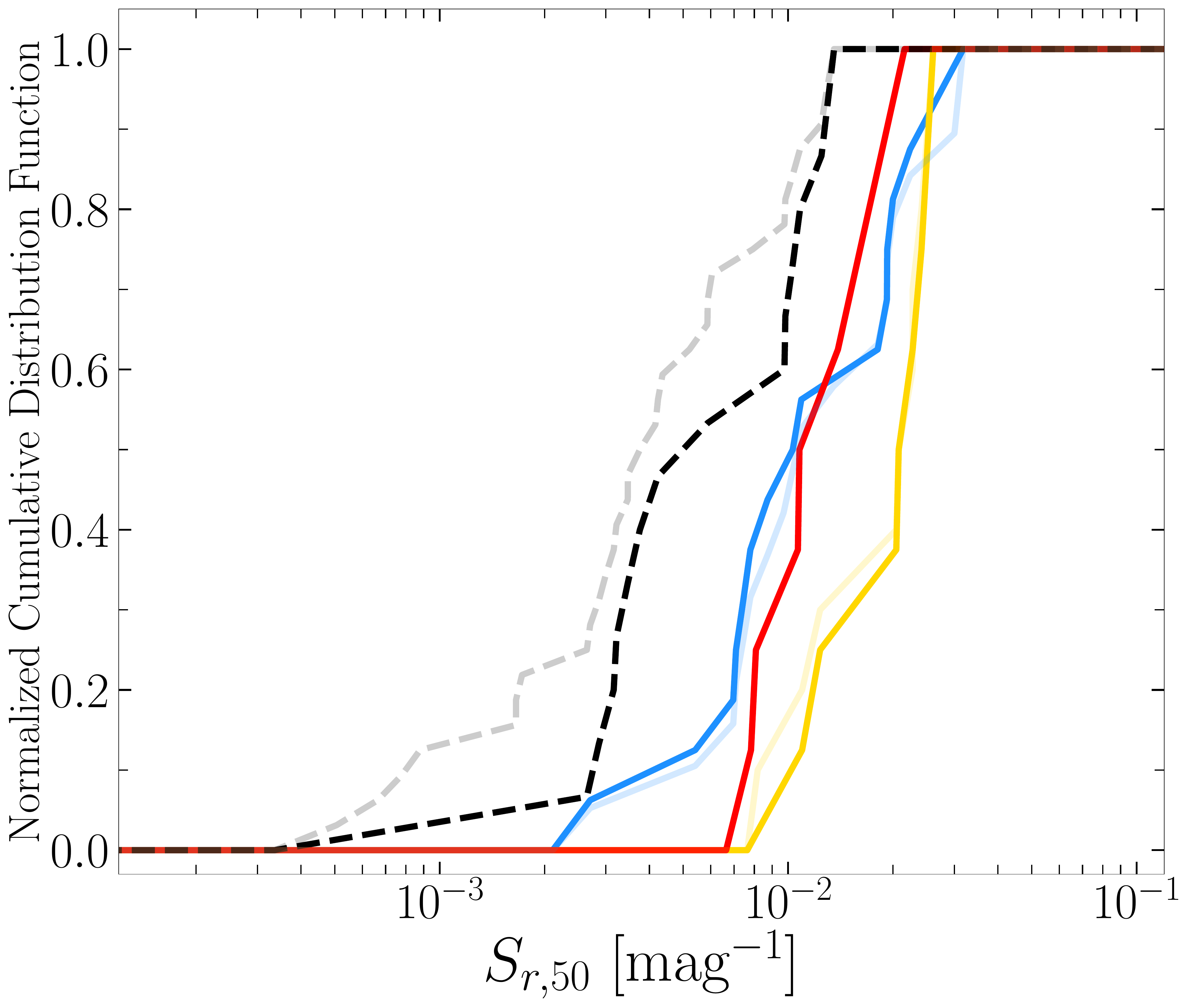}}\\
\subfigure{\includegraphics[width=0.34\textwidth]{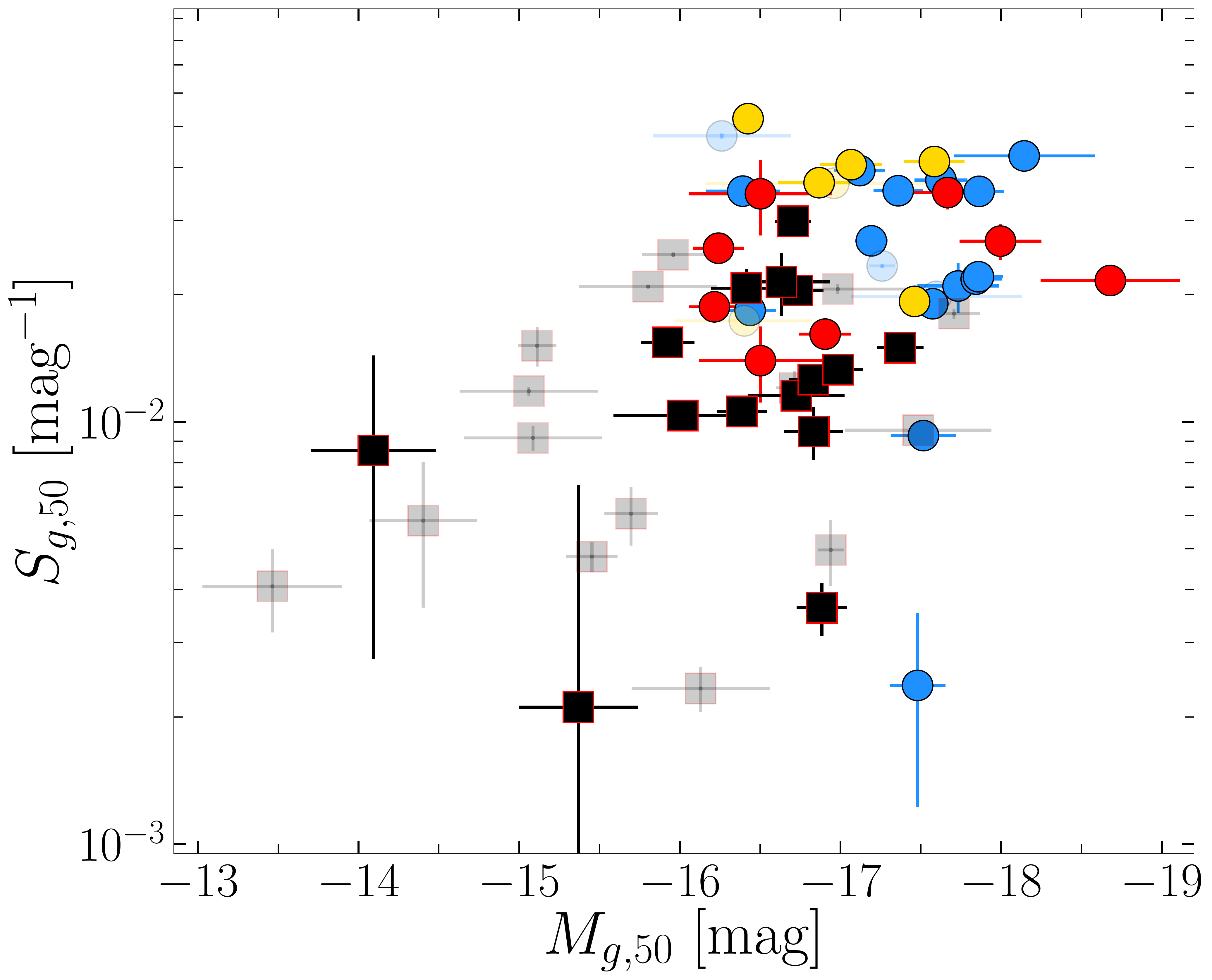}}
\subfigure{\includegraphics[width=0.32\textwidth]{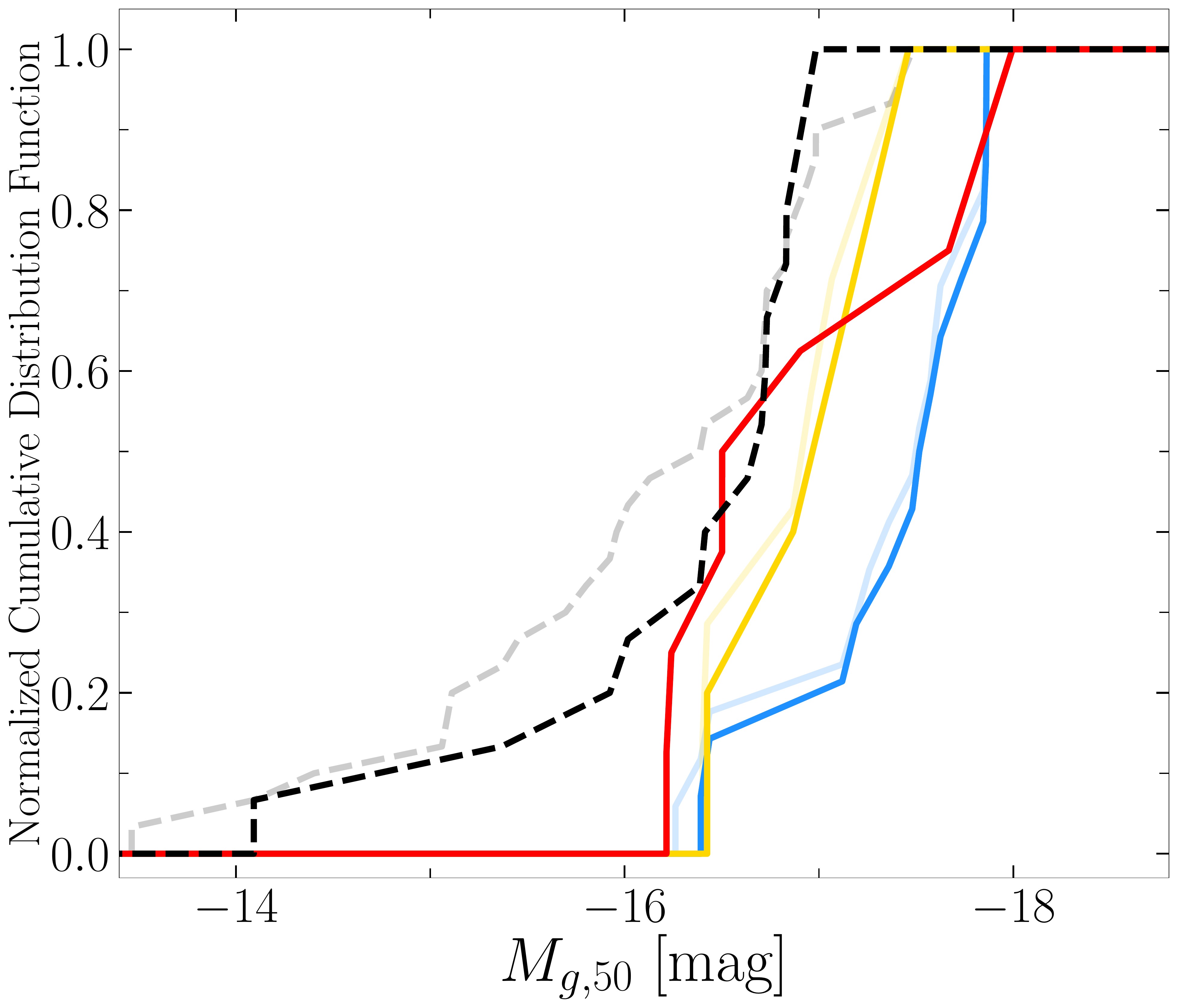}}
\subfigure{\includegraphics[width=0.325\textwidth]{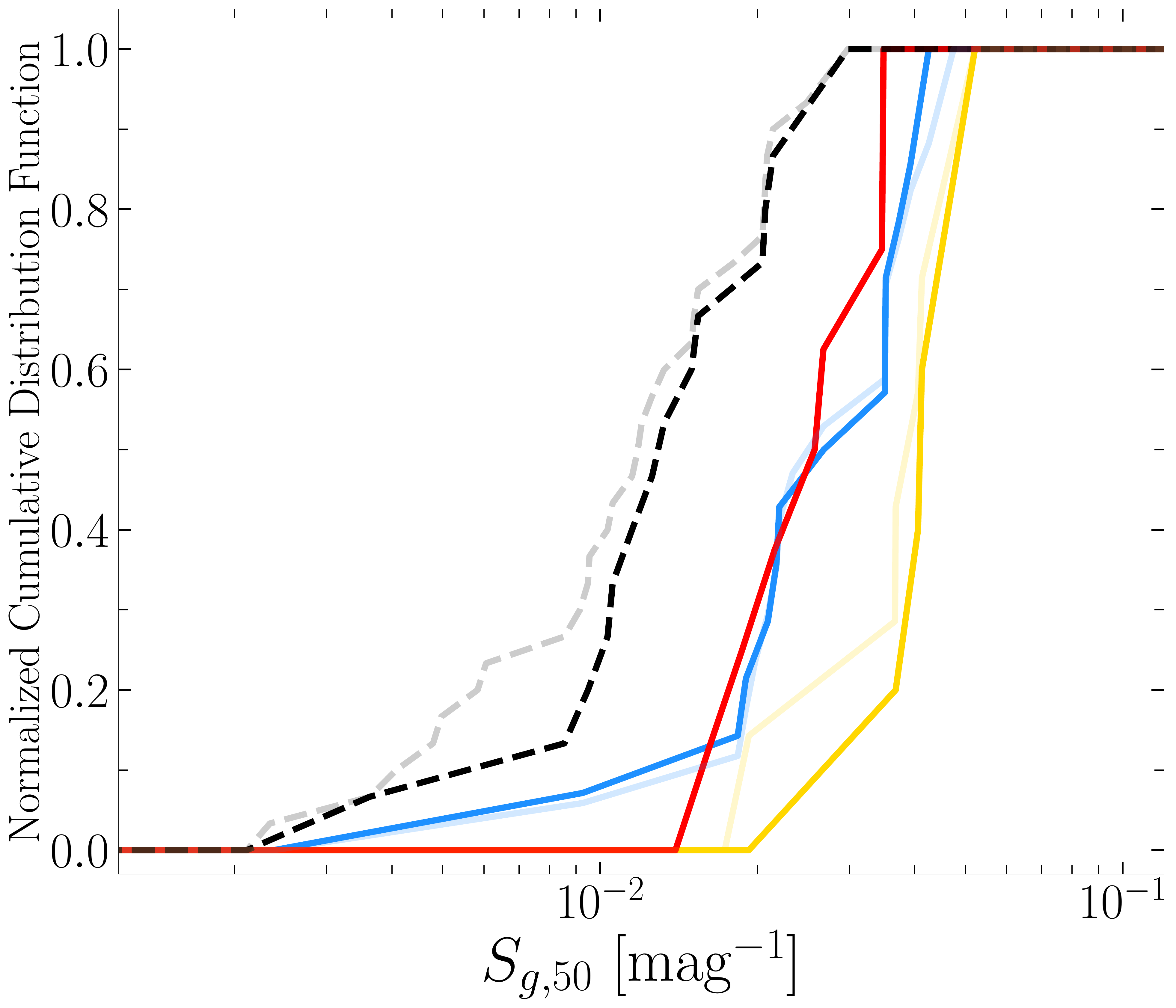}}
\caption{ Light-curve decline rate versus absolute magnitude at 50 days for $r$- (top panel) and $g$-band (bottom panel) light curves. Solid colored points/curves represent the subsample of objects at $D>40$~Mpc. Cumulative distributions of $r$- (top panel) and $g$-band (bottom panel) light-curve parameters shown for comparison- (black dashed lines) and gold/silver-sample (solid blue, yellow and red lines) objects. SNe~II with IIn-like features have a distinct distribution of $S_{50}$ and $M_{\rm 50}$ compared to comparison-sample objects.  
\label{fig:Mrg_S50} }
\end{figure*}

\begin{figure}[t!]
\centering
\includegraphics[width=0.49\textwidth]{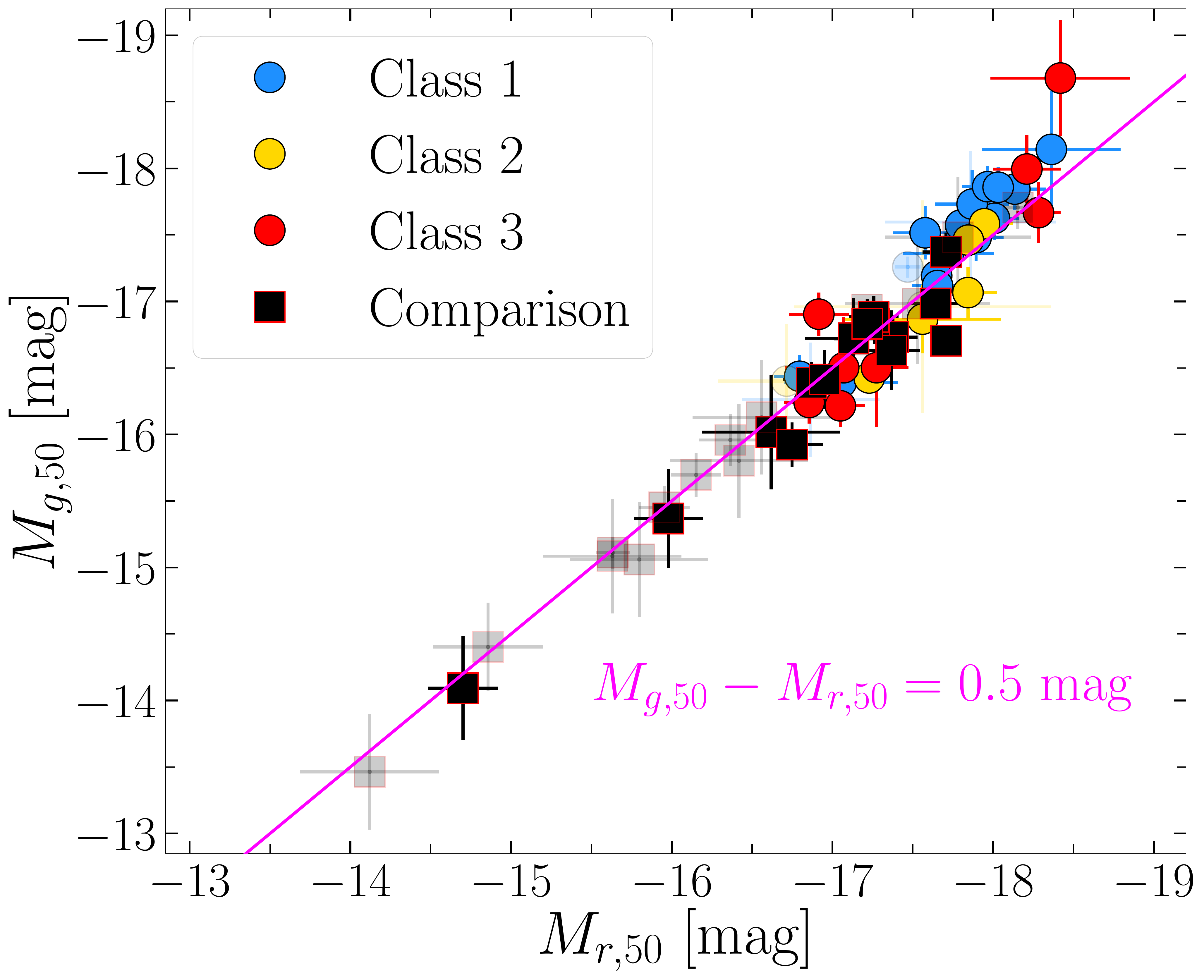}
\caption{ Absolute $g$- and $r$-band magnitudes at +50~days for gold/silver- (blue, yellow, and red circles) and comparison-sample (black squares) objects. $M_g - M_r = 0.5$~mag color difference shown as magenta line. Some brighter SNe~II with IIn-like features show bluer colors during the plateau phase, possibly revealing extra heating from ongoing CSM interaction.
\label{fig:MrMg} }
\end{figure}

\begin{figure*}[t!]
\centering
\subfigure{\includegraphics[width=0.33\textwidth]{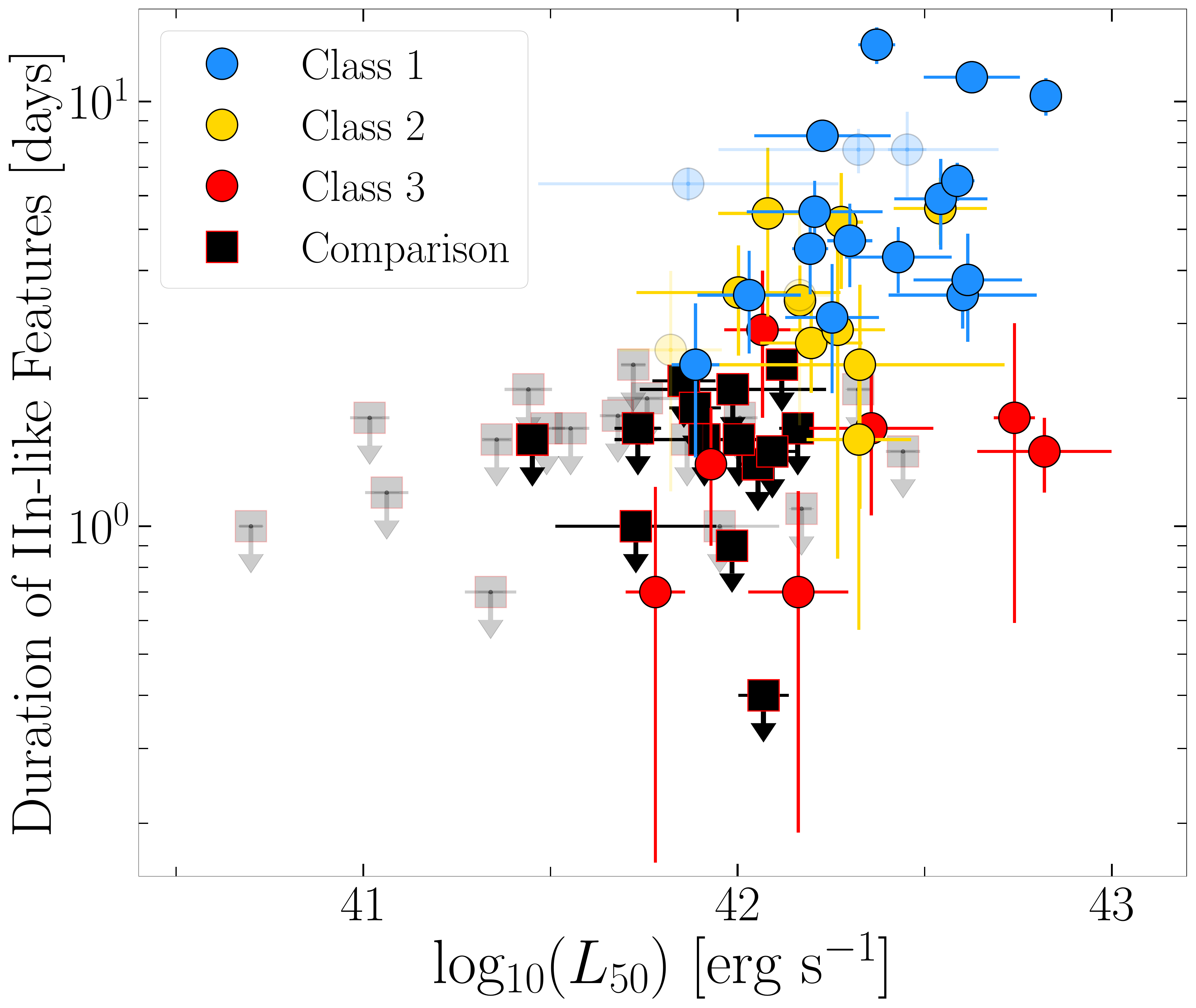}}
\subfigure{\includegraphics[width=0.33\textwidth]{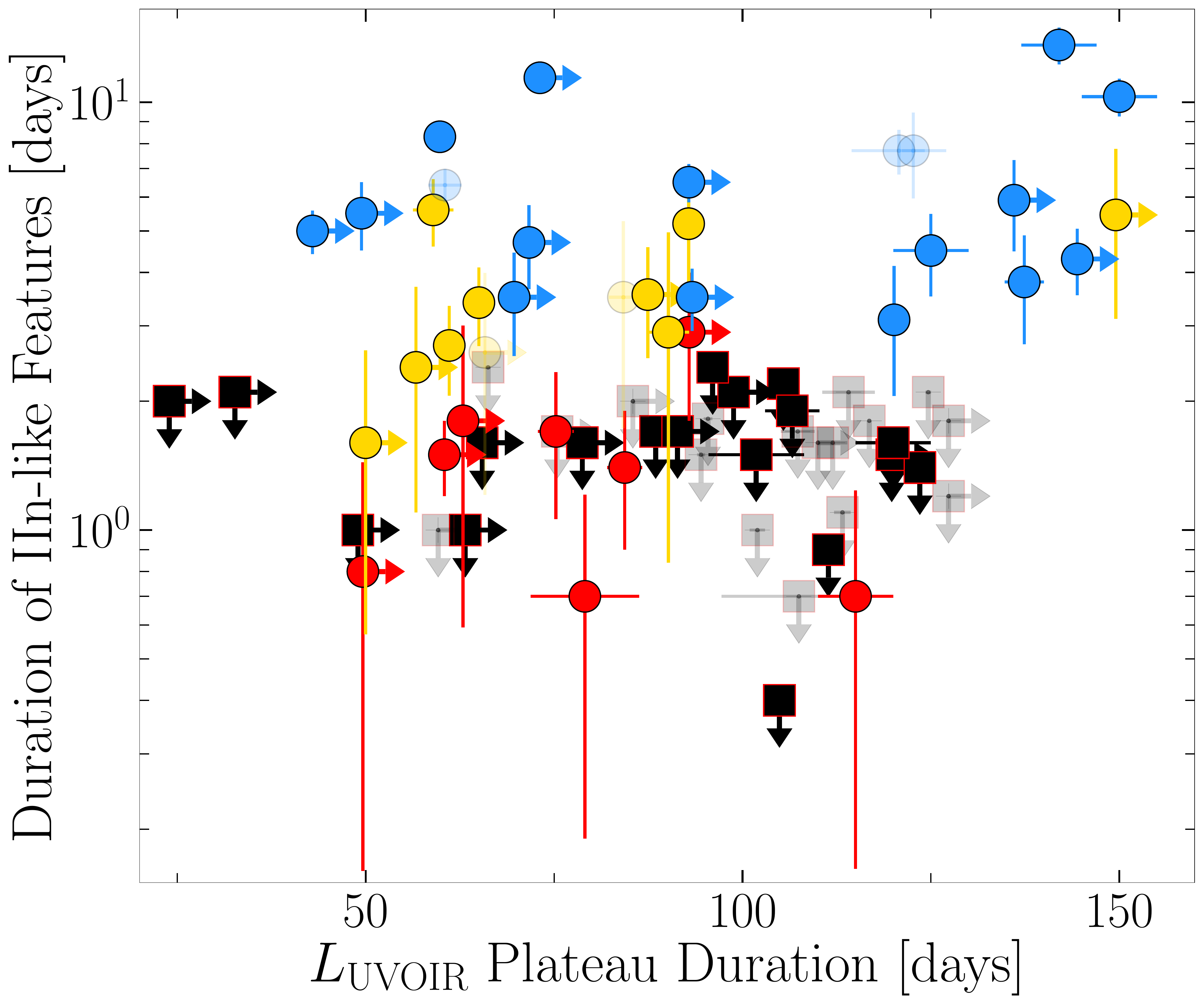}}
\subfigure{\includegraphics[width=0.33\textwidth]{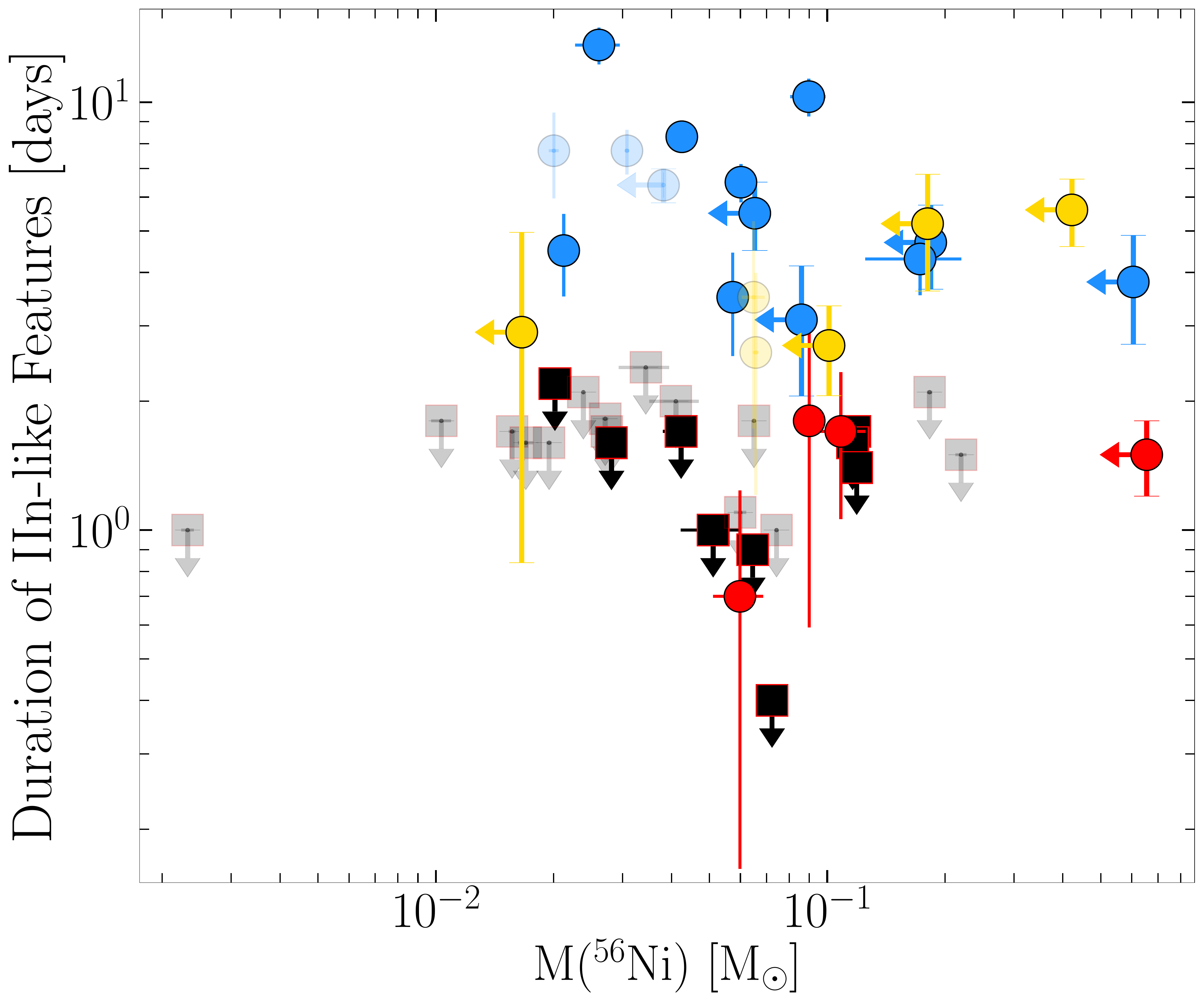}}\\
\subfigure{\includegraphics[width=0.33\textwidth]{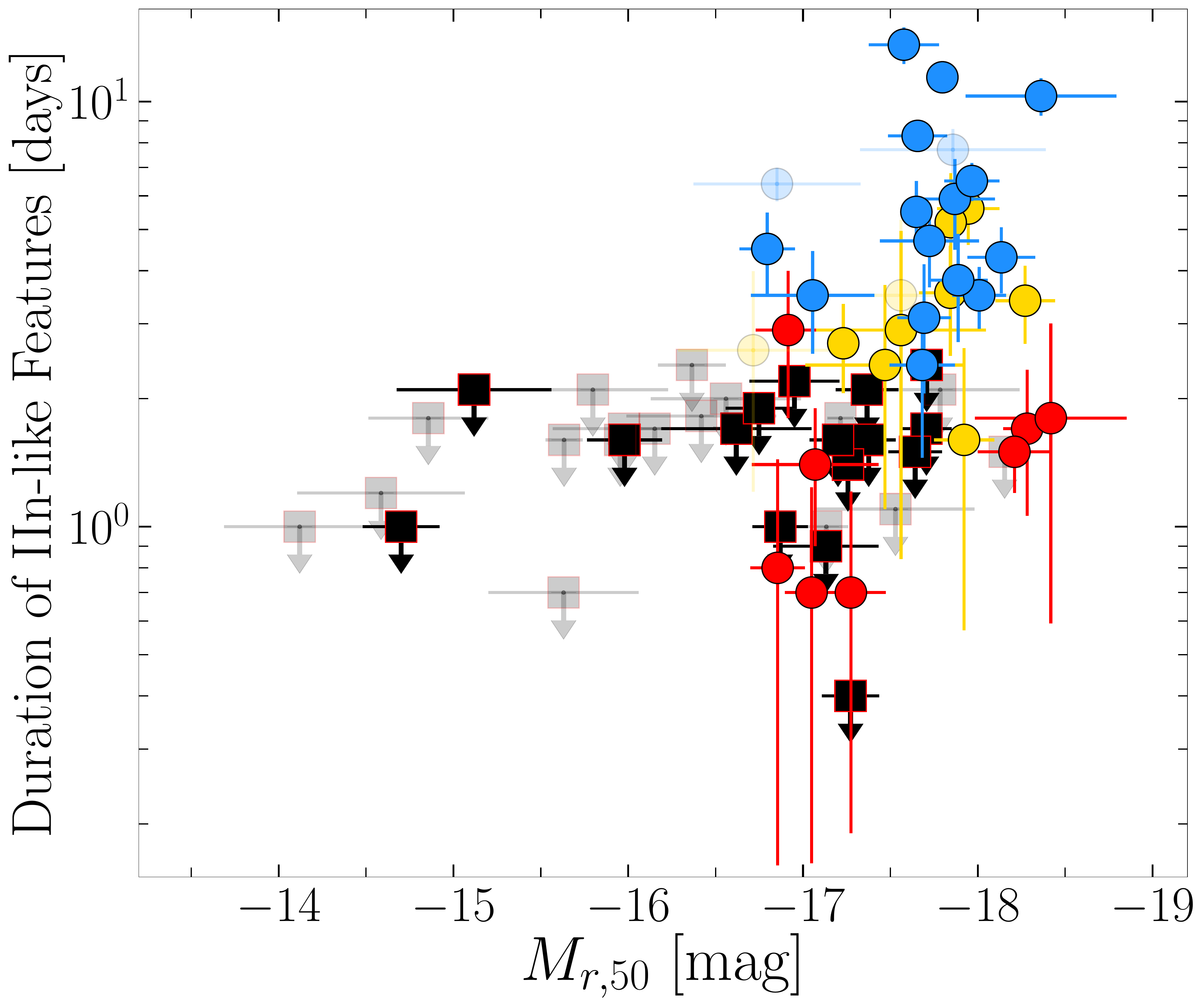}}
\subfigure{\includegraphics[width=0.33\textwidth]{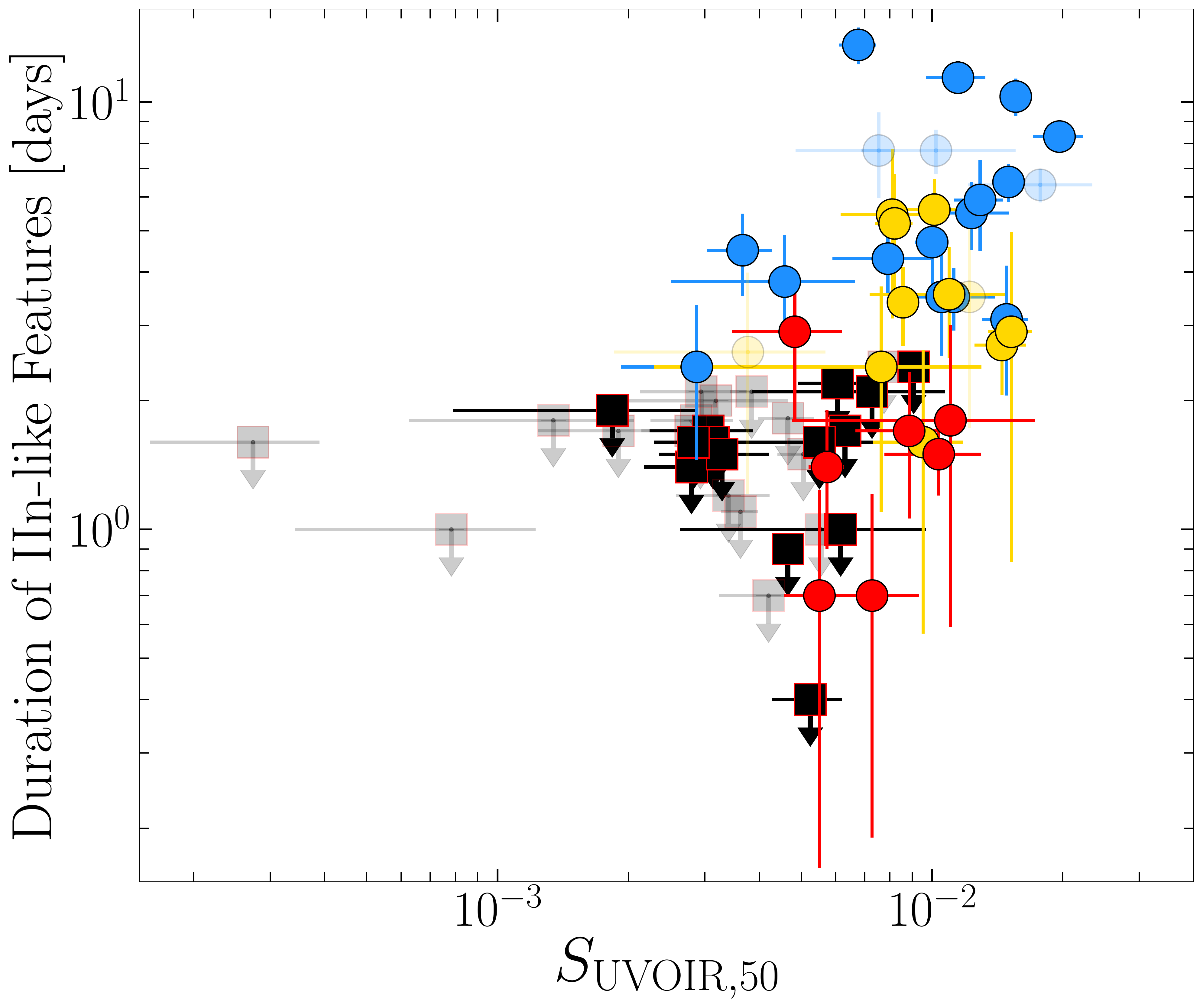}}
\subfigure{\includegraphics[width=0.33\textwidth]{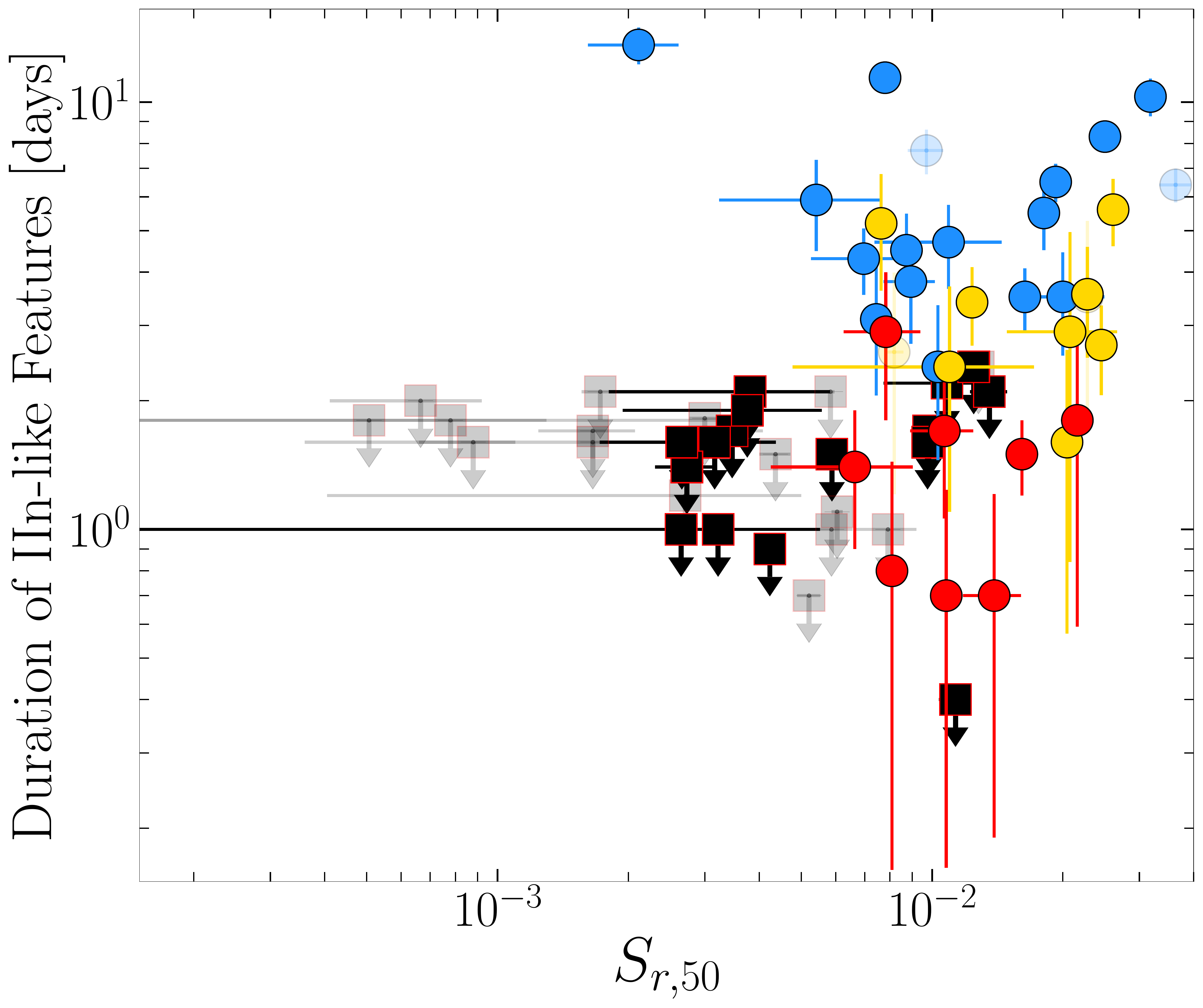}}
\caption{ Duration of IIn-like features for gold/silver- (blue, yellow, red circles) and comparison-sample (black squares) objects versus $L_{50}$, $t_{\rm pl}$, and $M(^{56}{\rm Ni})$ (top panel), and $M_{r,50}$, $S_{\rm UVOIR,50}$, and $S_{r,50}$  (bottom panel). SNe~II with IIn-like features have statistically distinct distributions for plateau brightness and decline rate but not for $^{56}$Ni mass or plateau duration.   
\label{fig:tIIn} }
\end{figure*}

Similar to the pseudobolometric light curves, we  measure the plateau brightness and decline rate in the $g$ (32 gold/silver and 30 comparison) and $r$ bands (37 gold/silver and 32 comparison). As shown in Figure \ref{fig:Mrg_S50}, there also exists a trend between the absolute magnitude and decline rate at $\delta t = 50$~days: gold/silver-sample objects are the most luminous and fast-declining during their plateau evolution compared to comparison-sample objects. We present cumulative distributions of the plateau brightness and decline rate at $\delta t = 50$~days in $g$ and $r$ in Figure \ref{fig:Mrg_S50}. As shown in Figure \ref{fig:MrMg}, all comparison-sample events have a similar $g-r$ color at $\delta t = 50$~days. While some gold/silver-sample objects also show consistent colors at this phase, a number of SNe~II with IIn-like features are bluer than normal SNe~II at the same phase, which could be the result of subdominant but still significant CSM-interaction power contributing to the SN spectral energy distribution (SED) at later-time phases \citep{dessart22}. Furthermore, we present all light-curve parameters discussed above with respect to the IIn-like feature timescale in Figure \ref{fig:tIIn}.

To test our null hypothesis of whether these sample observables come from the same parent distribution, we apply a logrank test for (i) gold/silver vs. comparison samples, (ii) gold/silver-sample Classes 1 \& 2 vs. 3, and (iii) gold/silver-sample Classes 1 vs. 3. These tests are only performed on the subsample of objects where $D>40$~Mpc. For (i), the chance probability that values of the gold/silver and comparison samples come from the same distribution is $<10^{-3}\%$ for $L_{50}$, $0.24\%$ for $M_{g, 50}$, and $0.11\%$ for $M_{r, 50}$. We find that $S_{50}$ for pseudobolometric, $g$, and $r$ between samples belong to the same distribution at the 0.01\%, $6\times10^{-4}$\%, and 0.02\% levels, respectively. For (ii), the null-hypothesis probability for pseudobolometric, $g$, and $r$  brightness (decline rate) at $\delta t = 50$~days is 76.8(33.9)\%, 13.7(2.9)\%, and 17.9(27.2)\%, respectively. For (iii), the null-hypothesis probability for pseudobolometric, $g$, and $r$  brightness (decline rate) at $\delta t = 50$~days is 47.1(60.2)\%, 13.0(11.9)\%, and 33.8(74.1)\%, respectively. Therefore, we confirm that all classes in the gold/silver samples are more luminous and decline faster on the plateau than comparison-sample objects. However, we find no statistically significant differences for these parameters between classes in the gold/silver samples. 

\subsection{$^{56}$Ni Masses} \label{sec:ni56}

In noninteracting SNe~II, the main power source for the ejecta is radioactive decay, of which the ${}^{56}\rm{Co} \rightarrow {}^{56}$Fe \citep{arnett82} chain dominates at the epochs of $\sim 100$--350~days studied here. In this section, we first assume that the only power source for the whole sample of objects is $^{56}$Ni decay power before demonstrating that this assumption breaks down for a number of objects with clear signs of late-time CSM interaction. In this framework of an exclusive decay power source, both the total mass of $^{56}$Ni present in the SN as well as the timescale of $\gamma$-ray escape ($t_{\gamma}$) can be measured by modeling the bolometric luminosity evolution post-plateau (e.g., \citealt{Cappellaro97, clocchiatti97,wheeler15, valenti08a, wjg21}). However, to properly utilize the analytic formalisms for radioactive-decay power, the bolometric light curve should be constructed with multiband photometry that covers ultraviolet (UV) through infrared (IR) wavelengths. Given that the pseudobolometric light curves presented in this Figure \ref{fig:Lbol_Mr} only include UV through optical/near-IR wavelengths, all resulting measurements of $M(^{56} \rm Ni)$ will be underestimates of the true amount of synthesized radioactive material powering the SN light curve. We do not construct the late-time bolometric light curve through fitting of the SED with a blackbody model because during post-plateau phases the SN spectrum will transition to the nebular phase (i.e., emission-line dominated) and therefore a blackbody approximation is no longer appropriate.


To measure $M(^{56} \rm Ni)$ and $t_{\gamma}$ in gold/silver- and comparison-sample objects, we compare to the pseudobolometric light curve of SN~1987A, as has been done in previous SN~II sample studies (e.g., \citealt{spiro14, valenti16}). To do this, we construct new pseudobolometric light curves using combinations of optical filters (e.g., $gri$, $UBVRI$, $BVgri$, etc.) and fit these light curves with the pseudobolometric light curve of SN~1987A derived with the same filter combination (see \citealt{Meza-Retamal24} for more information). This SN~1987A light-curve model is represented in the form 

\begin{equation}
    \frac{L_{\rm pbol}(t)}{L^{\rm 87A}_{\rm pbol}(t)} = \Bigg ( \frac{M_{\rm ^{56} Ni}}{0.075} \Bigg ) \Bigg ( \frac{1 - {\rm exp}(-(t_{\gamma}/t)^2)}{1 - {\rm exp}(-(540/t)^2)} \Bigg )\, ,
\end{equation}

\noindent
where $t$ is time in days post-explosion, $L_{\rm pbol}$ is the pseudobolometric luminosity, and the SN~1987A $\gamma$-ray trapping timescale is 540~days \citep{Suntzeff90}.

We present measured $M(^{56} \rm Ni)$ values for gold/silver (24 out of 39 objects) and comparison samples (24 out of 39 objects) in Figure \ref{fig:Lbol_tp_S50_L50} and Tables \ref{tab:sample_phot_gold}-\ref{tab:sample_phot_comp}. Similar to past studies (e.g., \citealt{Hamuy03, spiro14, Pejcha15, Valenti15}), there is a visible trend between $M(^{56} \rm Ni)$ and $L_{50}$ where more luminous SNe~II also produce more $^{56} \rm Ni$. However, a number of SNe~II with IIn-like features in the gold/silver samples break this correlation with their high $L_{50}$ but low/moderate $M(^{56} \rm Ni)$ values. This might imply that there is another parameter related to the progenitor star and/or the explosion itself that can account for this deviation (e.g., see \S\ref{sec:scaling} for more discussion). As shown in Figure \ref{fig:Lbol_tp_S50_L50}, there are a number of SNe~II in both gold/silver and comparison samples that have $M(^{56} \rm Ni) > 0.1~\Msun$ --- this being consistent with theoretical predictions for SNe~II from extremely large ($>18~\Msun$) RSGs by \cite{Curtis21} and \cite{burrows24}. To confirm that these values are accurate, we explore the possibility that the pseudobolometric light curves were overcorrected for host-galaxy extinction (e.g., see Appendix discussion of WJG24a) by plotting $E(B-V)_{\rm host}$ values versus $M(^{56} \rm Ni)$ for all objects in Appendix Figure \ref{fig:EBV}. Similar to comparison of $E(B-V)_{\rm host}$ with $L_{50}$ and $M_{r,50}$ measurements, there is no obvious correlation between $M(^{56} \rm Ni)$ and the adopted host-galaxy extinction; SNe~II with high $M(^{56} \rm Ni)$ span a range of $E(B-V)_{\rm host}$ values. Nevertheless, because of the general uncertainty in accurately measuring host extinction, objects with both high $M(^{56} \rm Ni)$ and $E(B-V)_{\rm host}$ values should be interpreted with caution, and it is advisable to treat their $^{56} \rm Ni$ masses as upper limits. 

\begin{figure*}[t!]
\centering
\subfigure{\includegraphics[width=0.49\textwidth]{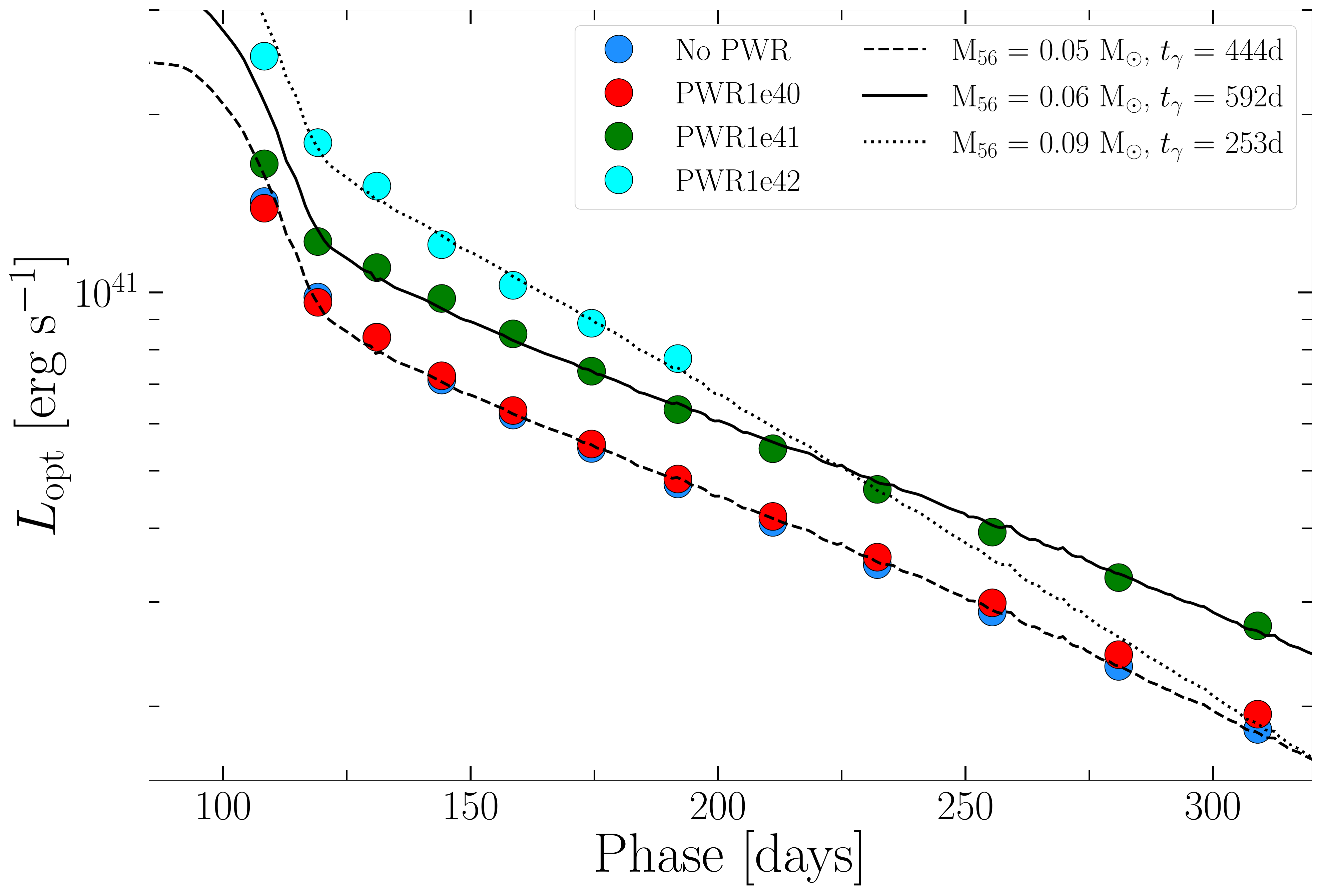}}
\subfigure{\includegraphics[width=0.49\textwidth]{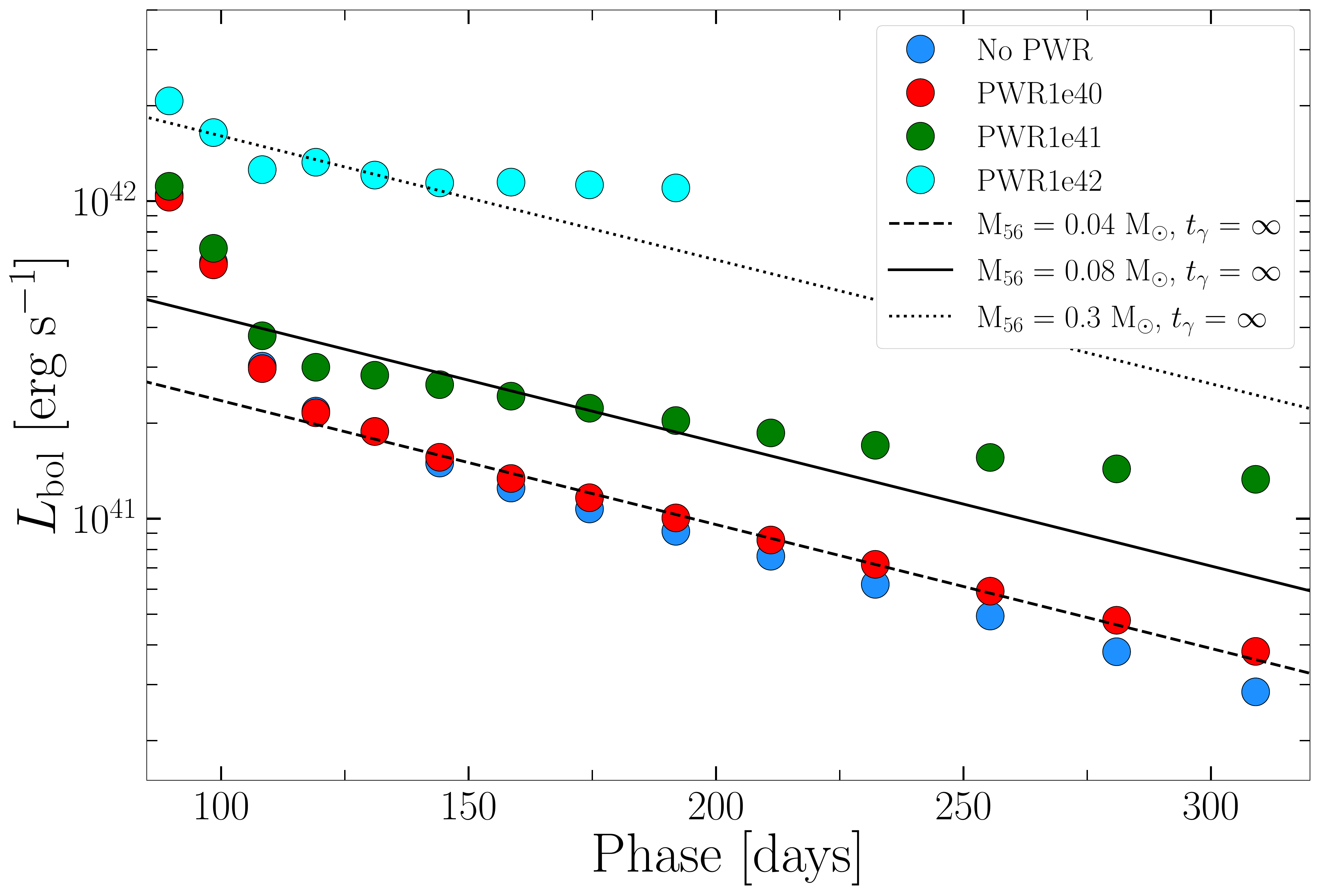}}
\caption{ {\it Left:} Post-plateau pseudo-bolometric light curves from \cite{dessart22} derived from 3000-10000~\AA (circles), all of which have the same M($^{56}$Ni) but with varying amounts of shock power. Fitting the analytic formalism for SN~1987A (e.g., Eq. 1) returns the model $M(^{56}$Ni) correctly for no shock power and $L_{\rm sh} = 10^{40}$~erg~s$^{-1}$, but the $M(^{56}$Ni) is overestimated if the shock power is $\geq 10^{41}$~erg~s$^{-1}$. {\it Right:} Complete bolometric light curves for shock-power models compared to radioactive-decay power models. The influence of shock power across the SN SED, in particular in the UV, will cause a significant overprediction of $M(^{56}$Ni) when modeling the late-time light curve. 
\label{fig:PWRLbol} }
\end{figure*}

Another source of uncertainty in measuring $M(^{56} \rm Ni)$ is the potential for non-negligible luminosity arising from ongoing CSM interaction. For a shock wave with velocity $v_{\rm sh}$ propagating through a steady-state wind (i.e., $\rho \propto r^{-2}$) with mass-loss rate $\dot M$ and velocity $v_w$, the power supply of the shocks goes as $L_{\rm sh} = \dot M v_{\rm sh}^3 / 2v_w$. Consequently, for even a weak RSG mass-loss rate of $\dot M \approx 10^{-6} ~ \Msun$~yr$^{-1}$ ($v_{\rm sh} = 10^{4}$~km s$^{-1}$, $v_w = 30$~km s$^{-1}$), the resulting shock power is quite large ($L_{\rm sh} \approx 10^{40}$~erg s$^{-1}$). This shock power is likely radiated by the reverse shock as it deposits energy into the cold dense shell (CDS) in the form of X-rays from free-free emission following conversion of kinetic to thermal energy in the post-shock gas \citep{Nymark06, Chevalier17, dessart22}. The CDS can be formed soon after shock breakout if there is confined, high density CSM (e.g., $\sim 0.1~\Msun$ is expected for SNe~II with IIn-like features) or later in the SN evolution as more CSM is swept up (e.g., $M_{\rm CDS} = 10^{-3}~\Msun$ at 1 year for $\dot M = 10^{-6}$~\mdot, $v_w = 10~\kms$ and $v_{\rm sh} = 10^{4}~\kms$; \citealt{dessart22}). If the distant CSM has sufficiently high density, the shock power could contribute significantly to the pseudobolometric luminosities used to measure $M(^{56} \rm Ni)$ and the derived masses would only be considered upper limits on the true amount of radioactive material present in the SN. This is demonstrated in Figure \ref{fig:PWRLbol}, where we show that modeling pseudobolometric light curves derived from only optical photometry with Equation 1 can lead to a significant overestimation of the $^{56}$Ni mass if a shock power of $>10^{40}$~erg~s$^{-1}$ is present \citep{dessart22}.

To explore this effect in more detail, we examine the late-time photometric and spectroscopic behavior of all sample objects with large $^{56}$Ni masses (e.g., $M(^{56} \rm Ni) > 0.08~\Msun$). Among the gold/silver events with high $M(^{56} \rm Ni)$, SNe~2020abjq, 2021zj, 2021can, 2022dml, 2021ont, and PTF11iqb show spectroscopic signatures of significant shock power in both their photospheric and late-time spectra (Figs. \ref{fig:neb_spec} and \ref{fig:neb_comp}), which will be responsible for overestimating $M(^{56} \rm Ni)$. Consequently, we present $M(^{56} \rm Ni)$ for these objects as upper limits (e.g., Fig. \ref{fig:Lbol_tp_S50_L50}). Additionally, it is worth noting that other events (e.g., SNe~2017ahn, 2020pni, 2014G) with more typical $M(^{56} \rm Ni)$ estimates also exhibit broad/boxy H$\alpha$ emission during their radioactive-decay decline phase. This indicates that $M(^{56} \rm Ni)$ could be overestimated in many events due to a nonnegligible contribution from shock power, which can only be confirmed by continuous spectroscopic monitoring out to late-time phases. More discussion of the detection of ongoing, late-time CSM interaction is presented in \S\ref{sec:highmdot}.


\subsection{$\gamma$-ray Trapping} \label{sec:tgamma}

In addition to $M(^{56} \rm Ni)$, we are able to measure $t_{\gamma}$ values that indicate incomplete $\gamma$-ray trapping in a number of sample objects. As shown in Tables \ref{tab:sample_phot_gold}--\ref{tab:sample_phot_comp}, most gold/silver- and comparison-sample objects with measured $M(^{56} \rm Ni)$ have trapping timescales $>450$~days and/or are consistent with complete $\gamma$-ray trapping. However, there are eight gold/silver and eight comparison-sample objects with constrained $t_{\gamma}$ values of $<400$~days, some even as low as $\sim 200$~days. Incomplete $\gamma$-ray trapping has interesting implications for SN ejecta mass and kinetic energy as shown in the analytic formalism from \cite{clocchiatti97}, which is typically applied to Type Ib/c or Ia SNe. Here, $t_{\gamma} = (C(\eta)\kappa_{\gamma}M_{\rm ej}^2E_{\rm k}^{-1})^{0.5}$, where $C(\eta)$ is the density function, $\kappa_{\gamma}$ is the $\gamma$-ray opacity (0.048~cm$^2$~g$^{-1}$ for H-rich ejecta), $M_{\rm ej}$ is the ejecta mass, and $E_{\rm k}$ is the kinetic energy. Assuming a flat density profile (i.e., $\eta = 0$) and a fiducial kinetic energy of 1~B, $t_{\gamma}$ values in the range 200--400~days would imply ejecta masses of $\sim 4.7$--9.3~$\Msun$. However, the analytic expression from \cite{clocchiatti97} is a very rough approximation that makes assumptions about spherical symmetry and the ejecta density profile that ignores known complexities about the ``Nickel bubble effect'' and clumping (e.g., \citealt{Basko94, Dessart19, Dessart21, Gabler21}). Nevertheless, sample objects with $t_{\gamma} < 250$~days implies smaller ejecta masses than what is expected from SNe~II with minimal envelope removal prior to core collapse. Notably, SNe~2014G, 2017ahn, 2021can, and 2020jfo have $t_{\gamma} < 200$~days and also have short plateau timescales ($<85$~days), both of which being consistent with lower ejecta mass values.

\begin{figure}[t!]
\centering
\includegraphics[width=0.47\textwidth]{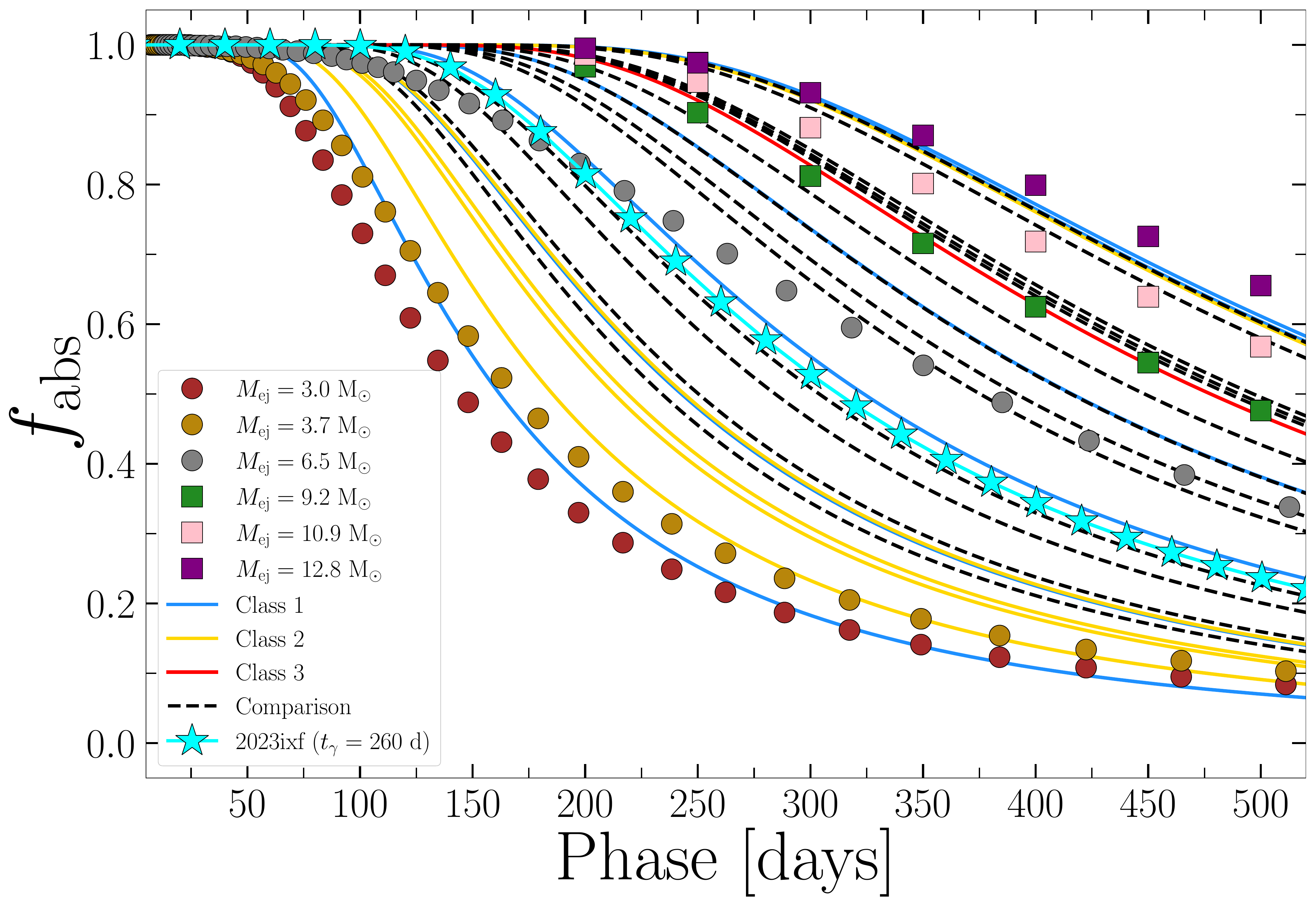}
\caption{ Fraction of radioactive-decay power absorbed in the SN ejecta ($f_{\rm abs}$) calculated for CMFGEN models of varying ejecta masses (circles and squares) compared to gold/silver- and comparison-sample objects. $f_{\rm abs}$ is estimated for the sample objects using their calculated $\gamma$-ray trapping timescales ($t_{\gamma}$). SN~2023ixf is shown for reference as cyan stars.
\label{fig:fabs} }
\end{figure}

We explore this in more detail by calculating the fraction of absorbed radioactive-decay power ($f_{\rm abs}$) for different SN ejecta masses in CMFGEN models for SNe~II out to late-time phases. In Figure \ref{fig:fabs}, we plot the $f_{\rm abs}$ for SN~II models from \cite{Dessart24binary} that have ejecta masses of 3--6.5~$\Msun$ owing to substantial removal of the H envelope from binary interaction --- these models have an initial mass of $M_{\rm ZAMS} = 12.6~\Msun$. Additionally, we plot more typical SN~II models that arise from RSGs with $M_{\rm ZAMS} = 11$--15~$\Msun$, which have $M_{\rm ej} = 9.2$--12.8~$\Msun$. Both sets of models have kinetic energies of 1--1.3~B. As shown in Figure \ref{fig:fabs}, models with low ejecta masses ($<6.5~\Msun$) best match sample objects with $t_{\gamma} = 200$--300~days, while models with intermediate ejecta masses (6.5--9.2~$\Msun$) more resemble SNe~II with $t_{\gamma} = 300$--400~days. Finally, all sample objects consistent with more complete $\gamma$-ray trapping (e.g., $t_{\gamma} > 500$~days) at phases $\delta t < 300$~days are likely best matched by more typical ejecta masses $>11~\Msun$. However, as discussed in \S\ref{sec:ni56}, there might be significant influence from shock power on the post-plateau decline rate, which could induce additional uncertainty on the $t_{\gamma}$ estimates in some gold/silver-sample objects. Furthermore, as shown in Figure \ref{fig:Lbol_Mr}, the late-time light curves of most sample objects only extend to $\delta t <300$~days --- ideally there would be complete, multiband coverage out to $\sim 500$--600~days post-explosion in order to robustly constrain the $\gamma$-ray trapping timescales.

\subsection{Nebular Spectroscopy} \label{sec:nebspec}

\begin{figure*}[t!]
\centering
\includegraphics[width=\textwidth]{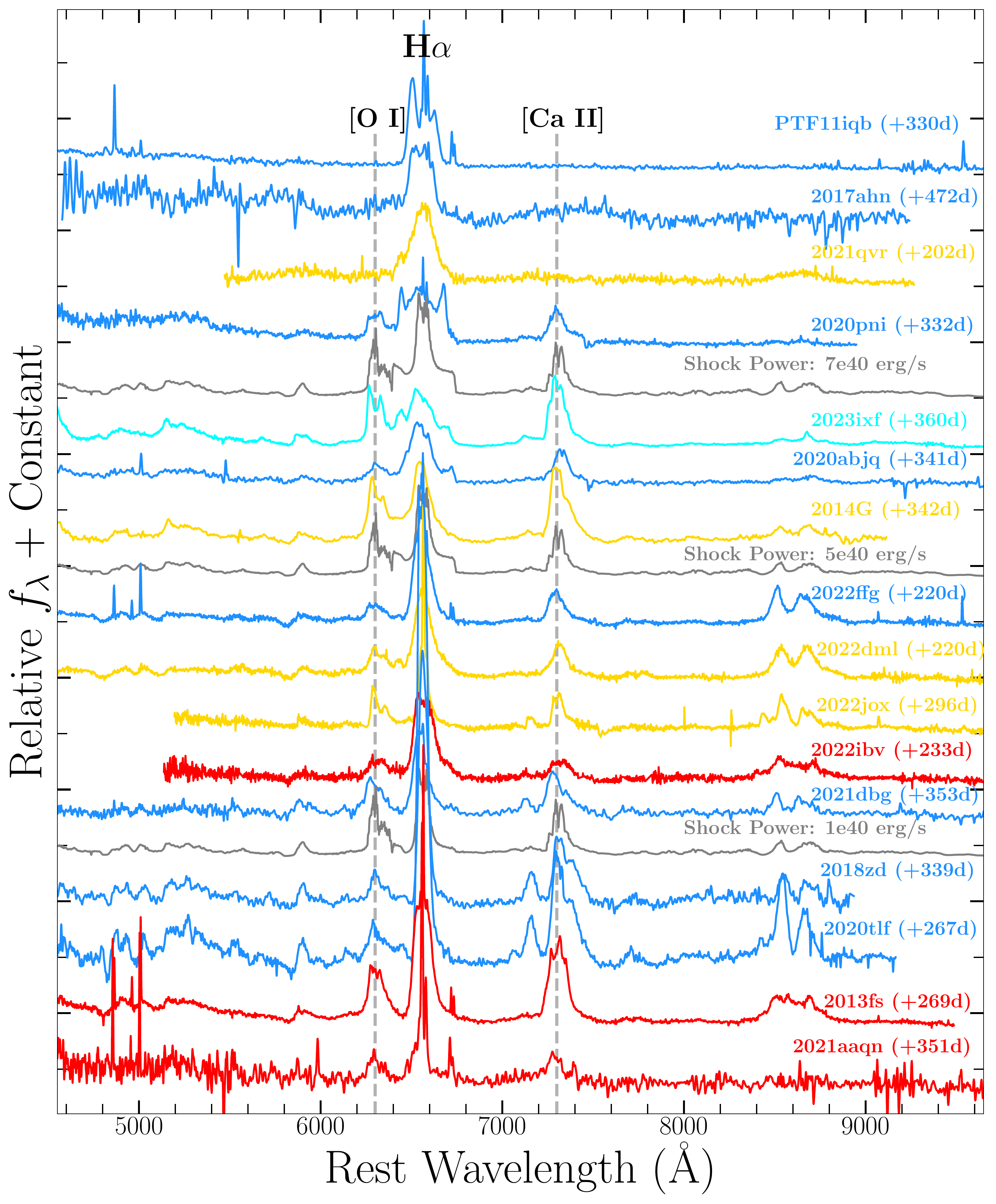}
\caption{ Nebular spectra of gold/silver-sample objects (blue, yellow, red), SN~2023ixf (cyan), and \cmfgen\ shock-power models from \cite{dessart22}. Spectra are presented in order of decreasing spectroscopic evidence for ongoing CSM-interaction (e.g., boxy H$\alpha$ emission) from top to bottom. 
\label{fig:neb_spec} }
\end{figure*}

We present late-time/nebular spectroscopy of gold/silver-sample objects at phases $\delta t = 200$--500~days in Figure \ref{fig:neb_spec}. We present a log of unpublished nebular spectra in Table \ref{tab:spec_all} and also include previously published late-time spectra for the following objects: SNe PTF11iqb \citep{smith15}, 2013ab \citep{Bose15}, 2013am \citep{Zhang14, Tomasella18}, 2013fs \citep{Gal-Yam17}, 2014G \citep{Terreran16}, 2015bf \citep{Lin21}, 2016X \citep{Huang18}, 2016aqf \citep{MullerBravo20}, 2017ahn \citep{Tartaglia21}, 2017eaw \citep{Szalai19, Weil20}, 2018zd \citep{zhang20-1}, 2018cuf \citep{Dong21}, 2018lab \citep{Pearson23}, 2020fqv \citep{Tinyanont22}, 2020jfo \citep{Kilpatrick23b}, 2020tlf \citep{wjg22}, 2021gmj \citep{Meza-Retamal24}, and 2022jox \citep{Andrews23}. In total, we this study includes nebular spectroscopy for 19 gold/silver- and 13 comparison-sample objects. We also include late-time spectra of SN~2023ixf (Jacobson-Gal\'an et al., in prep) for comparison to gold/silver-sample objects. Similar to nebular observations of SNe~II without IIn-like features, many early-time CSM-interacting SNe~II display strong forbidden emission lines such as [\ion{O}{i}] $\lambda\lambda6300$, 6364 and [\ion{Ca}{ii}] $\lambda\lambda$7291, 7323, as well as prominent H$\alpha$ emission. Despite the small number of gold/silver-sample objects with nebular spectra, these SNe display significant spectral diversity at late times, in particular in their [\ion{O}{i}]/H$\alpha$ complex in the wavelength range 6200--6800~\AA. 

\begin{figure*}[t!]
\centering
\subfigure{\includegraphics[width=0.25\textwidth]{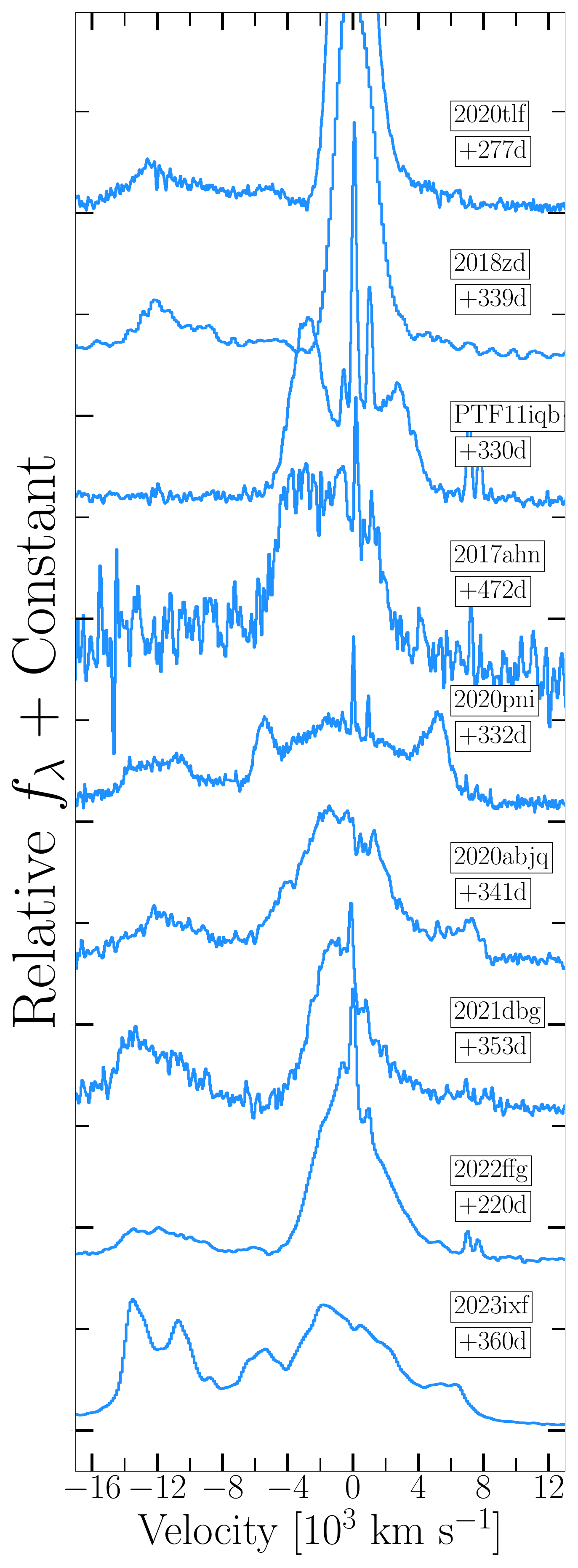}}
\subfigure{\includegraphics[width=0.2285\textwidth]{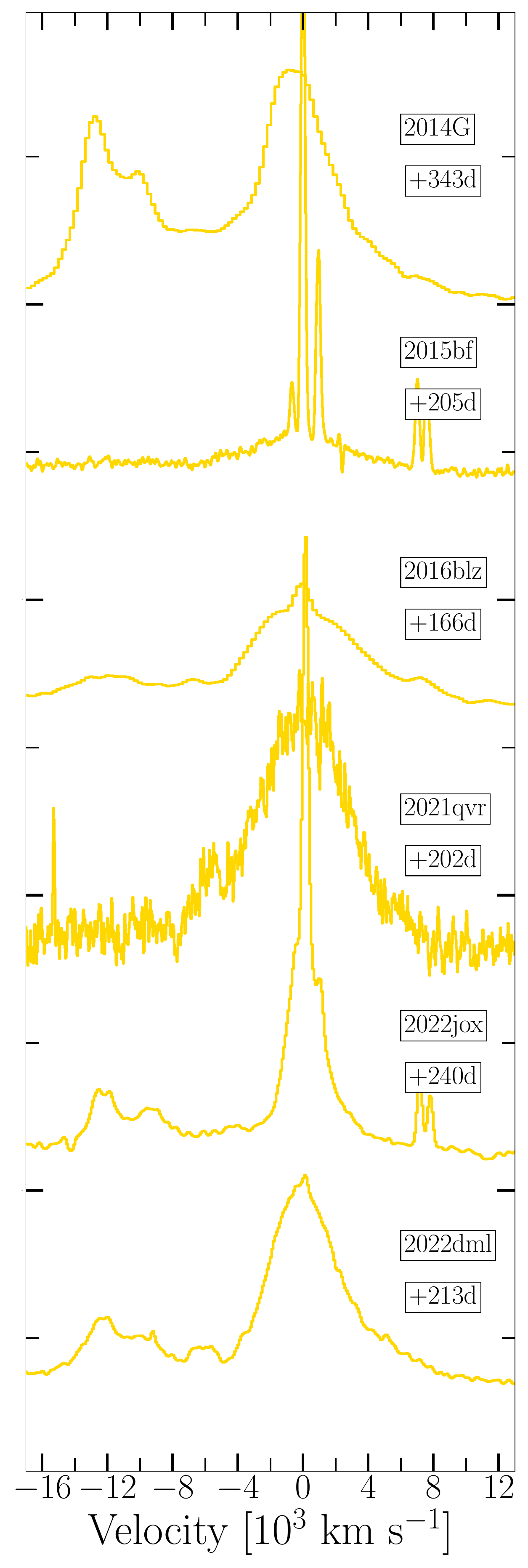}}
\subfigure{\includegraphics[width=0.2285\textwidth]{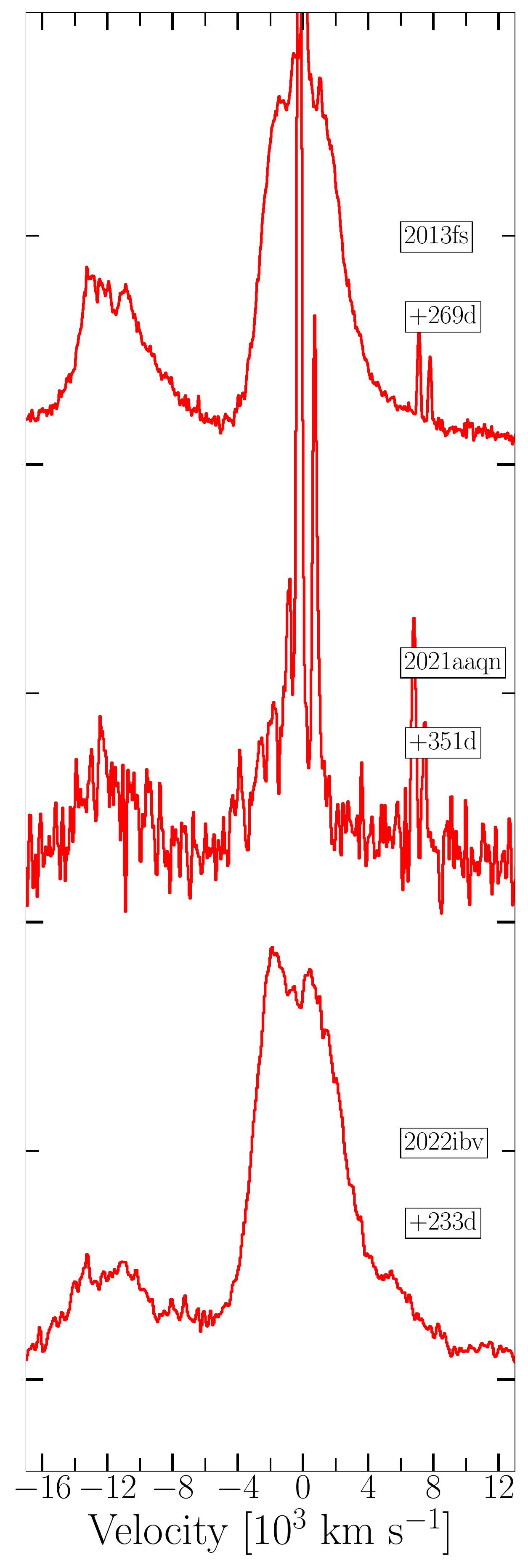}}
\subfigure{\includegraphics[width=0.2285\textwidth]{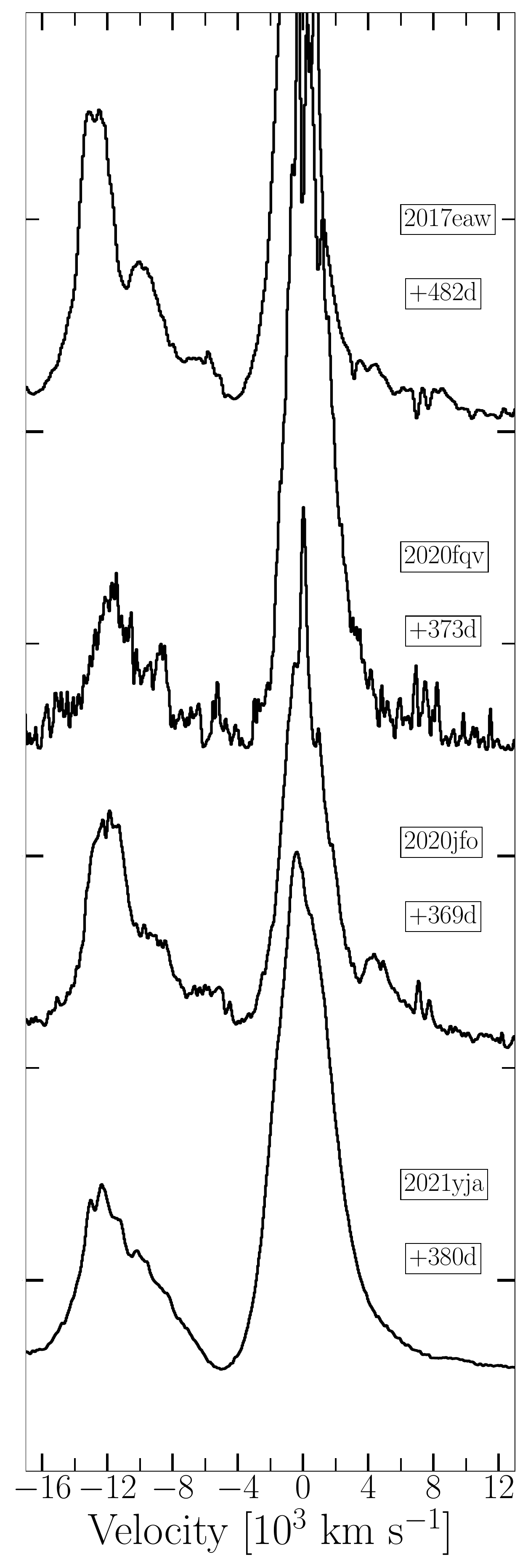}}
\caption{ H$\alpha$/[\ion{O}{i}] spectral region shown in velocity space relative to H$\alpha$ rest wavelength for gold/silver-sample objects shown in blue, yellow, and red as well as four comparison-sample objects shown in black for reference. In addition to the [\ion{O}{i}] and H$\alpha$ emission from the inner ejecta (narrower components), many SNe~II with IIn-like features develop an underlying broad, boxy H$\alpha$ profile from emission in a dense shell formed between the forward and reverse shocks. The detection of the dense-shell emission indicates significant contribution from shock power at late times and higher CSM densities (e.g., $> 10^{-6}$~\mdot) at large radii (e.g., $>10^{16}$~cm).
\label{fig:neb_vels} }
\end{figure*}

\begin{figure*}[t!]
\centering
\includegraphics[width=\textwidth]{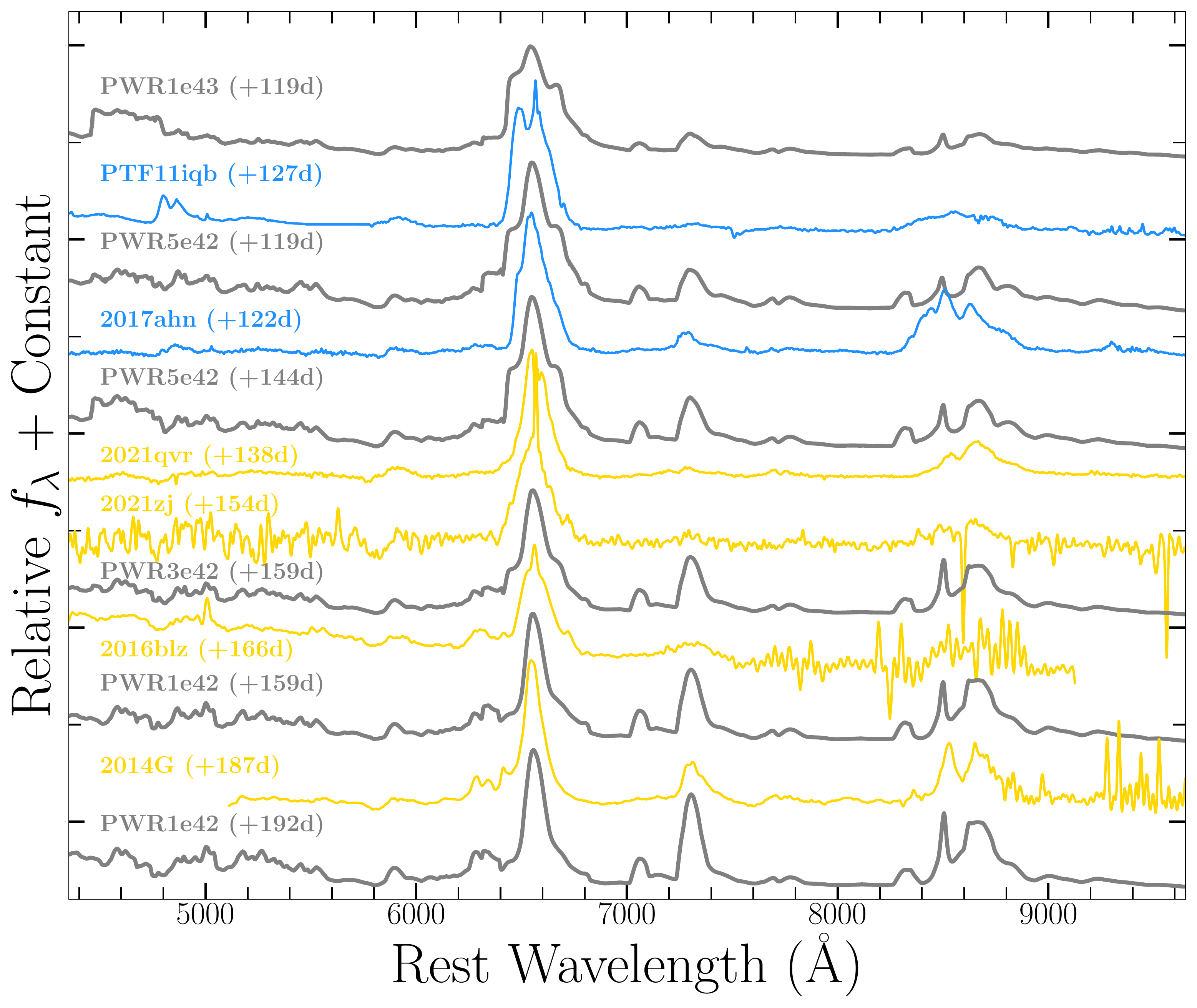}
\caption{ Late-time spectra of gold/silver-sample objects with signatures of significant CSM-interaction (blue and yellow spectra) compared to \cmfgen\ shock power models from \cite{dessart22}. These SNe~II show very weak or no forbidden line emission and their spectra are dominated by intermediate width H$\alpha$ emission from the dense shell formed as the SN shock continues to sweep up significantly high density CSM at large radii.      
\label{fig:neb_spec_pwr} }
\end{figure*}

\begin{figure}[t!]
\centering
\includegraphics[width=0.45\textwidth]{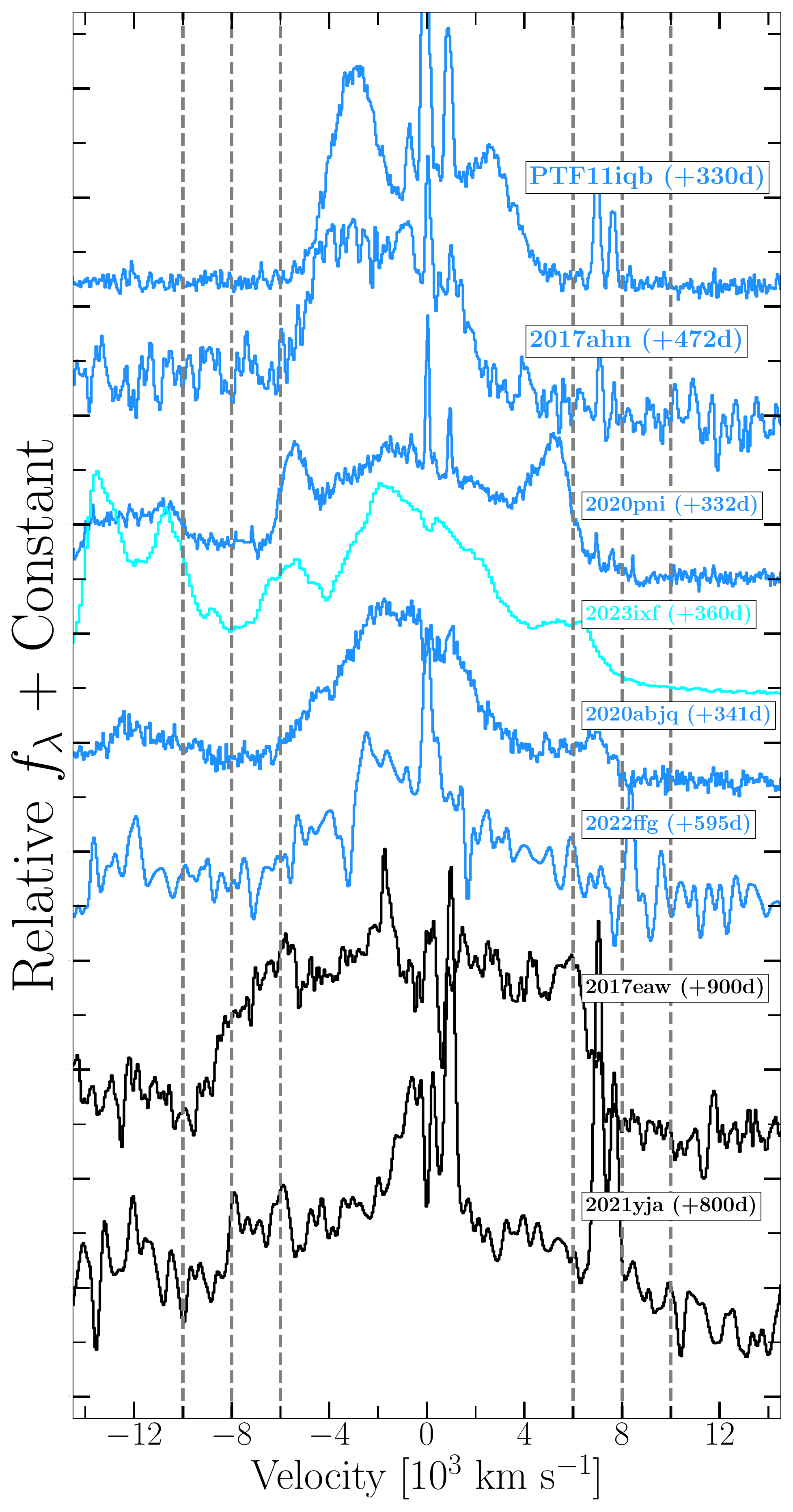}
\caption{ Late-time H$\alpha$ velocity for gold/silver- and comparison-sample objects that develop boxy emission from shock power reprocessed by a cold dense shell (CDS). The timescales over which this broad, boxy H$\alpha$ emission emerges is a tracer of the shock power derived from more distant ($R>10^{16}$~cm) CSM that is being consistently swept up into the CDS.     
\label{fig:CDS} }
\end{figure}

As shown in Figure \ref{fig:neb_vels}, the [\ion{O}{i}] profile in gold/silver-sample objects is quite diverse: some events show weak/no emission (e.g., SNe PTF11iqb, 2015bf, 2017ahn, 2018zd, 2020tlf, 2020pni, 2021qvr, and 2022ffg), while others exhibit strong, double-peaked profiles (e.g., SNe~2014G, 2022jox, 2022dml, 2013fs, and 2022ibv). For reference, we also display a zoom-in of the [\ion{O}{i}] profile of a subsample of comparison objects (SNe 2017eaw, 2020fqv, 2020jfo, and 2021yja), all of which have prominent forbidden-line emission. Examining the H$\alpha$ profile in more detail, we see that there is a potential continuum of late-time CSM-interaction signatures within the Class 1, 2, and 3 objects. For example, SNe like PTF11iqb and SN~2017ahn only show boxy H$\alpha$ emission ($v_{\rm max} \approx 4000~\kms$), which is similar, but likely more extreme, than events like SNe~2016blz and 2021qvr that have both boxy emission ($v_{\rm max} \approx 8000~\kms$) and a narrower Gaussian profile typical of nebular phases. These objects have weak and/or no detectable forbidden emission lines. Then there are events like SNe~2020pni, 2020abjq, 2022ffg, 2014G, 2022jox, 2022dml, and 2022ibv that display standard forbidden lines but show an underlying boxy component within H$\alpha$ ($v_{\rm max} \approx 8000~\kms$) --- this is similar to the nebular evolution of SN~2023ixf (\citealt{Kumar25, Folatelli25, Li2025}, Jacobson-Gal\'an et al, in prep). We discuss the implications for detecting late-time dense-shell emission in \S\ref{sec:highmdot} and illustrate more early-time examples of these interaction signatures in Figure \ref{fig:neb_spec_pwr}, as well as those at later-time phases in Figure \ref{fig:CDS}. Finally, gold/silver-sample SNe~2013fs, 2018zd, 2020tlf, 2021dbg, and 2021aaqn do not show any underlying boxy emission in H$\alpha$, similar to comparison-sample SNe at nebular times. 

\begin{figure*}[t!]
\centering
\subfigure{\includegraphics[width=0.49\textwidth]{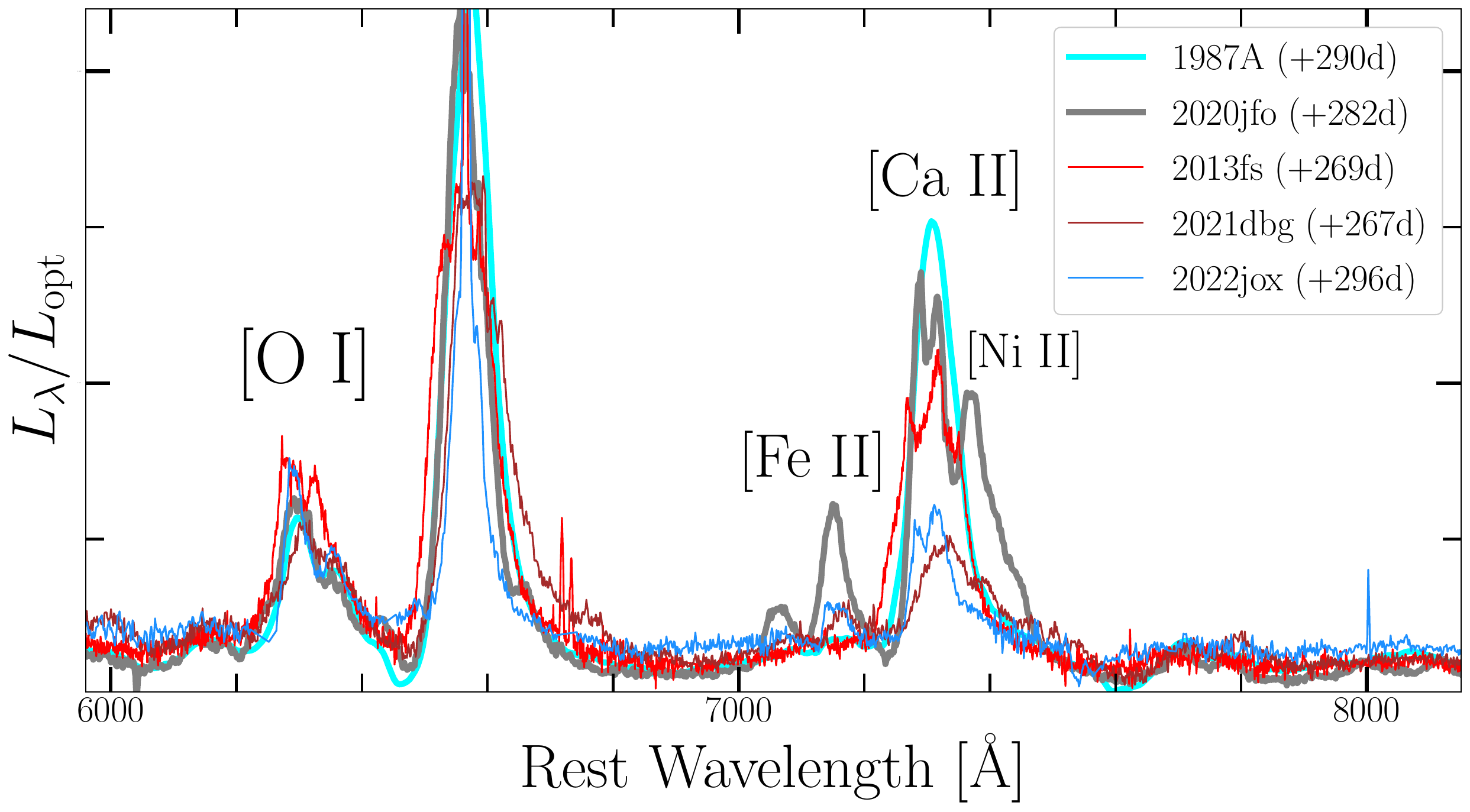}}
\subfigure{\includegraphics[width=0.49\textwidth]{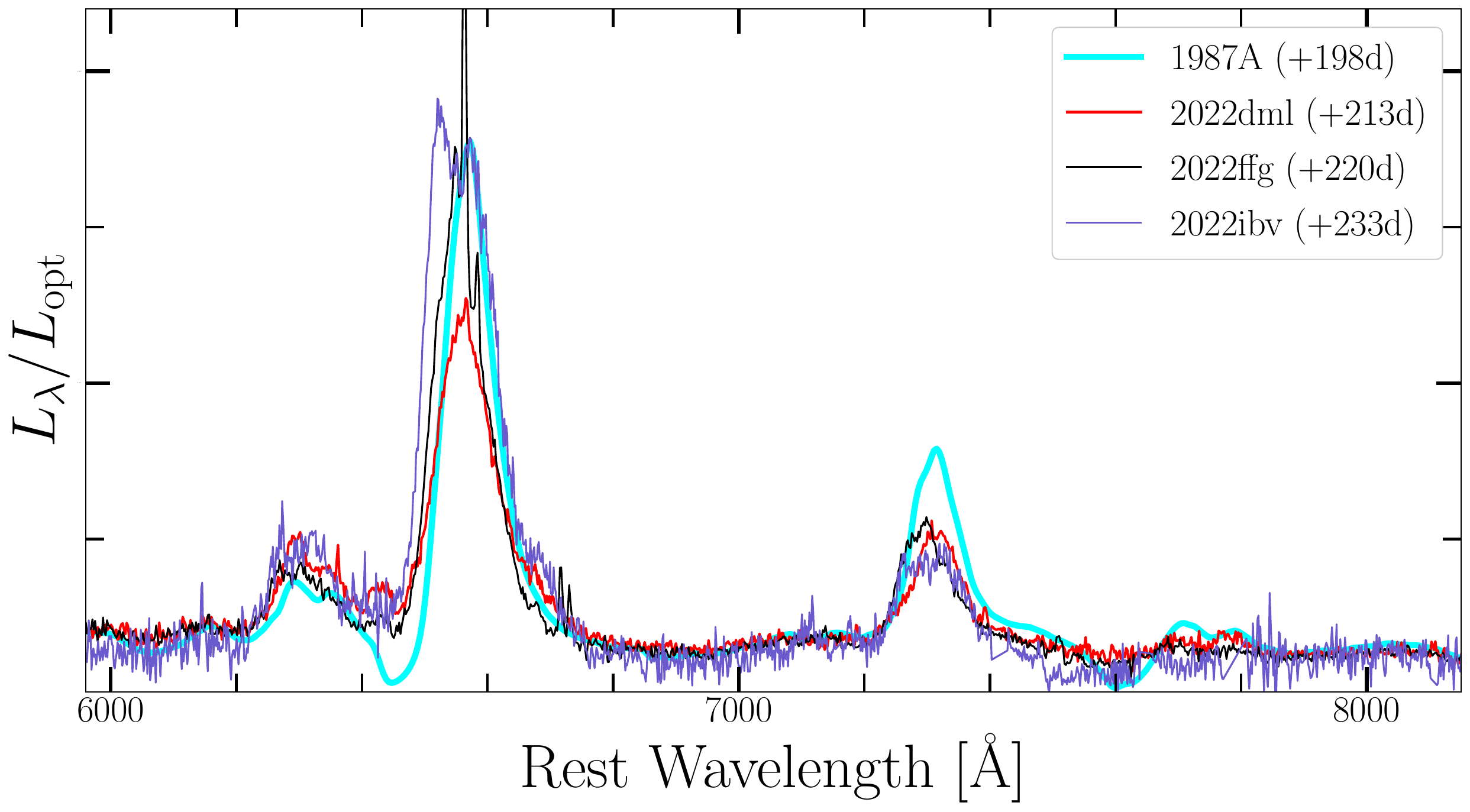}}\\
\subfigure{\includegraphics[width=0.49\textwidth]{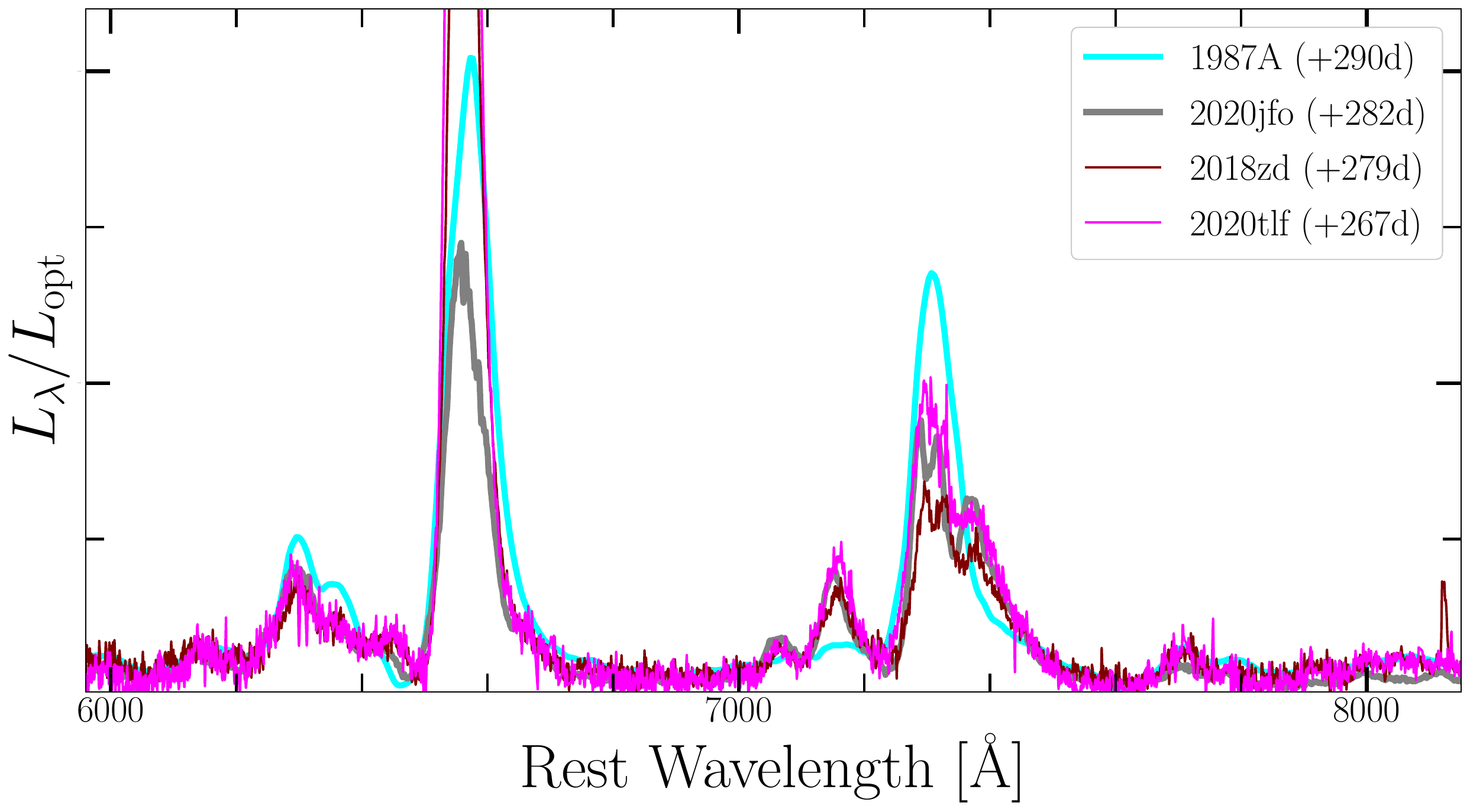}}
\subfigure{\includegraphics[width=0.49\textwidth]{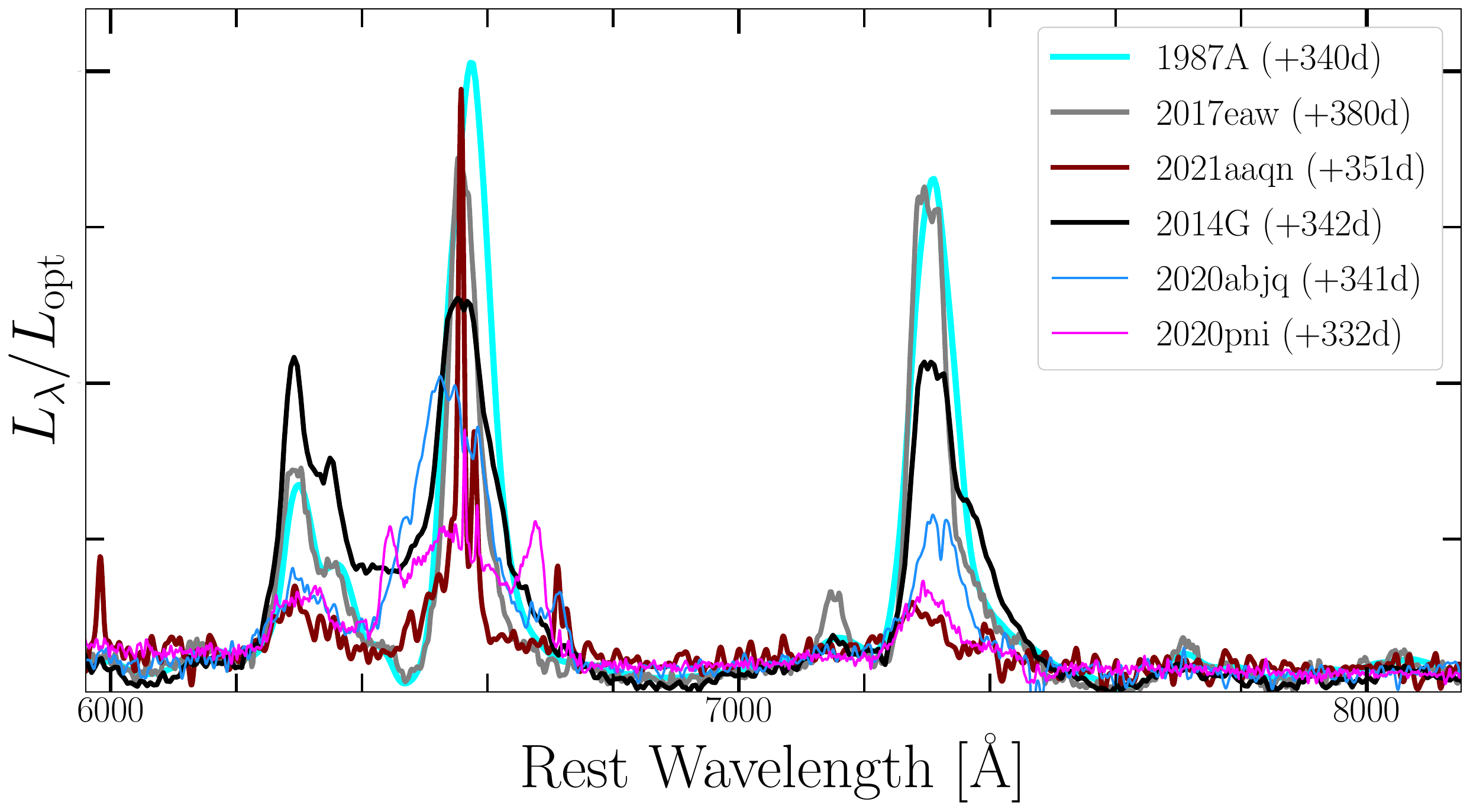}}
\caption{ {\it Upper left:} Nebular spectra comparison of SNe~1987A (cyan) and 2020jfo (gray) with gold/silver-sample SNe 2021dbg (brown), 2022jox (blue), and 2013fs (red). All spectra have been normalized by the integral of the optical spectrum in the range 5000--9000\,\AA. {\it Upper right:} SN~1987A (cyan) compared with gold/silver-sample SNe 2022dml (red), 2022ffg (black), and 2022ibv (turquoise). {\it Lower left:} SNe~1987A (cyan) and 2020jfo (gray) compared with gold/silver-sample SNe 2018zd (purple) and 2020tlf (magenta). Notably, these gold/silver-sample objects have weaker [\ion{O}{i}] emission than comparison SNe. {\it Lower right:} SNe~1987A (cyan) and 2017eaw (gray) compared with gold/silver-sample SNe 2014G (black), 2020abjq (blue), 2020pni (magenta), and 2021aaqn (brown). These gold/silver-sample objects are examples where the dense-shell emission is forming a ``boxy'' profile within the H$\alpha$ complex, which is a product of ongoing interaction with distant, relatively dense CSM.   
\label{fig:neb_comp} }
\end{figure*}


In Figure \ref{fig:neb_comp} we present nebular spectra normalized by their optical luminosities between 5000-9000~\AA. For gold/silver sample objects, we compare forbidden line emission (e.g., [\ion{O}{i}] and [\ion{Ca}{ii}]) to reference objects SNe~1987A, 2017eaw, and 2020jfo at similar phases and find significant spectral diversity. These comparison objects were chosen given the lack of CSM-interaction signatures and their variety of [\ion{O}{i}] emission line strengths. For example, only SNe~2013fs, 2022jox and 2014G have significantly larger [\ion{O}{i}] emission than SNe~1987A or 2020jfo, while the rest of the sub-sample have comparable or weaker [\ion{O}{i}] luminosities. Applying the common assumption that [\ion{O}{i}] is a metric for ZAMS mass (i.e., He-core mass), then it is possible that many CSM-interacting SNe~II in this come from lower mass stars ($<12~\Msun$). Notably, some of these objects such as SNe~2014G, 2020abjq, 2020pni, 2022dml, 2022jox and 2022ibv also show strong broad, boxy H$\alpha$ from emission from the dense shell, which strongly contrasts the typical absorption seen in H$\alpha$ of SNe~1987A and 2020jfo. Furthermore, it is intriguing to see similarity between a comparison-sample object such as SN~2020jfo that showed no IIn-like features and gold/silver-sample objects such as SNe~2020tlf and 2018zd -- all three objects have nearly perfectly matched [\ion{O}{i}]/[\ion{Ca}{ii}] and even show prominent [\ion{Ni}{II}] $\lambda7378$ emission, which is proposed to be a signature of lower mass ($<12~\Msun$ progenitors) (e.g., \citealt{Jerkstrand15}). However, the nebular spectra of SNe~2018zd and 2020tlf deviate from SN~2020jfo in their incredibly strong H$\alpha$ emission, which may also be a signature of a low mass progenitor star (e.g., \citealt{Dessart21}). SN~2018zd's consistency with the nebular properties of other SNe~II, both with and without IIn-like signatures, as well as the inconsistency with very low mass model predictions (e.g., $\leq9.5~\Msun$) makes it more ambiguous whether SN~2018zd is a true electron-capture SN (e.g., \citealt{Hiramatsu21}) or just the core-collapse of a low mass (e.g., $10-12~\Msun$) RSG.

\subsection{Progenitor Mass Estimation} \label{sec:zams}

\begin{figure}[t!]
\centering
\includegraphics[width=0.49\textwidth]{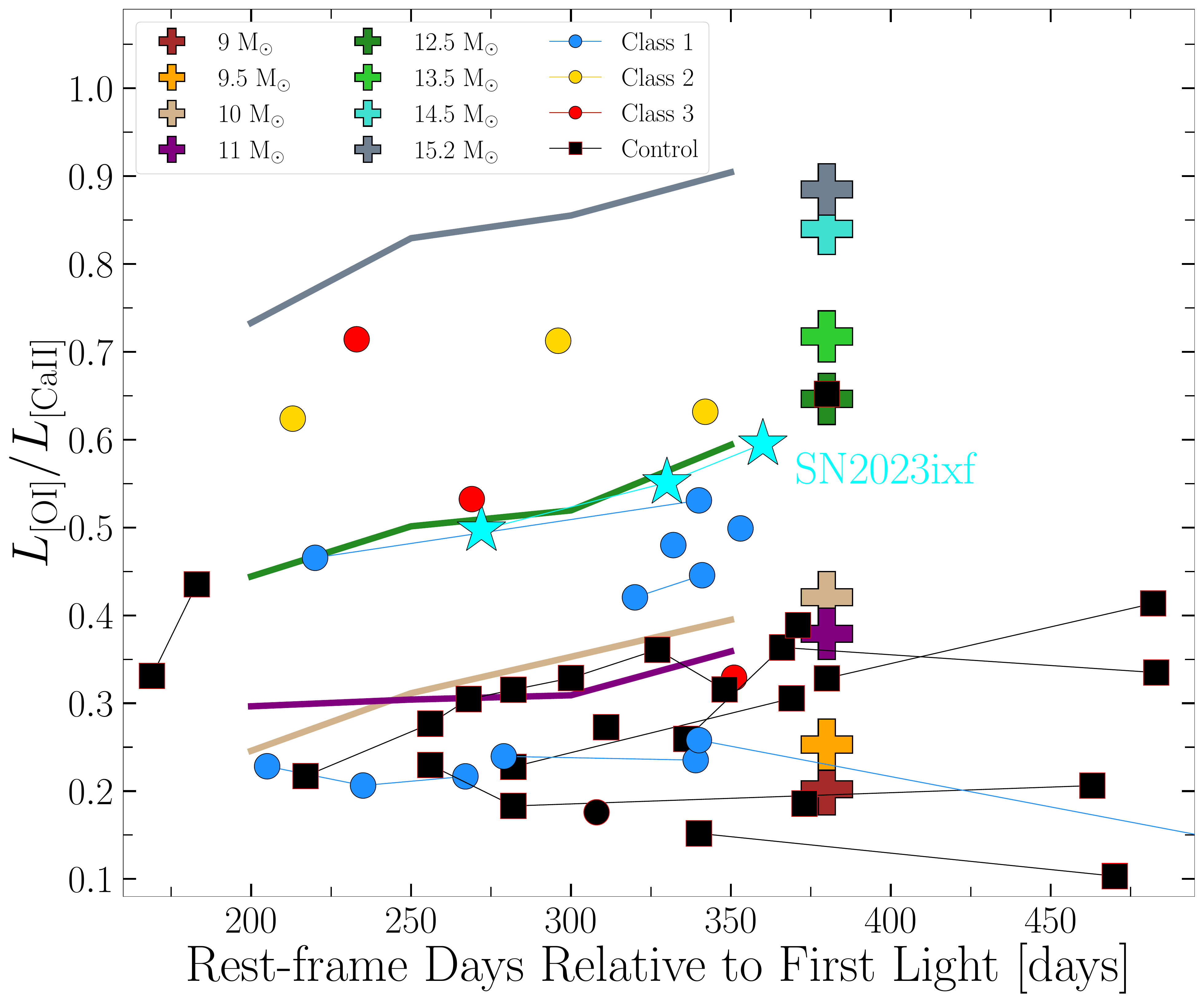}
\caption{ Ratio of [\ion{O}{i}] and [\ion{Ca}{ii}] luminosities for gold/silver- (blue, yellow, red circles) and comparison-sample (black squares) objects. SN~2023ixf is shown as cyan stars. Line ratios for 9--15.2~$\Msun$ ZAMS models are shown as lines and plus signs \citep{Dessart21}. Compared to model predictions, the majority of SNe~II in this sample are consistent with ZAMS masses of $\leq13~\Msun$.  \label{fig:LOICaII} }
\end{figure}

Nebular spectroscopy of SNe~II provides an additional constraint on the mass of the RSG progenitor star (e.g., \citealt{Jerkstrand14, valenti16,Dessart21}). In Figure \ref{fig:LOICaII}, we present the [\ion{O}{i}]/[\ion{Ca}{ii}] line-luminosity ratio as a function of phase for all sample objects with nebular-phase spectroscopy compared to the same measurement for a grid of models for progenitor ZAMS masses of 9--15.2~$\Msun$ from \cite{Dessart21}. For both models and data, we calculate line luminosities by first modeling and subtracting off the continuum emission before measuring the integrated line emission. For objects with boxy H$\alpha$ emission that extends into the wavelength range of [\ion{O}{i}], we model the forbidden-line emission using two Gaussian profiles centered at 6300 and 6364\,\AA. We also include line-ratio measurements for SN~2023ixf at late-time phases for comparison to sample objects. 

Interestingly, most comparison-sample SNe are consistent with ZAMS masses of 9--11~$\Msun$, as are Class 1 SNe~2018zd and 2020tlf and Class 3 SN 2021aaqn. Similarly, gold/silver-sample SNe~2013fs, 2014G, 2020abjq, 2020pni, 2021dbg, and 2022ffg, in addition to SN~2023ixf and comparison-sample SNe~2013ab and 2021yja, show consistency with $M_{\rm ZAMS} = 10$--12.5~$\Msun$ models. Finally, only three objects, all in Classes 2 and 3, are consistent with slightly higher masses of 13.5--14.5~$\Msun$. Intriguingly, we find no SNe~II in our sample that are consistent with the high-mass progenitor model using this ZAMS mass estimation metric. Furthermore, we directly compare synthetic model spectra from \cite{Jerkstrand14}, \cite{Jerkstrand18}, and \cite{Dessart21} with sample data at similar phases after normalizing by the integrated optical luminosity. As shown in Appendix Figures \ref{fig:neb_models_gold}--\ref{fig:neb_models_comp}, the best-matched $M_{\rm ZAMS}$ models for this smaller sample of objects are consistent with the masses derived from forbidden-line ratios. 

\subsection{Type II-P SN Scaling Relations}\label{sec:scaling}

There are well-established scalings that relate the SN II plateau luminosity at +50~days to progenitor/explosion properties (e.g., ejecta mass, kinetic energy, and progenitor radius), which have also been explored extensively with various stellar evolutionary codes \citep{Kasen06, kasen09, dessart13, Goldberg19}. Overall, increasing $L_{50}$ requires decreasing ejecta mass, increasing kinetic energy, and/or increasing progenitor radius. As discussed in \S\ref{sec:plateau}, the plateau luminosities of SNe~II with IIn-like features are significantly larger than those of the comparison-sample objects, suggesting a variation in the above progenitor/explosion properties. We show in \S\ref{sec:ni56} that some gold/silver-sample events may have low ejecta masses, which could explain the enhanced $L_{50}$ values for those objects specifically. However, if the remaining objects have typical $M_{\rm ej}$, then it is possible that kinetic energy and/or progenitor radius play a role. 

While a true estimate of $E_{\rm k}$ is challenging for SNe~II, we can gain some insight by examining the evolution of the photospheric velocities measured throughout the plateau phase. As shown in Figure \ref{fig:vphot}, the spread in \ion{Fe}{ii} velocities at $\sim50$~days is similar across both gold/silver  and comparison samples. We then compare the velocity measurements to model predictions for varying explosion energies (0.1--3~B) from \cite{dessart10b} and ejecta masses from \cite{Hillier19}. While an extreme kinetic energy of 3~B is needed to match the fastest velocities measured at +50\,d, lowering the ejecta mass in a standard SN~II explosion model can produce a similar effect, suggesting that increased $E_{\rm k}$ is unlikely to be the main factor causing enhanced $L_{50}$ in SNe~II with IIn-like signatures. Furthermore, some gold/silver events have $v_{\rm ph} < 3000~\kms$ yet $L_{\rm 50} > 3\times 10^{42}$~erg~s$^{-1}$, suggesting that an additional component is relevant for increasing their plateau luminosity. Therefore, one possibility is that many SNe~II with IIn-like signatures arise from RSGs with extended radii ($>800~\Rsun$), which could be related to their enhanced mass loss in the final years before explosion. Alternatively, the plateau luminosity could be boosted by ongoing interaction with more distant, yet still sufficiently dense, CSM.

\begin{figure}[t!]
\centering
\includegraphics[width=0.49\textwidth]{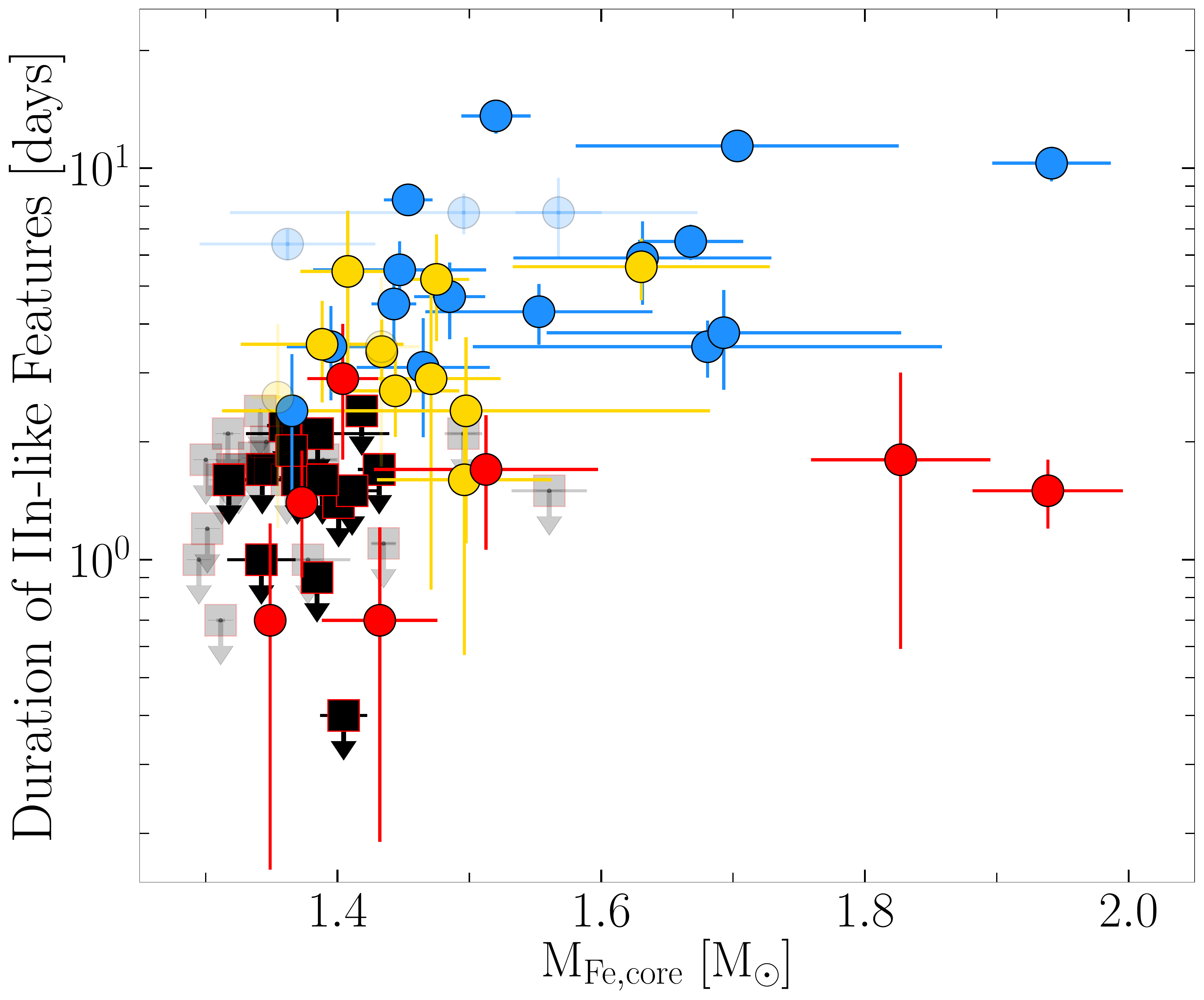}
\caption{ Iron-core mass estimated from $L_{50}$ using the analytic relation derived by \cite{Barker22} for gold/silver- (blue, yellow, red circles) and comparison-sample (black squares) objects versus the IIn-like feature duration. The incredibly large $M_{\rm Fe,core}$ masses inferred for many gold/silver-sample objects demonstrates that this relation is \emph{not} physical --- high plateau luminosities are much more likely the result of larger progenitor radii and/or smaller $M_{\rm ej}$ rather than extremely massive progenitor stars ($M_{\rm ZAMS} > 25~\Msun$). 
\label{fig:MFe} }
\end{figure}

We further explore the ambiguity of inferring progenitor properties for CSM-interacting SNe~II by applying the relation derived by \cite{Barker22} that relates $L_{50}$ and Fe core mass. As shown in Figure \ref{fig:MFe}, the use of plateau brightness, especially for SNe~II with IIn-like features, yields Fe core-mass estimates that are only consistent with extremely massive progenitor stars ($>25~\Msun$) as well as the largest neutron stars in the Universe. While these estimates are not only nonphysical, they are also invalidated by other progenitor mass probes such as $^{56}$Ni mass, nebular emission, and/or plateau duration. Consequently, we strongly caution the use of relationships derived from simulations that relate core and envelope properties given that applications of these relations typically only rely on photometric information and/or do not consider that a significant fraction of SNe~II with early-time CSM interaction could be mistaken as standardizable SNe~II. 

\section{Discussion}\label{sec:dis}
\subsection{Connecting Properties of the CSM, Progenitor, and Explosion }\label{sec:CSMZAMS}

\begin{figure*}[t!]
\centering
\subfigure{\includegraphics[width=0.49\textwidth]{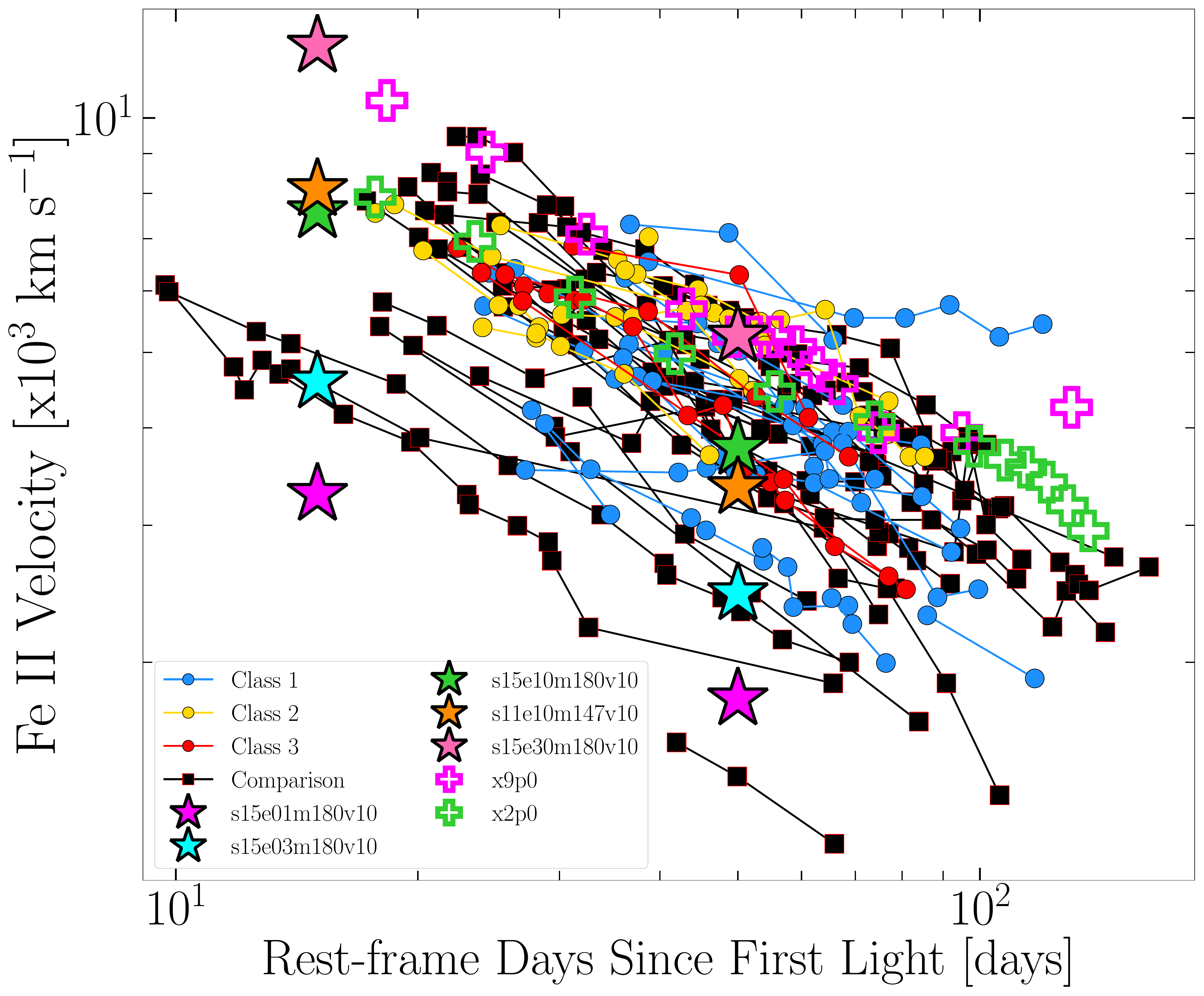}}
\subfigure{\includegraphics[width=0.49\textwidth]{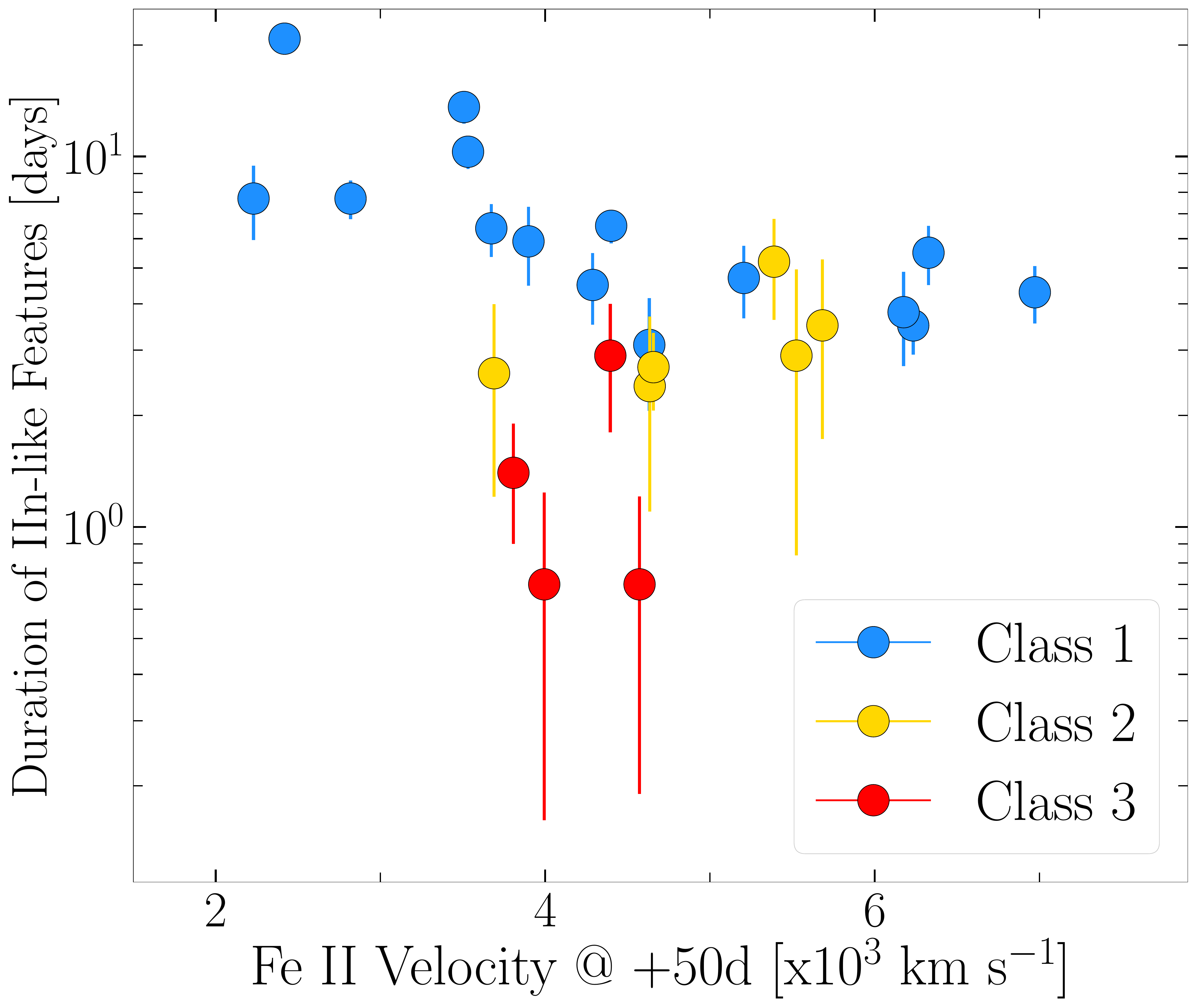}}
\caption{ {\it Left:} \ion{Fe}{ii} velocity measurements for gold/silver- (blue, yellow, red circles) and comparison-sample (black squares) objects during and after the plateau light-curve phase. Model predictions with varying explosion energies (0.1--3~B) are shown as stars \citep{dessart10b} and varying ejecta masses as plus signs \citep{Hillier19}. Higher photospheric velocities in gold/silver-sample objects are likely due to lower ejecta masses than extreme kinetic energies. {\it Right:} Measured photospheric \ion{Fe}{ii} velocities relative to IIn-like feature duration for gold/silver-sample objects. A possible correlation exists where objects with longer $t_{\rm IIn}$ also have lower photospheric velocities, suggesting a link between CSM-interaction timescales and explosion kinetic energy.
\label{fig:vphot} }
\end{figure*}

A fundamental question in the study of SNe~II is how the RSG progenitors of objects with IIn-like features differ from those without this early-time phenomenon. Ideally, we would like to know if there is any correlation between RSG ZAMS (or He core) mass and the creation of confined, high-density CSM capable of producing IIn-like features in some SNe~II. As discussed in \S\ref{sec:scaling}, the enhanced plateau luminosities of many gold/silver-sample events are consistent with larger progenitor radii and/or low ejecta masses, both of which connect to significant late-stage mass loss. However, among gold/silver-sample SNe, there is a large spread (e.g., 0.01--0.2~$\Msun$) in the estimated $^{56}$Ni masses, which covers model predictions for $M_{\rm ZAMS} = 9.5$--25~$\Msun$. Finally, gold/silver-sample objects are unlikely to be highly energetic explosions (e.g., \S\ref{sec:scaling}), with some SNe displaying quite slow expansion velocities during and after the plateau (e.g., $v_{\rm ph} < 3000~\kms$). As shown in the left panel of Figure \ref{fig:vphot}, this has interesting implications for the survival time of IIn-like features because $t_{\rm IIn}$ can be increased for the same CSM density profile if the shock speed is smaller than average (e.g., $<10^4~\kms$). 

\begin{figure*}[t!]
\centering
\subfigure{\includegraphics[width=0.49\textwidth]{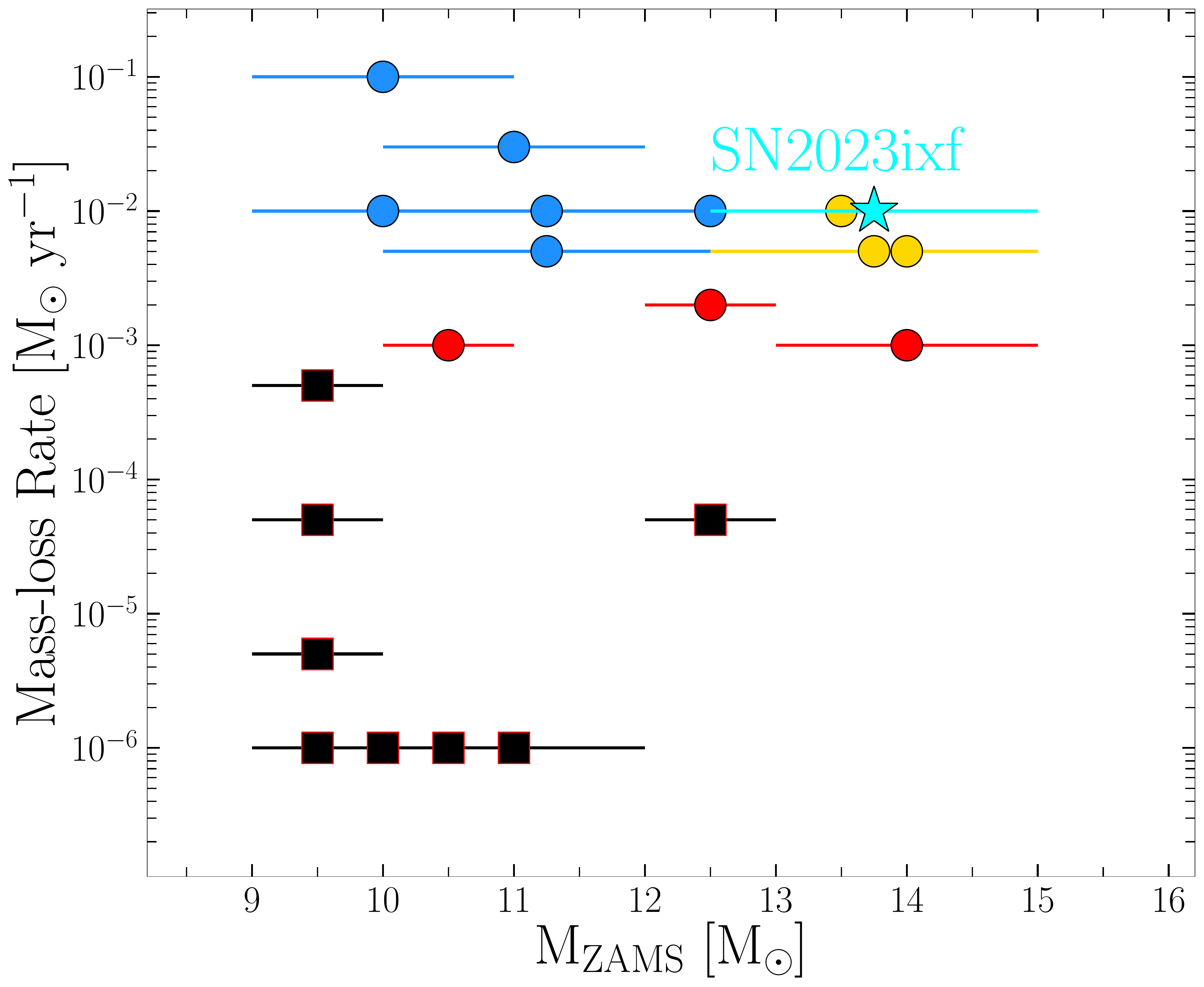}}
\subfigure{\includegraphics[width=0.49\textwidth]{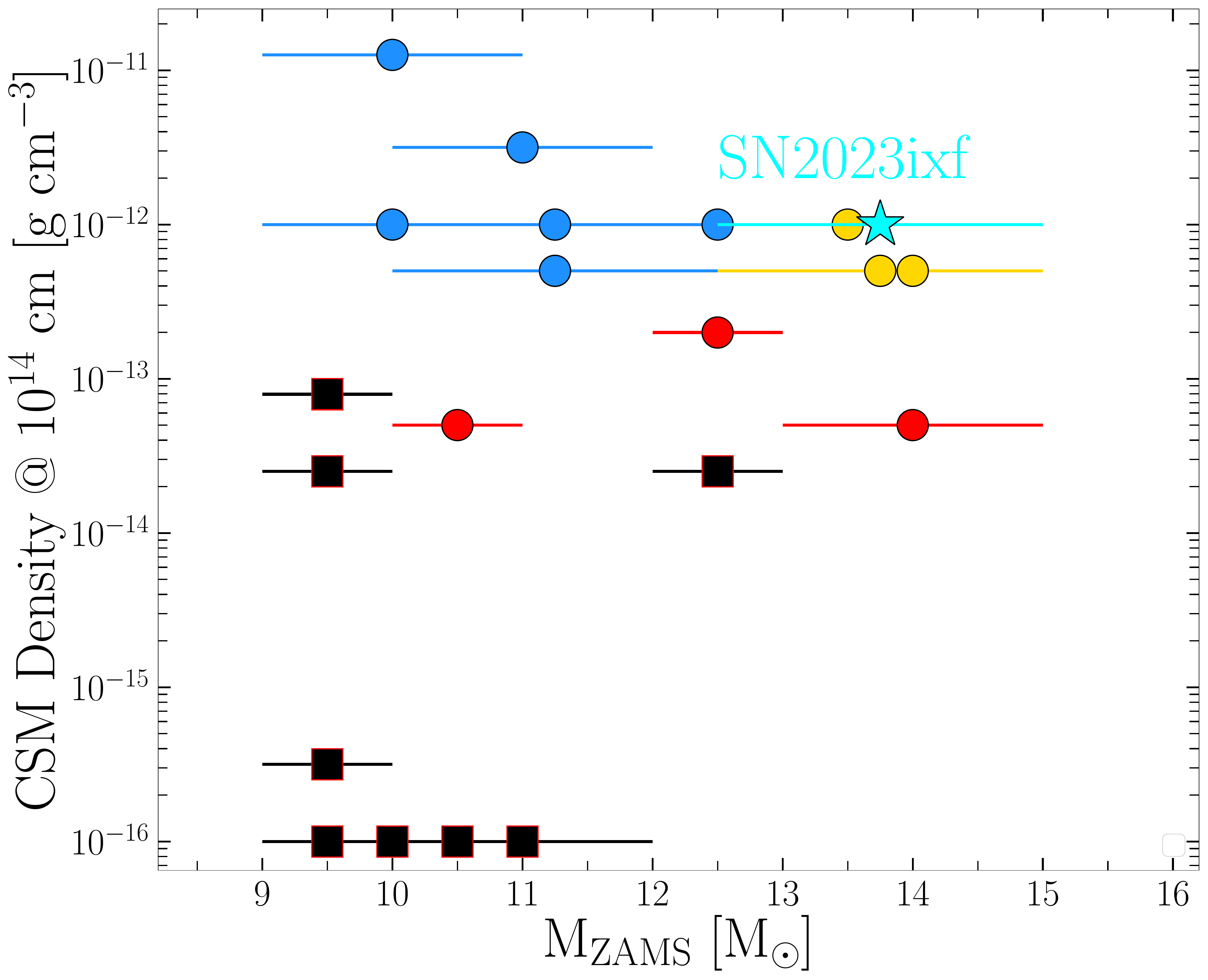}}\\
\subfigure{\includegraphics[width=0.49\textwidth]{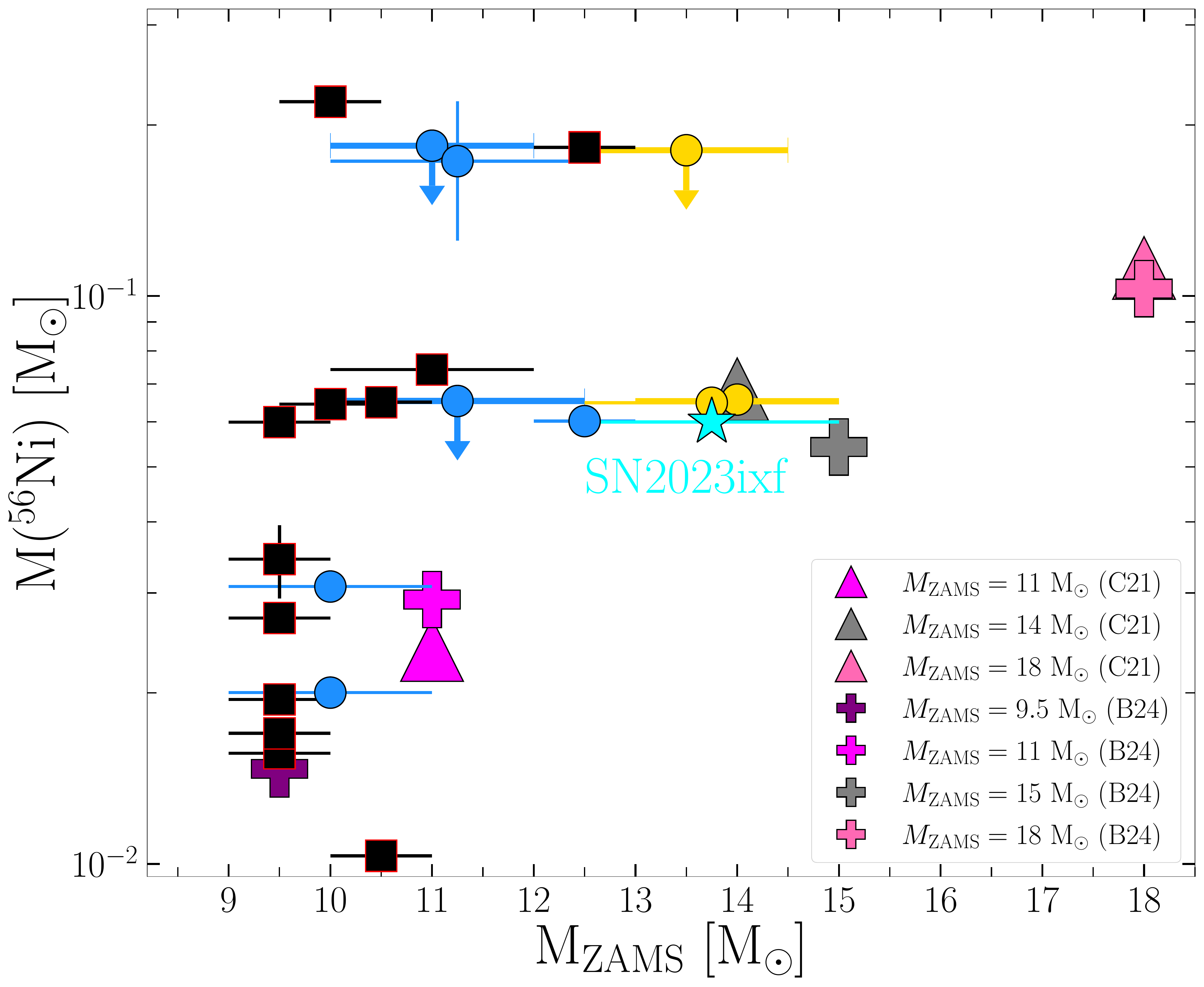}}
\subfigure{\includegraphics[width=0.49\textwidth]{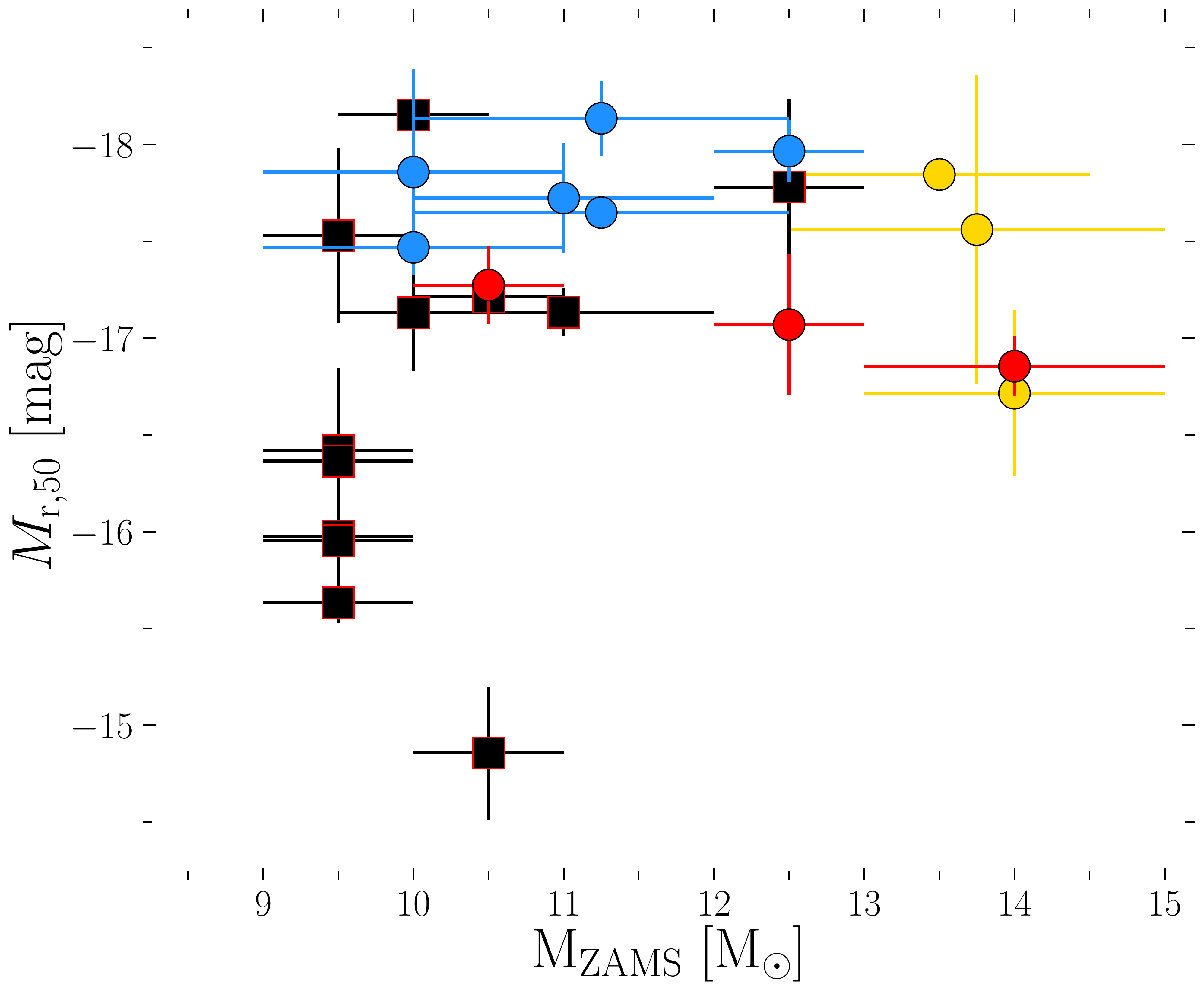}}
\caption{ {\it Upper Panel:} Mass-loss rate and CSM density at $10^{14}$~cm estimated from early-time spectra (WJG24a) compared to progenitor ZAMS masses inferred from nebular spectra (\S\ref{sec:zams}) for gold/silver (blue, yellow, red circles) and comparison-sample (black squares) objects. SN~2023ixf is shown as a cyan star for reference. {\it Lower Panel:} $^{56}$Ni mass and $M_{r,50}$ values compared to ZAMS masses estimated from nebular spectra. $^{56}$Ni mass predictions for different massive star core-collapse simulations from \cite{Curtis21} and \cite{burrows24} are shown as triangles and plus signs, respectively. 
\label{fig:ZAMS} }
\end{figure*}

To visualize a potential link between progenitor mass and confined CSM properties, we plot the mass-loss rate and CSM density at $10^{14}$~cm derived from early-time spectroscopic and photometric modeling in WJG24a with respect to the ZAMS mass derived from matching observed nebular spectra to models from \cite{Dessart21}. We estimate the range of best-matched ZAMS masses for a given object based on the closest forbidden line ratio (e.g., Fig. \ref{fig:LOICaII}) at all available spectral epochs, as well as the most consistent model to the complete optical nebular spectrum (e.g., Fig. \ref{fig:neb_models_gold}). As shown in Figure \ref{fig:ZAMS}, mass-loss rates and CSM densities range from $10^{-6}$--$10^{-1}$~\mdot\ and $10^{-16}$--$10^{-11}$~g~cm$^{-3}$, respectively, for $M_{\rm ZAMS}$ estimates of $\sim 9$--12~$\Msun$. For this subsample, we find only one comparison-sample object, SN~2021yja, that has a higher estimated ZAMS mass above $12~\Msun$ in addition to six gold/silver-sample objects, as well as SN~2023ixf (shown for reference). Furthermore, it is intriguing that there is a visible scarcity of higher mass ($>12~\Msun$) progenitors with low CSM densities at $\sim 10^{14}$~cm. However, we caution that this inference is made for a small sample of only 22 events with both early-time CSM estimates and nebular spectroscopy, which is far from a complete sample (i.e., not controlled). Nevertheless, it motivates the continuous monitoring of SNe~II in volume-limited surveys where there are robust constraints on the presence of dense, confined CSM. 


Examining potential diversity in SN~II progenitor masses in the context of mass-loss mechanisms can directly inform the theoretical picture for the creation of CSM around RSGs. One proposed model for explaining the range of CSM properties is that of convection-driven mass loss (``boil off''), which creates a chromosphere or effervescent zone of static, high-density material at $<5\,R_{\star}$ \citep{dessart17, soker21, Fuller24}. As shown by the modeling of \cite{Fuller24}, high-density material, capable of producing early-time IIn-like spectral features, can be created within the immediate vicinity of the RSG during the final years before explosion. Furthermore, while this mechanism is viewed as being stochastic, the amount of material present in this chromosphere is dependent on the convective velocity and, consequently, the mass of the RSG. Therefore, it may be the case that larger RSGs create higher density confined CSM and could then represent the progenitor systems of some SNe~II with IIn-like features at early times. Intriguingly, this could explain the general trend seen in Figure \ref{fig:ZAMS} between $M_{\rm ZAMS}$ and CSM density, but notably it cannot explain gold/silver-sample SNe with both large CSM densities and smaller inferred progenitor masses (e.g., SNe~2018zd, 2020tlf, and 2021aaqn). 

For outlier events with high $\dot M$ but low ZAMS mass, an additional mass-loss mechanism needs to be invoked. We now know that wave-driven mass loss is insufficient to produce the CSM densities needed to explain SNe~II with IIn-like signatures (e.g., \citealt{fuller17, Wu21}), but other core processes such as Si deflagrations leading to pre-SN outbursts could be invoked to explain the creation of this dense, confined material \citep{woosley15}. Notably, these core instabilities are specifically predicted for low-mass stars ($\sim 9$--11~$\Msun$), consistent with the ZAMS mass estimates of SNe~2018zd, 2020tlf, and 2021aaqn. However, more modeling needs to be done in order to properly treat the deflagration in the O core and the consequent injection of energy into the convective RSG envelope (e.g., \citealt{Tsang22}). Beyond this, another possibility is binary interaction that leads to the formation of confined, high-density CSM in the final years before explosion. This has been explored for SN~II progenitors by \cite{Ercolino24} wherein they model the Case C mass transfer involving an RSG with a main-sequence companion. These simulations show that mass-transfer rates up to $\sim 0.01$~\mdot\ can occur during late-stage evolution, leading to CSM density profiles that are compatible with SNe~II having IIn-like features. Furthermore, explosion modeling of these systems reveals significant diversity among the light-curve and spectral properties of SNe~II \citep{Dessart24binary}.



\subsection{Shock-Power Emergence at Late Times}\label{sec:highmdot}

While it is clear that the progenitors of gold/silver-sample objects undergo enhanced mass loss in their final years to decades, what is the nature of the more distant CSM ($>10^{16}$~cm) that traces pre-explosion mass loss in the final centuries to millennia before explosion? As discussed in \S\ref{sec:nebspec}, some gold/silver-sample objects show evidence for boxy emission within the H$\alpha$/[\ion{O}{i}] complex, which is commonly associated with a radiative reverse shock injecting power into a CDS; the bulk of this mass is formed in the first few days post-explosion as the SN shock sweeps up the densest, most local CSM \citep{Chevalier94, Chugai07, Chevalier17, dessart17, dessart22}. Notably, the kinetic shock power scales with the mass-loss rate as $L_{\rm sh} = \dot M v_{\rm sh}^3/2v_{\rm w}$ and any luminosity from free-free emission of the cooling reverse-shocked gas goes as $L_{\rm RS} \propto (\dot M/v_{\rm w})^2$. Therefore, the detection of broad, boxy H$\alpha$ emission from the CDS in late-time gold/silver-sample object spectra confirms ongoing CSM interaction with sufficiently dense CSM and larger mass-loss rate (e.g., $>10^{-6}$~\mdot) at radii $>10^{16}$~cm. Furthermore, it has been shown that the range in H$\alpha$ widths follows the range in CDS velocity, which scales with total CDS mass and the strength of the CSM interaction (e.g., see \citealt{Dessart23}). 

Interestingly, the boxy emission observed in some SNe~II is not present in comparison-sample objects until phases $>800$~days (e.g., SNe~2017eaw, 2021yja; Fig. \ref{fig:CDS}) and has been modeled as the emergence of shock power as the primary luminosity source in SNe~II at very late times when radioactive-decay power becomes subdominant \citep{Dessart23PWR}. These observations have interesting implications for late-stage RSG mass loss because if the evolution of these SNe~II without IIn-like signatures can be modeled with $\dot M \lesssim 10^{-6}$~\mdot, then the fact that CDS emission is visible in SNe~II with IIn-like signatures as early as $\sim 200$--300~days implies $\dot M > 10^{-6}$~\mdot\ at a few hundred years before explosion. Furthermore, as demonstrated in Figure \ref{fig:neb_spec_pwr}, there are gold/silver-sample events that either show very weak or no forbidden emission lines, only displaying strong H$\alpha$ emission as early as $\sim 120$~days after first light. These spectra are consistent with models from \cite{dessart22} with shock powers of $10^{42}$--$10^{43}$~erg~s$^{-1}$, suggesting even higher mass-loss rates of $>10^{-5}$~\mdot. This spectral morphology/evolution has been discussed in the context of late-time CSM interaction for gold-sample object PTF11iqb by \cite{smith15} which, like these other SNe, resembles SN~II 1998S whose multiwavelength late-time emission suggested a mass-loss rate of $>10^{-4}$~\mdot\ \citep{Pooley02, Mauerhan12, Dessart16}. 

For SNe~II that exhibit strong CSM-interaction signatures within their first hundreds of days, the formation of dust in their CDS may significantly affect the line profiles arising from the dense shell as well as quench forbidden-line luminosity \citep{Gall14}. As shown in Figure \ref{fig:CDS}, there is a noticeable blue/red asymmetry in the H$\alpha$ profile for some of these events, suggesting that dust formation may be present in the CDS. Importantly, dust formed in the CDS will absorb emission arising from the innermost, metal-rich ejecta and therefore any measurement of forbidden-line flux (e.g., [\ion{O}{i}]) will be underestimated \citep{Dessart25dust}. This phenomenon may be contributing to the lack of forbidden-line emission in some of these gold/silver-sample events at $>1$~yr post-explosion, but it could also influence the nebular emission-line strength of SNe~II that display a blend of expected forbidden lines and a boxy H$\alpha$ profile. As shown in SN~2023ixf, the H$\alpha$ emission at only $\delta t = 364$~days is strongly affected by dust attenuation, which may be blocking forbidden-line emission from the innermost ejecta. This has significant implications for using the [\ion{O}{i}] luminosity as a proxy for progenitor ZAMS mass in SNe with CDS emission. Nevertheless, using forbidden-line ratios of [\ion{O}{i}] and [\ion{Ca}{ii}] to derive ZAMS masses is safe from this effect because both lines will be reduced approximately equally by dust attenuation. 

\section{Conclusions} \label{sec:conclusion}

In this paper, we present additional analysis and late-time observations of a sample of SNe~II with (``gold/silver'') and without (``comparison'') spectroscopic evidence for CSM interaction at very early-time phases ($\delta t <1$~week), originally presented by WJG24a. Below we summarize the primary observational findings from this work.  

\begin{itemize}
    \item We compile multiband and pseudobolometric light curves of all gold/silver- and comparison-sample objects with photometry extending beyond $\delta t = 50$~days (\S\ref{sec:plateau}). We measure significantly brighter plateau luminosities ($L_{50}$) and faster decline rates ($S_{50}$) at $\delta t = 50$~days for gold/silver-sample objects relative to comparison-sample objects. This statistically significant contrast in sample populations is also present in $g$- and $r$-band light-curve measurements. However, we find no statistical evidence that the plateau durations of gold/silver- and comparison-sample objects come from separate distributions. 

    \item We derive $^{56}$Ni masses and $\gamma$-ray trapping timescales ($t_{\gamma}$) by fitting post-plateau pseudobolometric light curves with a analytic formalism for the radioactive-decay decline rate observed in SN~1987A (\S\ref{sec:ni56}). Similar to previous SN~II studies, we observe a general trend between $L_{50}$ and $M(^{56}$Ni), but with significant variance among gold/silver-sample objects (e.g., some events have relatively low $^{56}$Ni mass but large plateau luminosities). We find no statistically significant difference between the $M(^{56}$Ni) values derived for gold/silver versus comparison samples. Furthermore, we find evidence for incomplete $\gamma$-ray trapping ($t_{\gamma} < 250$~days) in five gold/silver- and three comparison-sample objects, implying low ejecta masses of $\sim 4$--6~$\Msun$ for these SNe.
    

    \item In \S\ref{sec:nebspec}, we analyze the nebular spectra of 19 gold/silver- and 13 comparison-sample objects at phases $\delta t >200$~days. We observe a continuum of late-time spectral properties, in particular within the H$\alpha$/[\ion{O}{i}] complex. For example, all comparison and five gold/silver SNe show expected forbidden transitions, while the remaining gold/silver-sample objects show ``boxy'' H$\alpha$ emission either blended with [\ion{O}{i}] or as the dominant emission line in the late-time spectrum. This emission is commonly associated with a CDS located between the forward and reverse shocks, containing mostly material swept up in the first few days post-explosion, and powered by ongoing ejecta interaction with significantly dense, distant ($r>10^{16}$~cm) CSM. Compared to {\tt CMFGEN} models for late-time shock power from CSM interaction in SNe~II by \cite{dessart22}, the boxy H$\alpha$ luminosities and emergence times suggest mass-loss rates as high as $\sim 10^{-4}$~\mdot\ at distances $>10^{16}$~cm (i.e., hundreds of years pre-explosion) in some fraction of SN~II progenitors. 

    \item In \S\ref{sec:zams}, we use the nebular SN~II models from \cite{Dessart21} to estimate progenitor-star ZAMS masses using both [\ion{O}{i}]/[\ion{Ca}{ii}] line ratios as well as direct spectral comparison for objects with $M(^{56}$Ni) measurements that can be used for normalization. Intriguingly, we find that most sample objects with nebular spectra can be matched by models for $M_{\rm ZAMS} \approx 9$--$11~\Msun$, while only 7 SNe~II are best described by models with $M_{\rm ZAMS} \geq 12.5~\Msun$. However, we find no sample objects consistent with progenitor masses $>15~\Msun$.

    \item Placing the plateau properties of gold/silver-sample objects in the context of SN~II scaling relations indicates that the large $L_{50}$ values may be related to either lower ejecta masses and/or higher kinetic energies and/or large progenitor radii. The photospheric velocities of SNe~II with IIn-like features confirm that they do not arise from intrinsically energetic explosions and suggest that either lower ejecta mass and/or envelope inflation is a primary cause of their large plateau luminosities. However, we note that significantly strong, ongoing CSM interaction may contribute to the brightness of some objects. Furthermore, we show that SN~II observables such as $L_{50}$ are highly inaccurate metrics for progenitor core properties (e.g., Fe-core mass). 

    \item We compare the mass-loss rates and CSM densities at $10^{14}$~cm derived from early-time observations (e.g., WJG24a) to the ZAMS masses inferred from nebular spectra in Figure \ref{fig:ZAMS}. SNe~II with mass-loss rates ranging from $10^{-6}$ to  $10^{-1}$~\mdot\ ($\rho_{14} = 10^{-16}$--$10^{-11}$~g~cm$^{-3}$) can all be described by low-mass progenitor models (e.g., 9--11~$\Msun$), while only SNe~II consistent with $M_{\rm ZAMS} > 12~\Msun$ models have CSM densities at the upper end of the continuum. If representative of all SN~II progenitors, it suggests multiple channels for enhanced mass loss in the final years before explosion: one that may be progenitor-mass dependent (e.g., convection-driven) and one that is more stochastic (e.g., eruptions, binary interaction). 

\end{itemize}

This study highlights the need for continuous, multiwavelength monitoring of SNe~II throughout their evolution. Furthermore, it is necessary to include knowledge gleamed from early-time observations (e.g., confined CSM densities) when analyzing large samples of SNe~II during their plateau and at late times. Obtaining larger spectroscopically and volumetrically complete samples of SNe~II with late-time light curves and nebular spectra is essential to better constrain the progenitors of all SNe~II, in particular those with large CSM densities at radii $<10^{15}$~cm. 

\section{Acknowledgments} \label{Sec:ack}

We thank Jared Goldberg and Jim Fuller for valuable discussions.

The Young Supernova Experiment (YSE) and its research infrastructure is supported by the European Research Council under the European Union's Horizon 2020 research and innovation programme (ERC Grant Agreement No.\ 101002652, PI K.\ Mandel), the Heising-Simons Foundation (2018-0913, PI R.\ Foley; 2018-0911, PI R.\ Margutti), NASA (NNG17PX03C, PI R.\ Foley), NSF (AST-1720756, AST-1815935, PI R.\ Foley), the David \& Lucille Packard Foundation (PI R.\ Foley), VILLUM FONDEN (project \#16599, PI J.\ Hjorth), and the Center for AstroPhysical Surveys (CAPS) at NCSA and the University of Illinois Urbana-Champaign.

W.J.-G. is supported by NASA through Hubble Fellowship grant HSTHF2-51558.001-A awarded by the Space Telescope Science Institute, which is operated for NASA by the Association of Universities for Research in Astronomy, Inc., under contract NAS5-26555. This research was supported in part by the NSF under grant PHY-1748958.  The Margutti team at UC Berkeley is partially funded by the Heising-Simons Foundation under grants \#2018-0911 and \#2021-3248 (PI R. Margutti). R.C. acknowledges support from NASA {\it Swift} grant 80NSSC22K0946. The TReX team at UC Berkeley is supported in part by the NSF under grants AST-2221789 and AST-2224255, and by the Heising-Simons Foundation under grant \#2021-3248 (PI R. Margutti).

C.D.K. gratefully acknowledges support from the NSF through AST-2432037, the HST Guest Observer Program through HST-SNAP-17070 and HST-GO-17706, and from JWST Archival Research through JWST-AR-6241 and JWST-AR-5441. 
C.G. and D.F are supported by a VILLUM FONDEN Young Investigator Grant (project \#25501) and VILLUM FONDEN Experiment grant (VIL69896).
This work was funded by ANID, Millennium Science Initiative, ICN12\_009.
The work of X.W. is supported by the National Natural Science Foundation of China (NSFC grants 12288102 and 12033003) and the New Cornerstone Science Foundation through the XPLORER PRIZE.
This work was granted access to the HPC resources of TGCC under the allocation 2021 -- A0110410554  and 2022 -- A0130410554 made by GENCI, France.
This research was supported by the Munich Institute for Astro-, Particle and BioPhysics (MIAPbP) which is funded by the Deutsche Forschungsgemeinschaft (DFG, German Research Foundation) under Germany's Excellence Strategy – EXC-2094 – 390783311.
K.A.B. is supported by an LSSTC Catalyst Fellowship; this publication was thus made possible through the support of grant 62192 from the John Templeton Foundation to LSSTC. The opinions expressed in this publication are those of the authors and do not necessarily reflect the views of LSSTC or the John Templeton Foundation.

A.V.F.'s research group at UC Berkeley acknowledges financial assistance from the Christopher R. Redlich Fund, as well as donations from Gary and Cynthia Bengier, Clark and Sharon Winslow, Alan Eustace and Kathy Kwan, William Draper, Timothy and Melissa Draper, Briggs and Kathleen Wood, and Sanford Robertson (W.Z. is a Bengier-Winslow-Eustace Specialist in Astronomy, T.G.B. is a Draper-Wood-Robertson Specialist in Astronomy, Y.Y. was a Bengier-Winslow-Robertson Fellow in Astronomy). Numerous other donors to his group and/or research at Lick Observatory include Douglas and Judith Adams, Raymond Adams, Lawrence Anderson, Charlie Baxter and Jinee Tao, Duncan Beardsley, Eric Behrens and Joyce Hicks, Barbara Berliner, Jack Bertges, Susan Broadston, Ruth Bromer and Joseph Huberman, Ann Brown, Patrick Bukowski and Eve Grossman-Bukowski, Michael and Sharon Burch, Tina Butler, Richard and Susan Cardwell, Alan and Jane Chew, Donna Clarke, Jim Connelly and Anne Mackenzie, Christopher and Patricia Cook, Laurence and Carole Coole, Curtis and Shelley Covey, Roger Cukras, Michael Danylchuk, Lisa Danylchuk, Robert and Margaret Davenport, Byron and Allison Deeter, James and Hilda DeFrisco, Darril dela Torre and Helen Levay, Paul and Margaret Denning, Jane and Patrick Donnelly, Barbara Edwards, Paul and Silvia Edwards, Christopher and Joan Ennis, Arthur and Cindy Folker, Peter and Robin Frazier, Heidi Gerster, Ernest Giachetti, Charles and Gretchen Gooding, Jeffrey Green, Richard and Carol Gregor, Misako and Dennis Griffin, Thomas and Dana Grogan, Judith and Timothy Hachman, Gregory Hirsch, Alan and Gladys Hoefer, Russell and Susan Holdstein, the Hugh Stuart Center Charitable Trust, George Hume,  Charles and Patricia Hunt, Stephen and Catherine Imbler, John and Virginia Johnson, Ross Jones and Jane Paul, Michael Kast and Rebecca Lyon, Joel Krajewski, Steven Kusnitz, Max Lacounte, Rudi Lindner, Katherine Lipka, Walter and Karen Loewenstern, Gregory Losito and Veronica Bayduza, Jerri Mariott and Michael Silpa, Herbert Masters III, Bruce Maximov and Susan Albert, Louisa McNatt, Joan and Oliver Mellows, Joseph Meyer and Karyn Chung, Bruce and Judith Moorad, Rand Morimoto and Ana Henderson, Eric and Patty Ng, Edward Oates, Doug and Emilie Ogden, Sandra Otellini, Angelo Paparella, Ned and Ellin Purdom, Charles Pyle, Richard Reeder, Jonathan and Susan Reiter, Richard Rissel, Paul Robinson, Catherine Rondeau, Ben Samman, Geraldine Sandor, Theodore Sarbin Jr., Laura Sawczuk and Luke Ellis, Stanley and Miriam Schiffman, Tom and Alison Schneider, Richard and Betsey Sesler, Ajay Shah and Lata Krishnan, Lauren and Jerry Shen, Hans Spiller, Richard and Shari Stegman, Justin and Seana Stephens, Ilya Strebulaev and Anna Dvornikova, Charles and Darla Stevens, Christopher Stookey and Sandra Yamashiro, Benjamin Sykes, Marie Teixeira, Eric Tilenius, Eudora and James Ting, David Turner III and Joanne Turner, Andrew Waterman, Kirk and Jacqueline Weaver, Gerald and Virginia Weiss, Janet Westin and Mike McCaw, Byron and Nancy Wood, Richard Wylie, Wen Yang, and others.

M.R.D. acknowledges support from the NSERC through grant RGPIN-2019-06186, the Canada Research Chairs Program, and the Dunlap Institute at the University of Toronto.
%
%
X.W. acknowledges support by the National Natural Science Foundation of China (NSFC grants 12288102 and 12033003).  
V.A.V. acknowledges support by the NSF under grant AST-2108676. C.R.A. was supported by grants from VILLUM FONDEN (project \#16599 and \#25501). 

Parts of this research were supported by the Australian Research Council Centre of Excellence for Gravitational Wave Discovery (OzGrav), through project number CE230100016.
The UCSC team is supported in part by NASA grant 80NSSC20K0953, NSF grant AST--1815935, the Gordon \& Betty Moore Foundation, the Heising-Simons Foundation, and by a fellowship from the David and Lucile Packard Foundation to R.J.F.

Based in part on observations made with the Nordic Optical Telescope, owned in collaboration by the University of Turku and Aarhus University, and operated jointly by Aarhus University, the University of Turku and the University of Oslo, representing Denmark, Finland and Norway, the University of Iceland and Stockholm University at the Observatorio del Roque de los Muchachos, La Palma, Spain, of the Instituto de Astrofisica de Canarias. Observations were obtained under program P62-507 (PI C. Angus).

This work includes data obtained with the Swope telescope at Las Campanas Observatory, Chile, as part of the Swope Time Domain Key Project (PI A. Piro; CoIs Coulter, Drout, Phillips, Holoien, French, Cowperthwaite, Burns, Madore, Foley, Kilpatrick, Rojas-Bravo, Dimitriadis, Hsiao). We thank Abdo Campillay, Yilin Kong-Riveros, Piera Soto-King, and Natalie Ulloa for observations on the Swope telescope.

Some of the data presented herein were obtained at the W. M. Keck Observatory, which is operated as a scientific partnership among the California Institute of Technology, the University of California, and NASA. The Observatory was made possible by the generous financial support of the W. M. Keck Foundation. The authors wish to recognize and acknowledge the very significant cultural role and reverence that the summit of Maunakea has always had within the indigenous Hawaiian community. We are most fortunate to have the opportunity to conduct observations from this mountain.
A major upgrade of the Kast spectrograph on the Shane 3~m telescope at Lick Observatory, led by Brad Holden, was made possible through generous gifts from the Heising-Simons Foundation, William and Marina Kast, and the University of California Observatories. Research at Lick Observatory is partially supported by a generous gift from Google.

Based in part on observations obtained with the Samuel Oschin 48-inch Telescope at the Palomar Observatory as part of the Zwicky Transient Facility project. ZTF is supported by the NSF under grant AST-1440341 and a collaboration including Caltech, IPAC, the Weizmann Institute for Science, the Oskar Klein Center at Stockholm University, the University of Maryland, the University of Washington, Deutsches Elektronen-Synchrotron and Humboldt University, Los Alamos National Laboratories, the TANGO Consortium of Taiwan, the University of Wisconsin at Milwaukee, and the Lawrence Berkeley National Laboratory. Operations are conducted by the Caltech Optical Observatories (COO), the Infrared Processing and Analysis Center (IPAC), and the University of Washington (UW).

The Pan-STARRS1 Surveys (PS1) and the PS1 public science archive have been made possible through contributions by the Institute for Astronomy, the University of Hawaii, the Pan-STARRS Project Office, the Max-Planck Society and its participating institutes, the Max Planck Institute for Astronomy, Heidelberg and the Max Planck Institute for Extraterrestrial Physics, Garching, The Johns Hopkins University, Durham University, the University of Edinburgh, the Queen's University Belfast, the Harvard-Smithsonian Center for Astrophysics, the Las Cumbres Observatory Global Telescope Network Incorporated, the National Central University of Taiwan, STScI, NASA under grant NNX08AR22G issued through the Planetary Science Division of the NASA Science Mission Directorate, NSF grant AST-1238877, the University of Maryland, Eotvos Lorand University (ELTE), the Los Alamos National Laboratory, and the Gordon and Betty Moore Foundation.

This work makes use of observations taken by the Las Cumbres Observatory global telescope network. The Las Cumbres Observatory Group is funded by NSF grants AST-1911225 and AST-1911151.
The new SALT data presented here were obtained through Rutgers University program 2022-1-MLT-004 (PI S. Jha).
Funding for the Lijiang 2.4~m telescope has been provided by the CAS and the People's Government of Yunnan Province.

\facilities{\emph{Neil Gehrels Swift Observatory}, Zwicky Transient Facility, ATLAS, YSE/PS1, Lick/Shane (Kast), Lick/Nickel, MMT (Binospec), Keck I/II (LRIS, DEIMOS), Las Cumbres Observatory, {\it TESS}, Gemini Observatory (GMOS)}

\software{IRAF (Tody 1986, Tody 1993), photpipe \citep{Rest+05}, DoPhot \citep{Schechter+93}, HOTPANTS \citep{becker15}, YSE-PZ \citep{Coulter22, Coulter23}, \cmfgen\ \citep{hillier12, D15_2n}, \heracles\ \citep{gonzalez_heracles_07,vaytet_mg_11,D15_2n}, HEAsoft (v6.33; HEASARC 2014), DRAGONS \citep{Labrie23}, Lpipe \citep{Perley19} }

\bibliographystyle{aasjournal} 
\bibliography{references} 


\clearpage
\appendix
\counterwithin{figure}{section}

\renewcommand\thetable{A\arabic{table}} 
\setcounter{table}{0}

\section{Supplementary Figures \& Tables}\label{subsec:suppl}

\begin{figure*}[t!]
\centering
\subfigure{\includegraphics[width=0.325\textwidth]{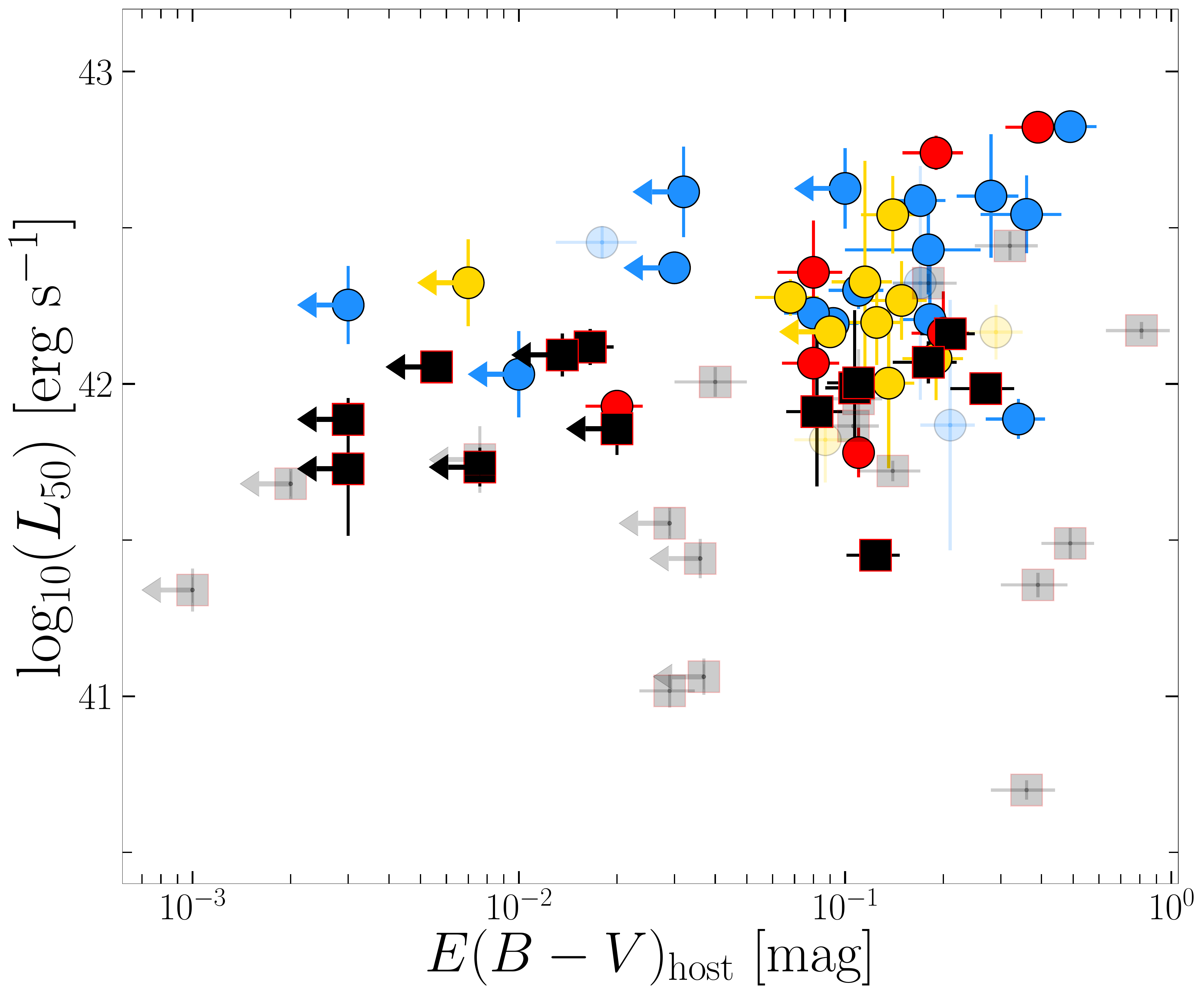}}
\subfigure{\includegraphics[width=0.33\textwidth]{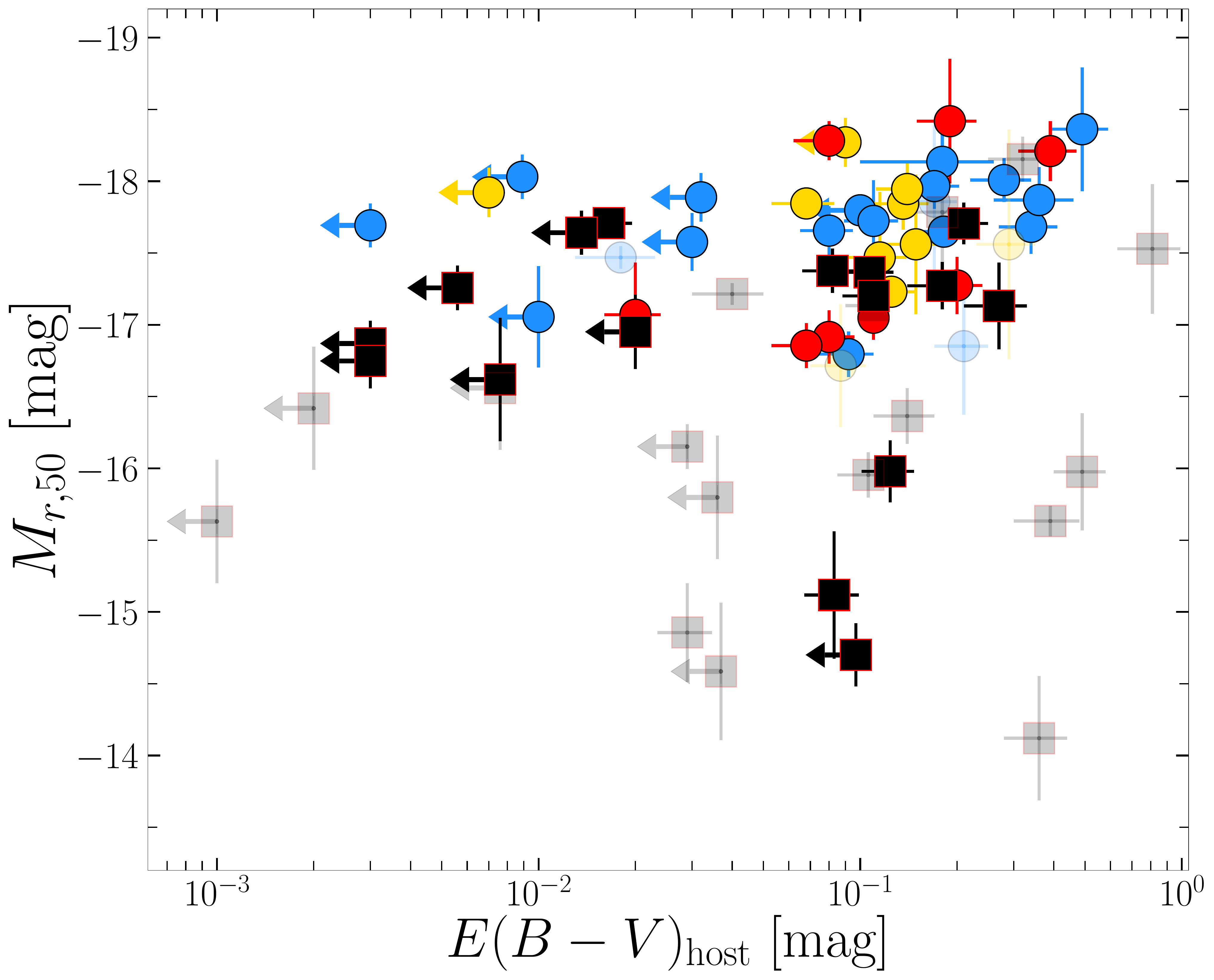}}
\subfigure{\includegraphics[width=0.33\textwidth]{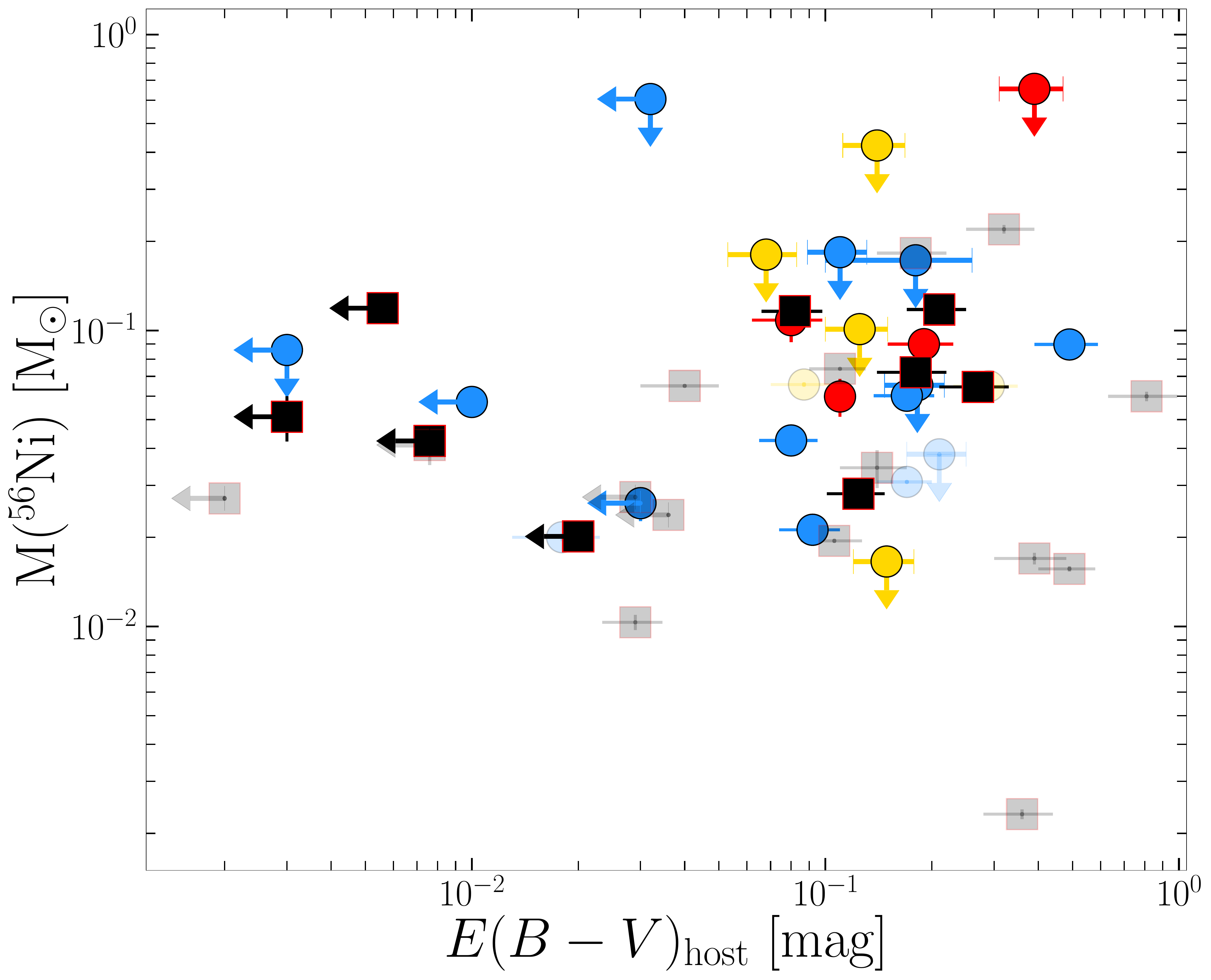}}
\caption{ Host-galaxy reddening for gold/silver- and comparison-sample objects compared to measured bolometric and $r$-band luminosities at +50~days as well as $^{56}$Ni mass.
\label{fig:EBV} }
\end{figure*}

\begin{figure*}[t!]
\centering
\subfigure{\includegraphics[width=0.44\textwidth]{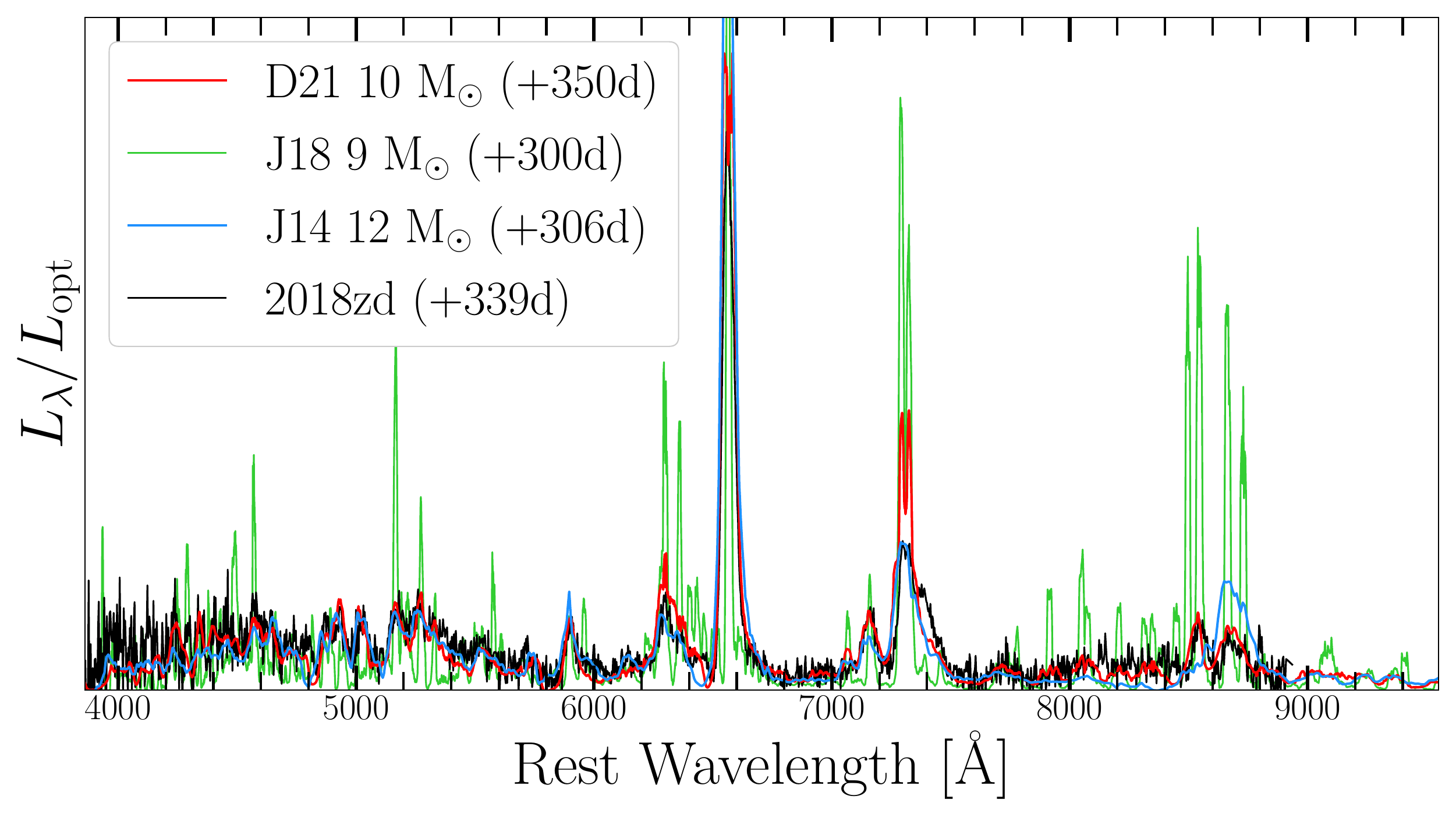}}
\subfigure{\includegraphics[width=0.44\textwidth]{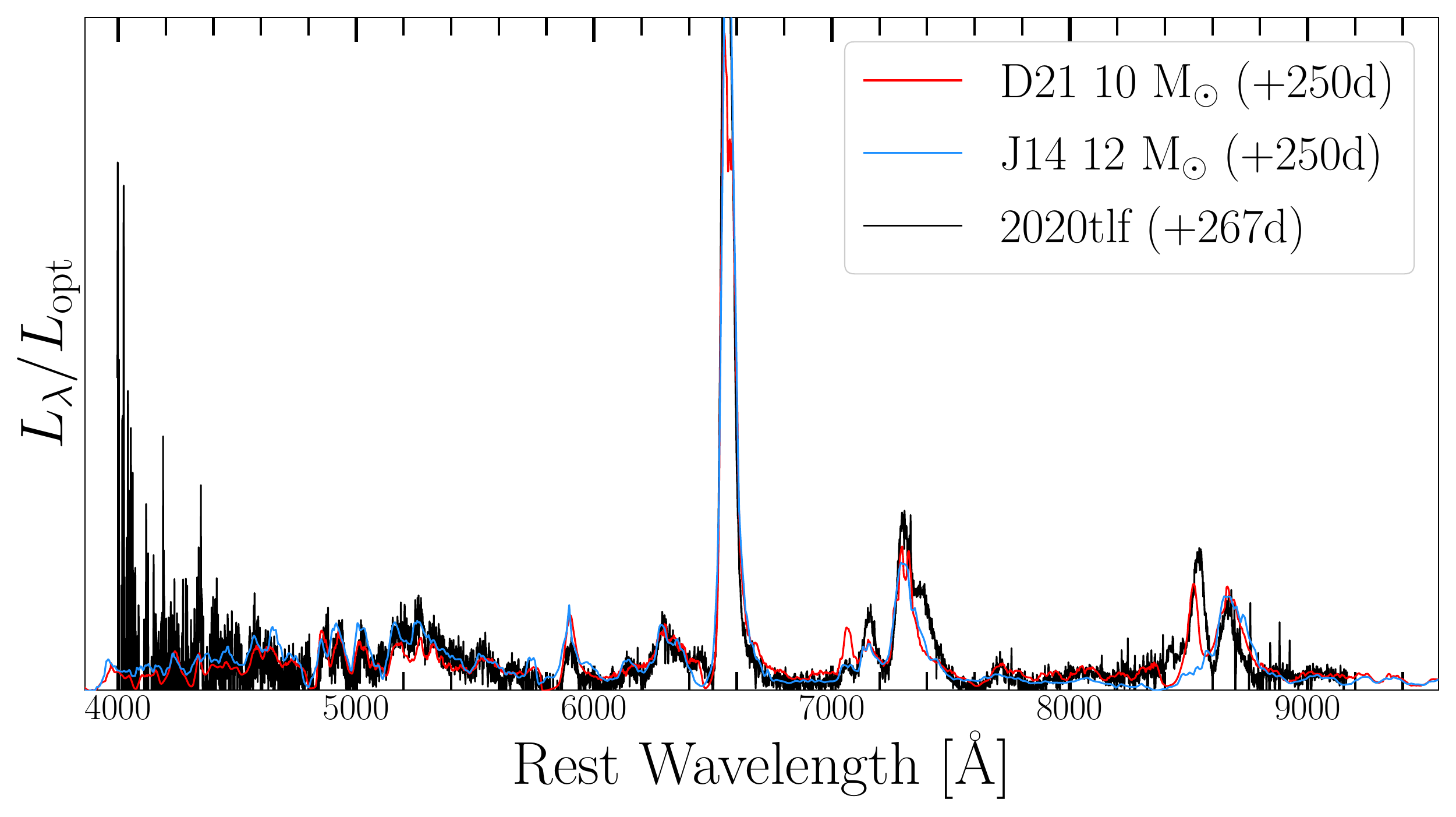}}\\
\subfigure{\includegraphics[width=0.44\textwidth]{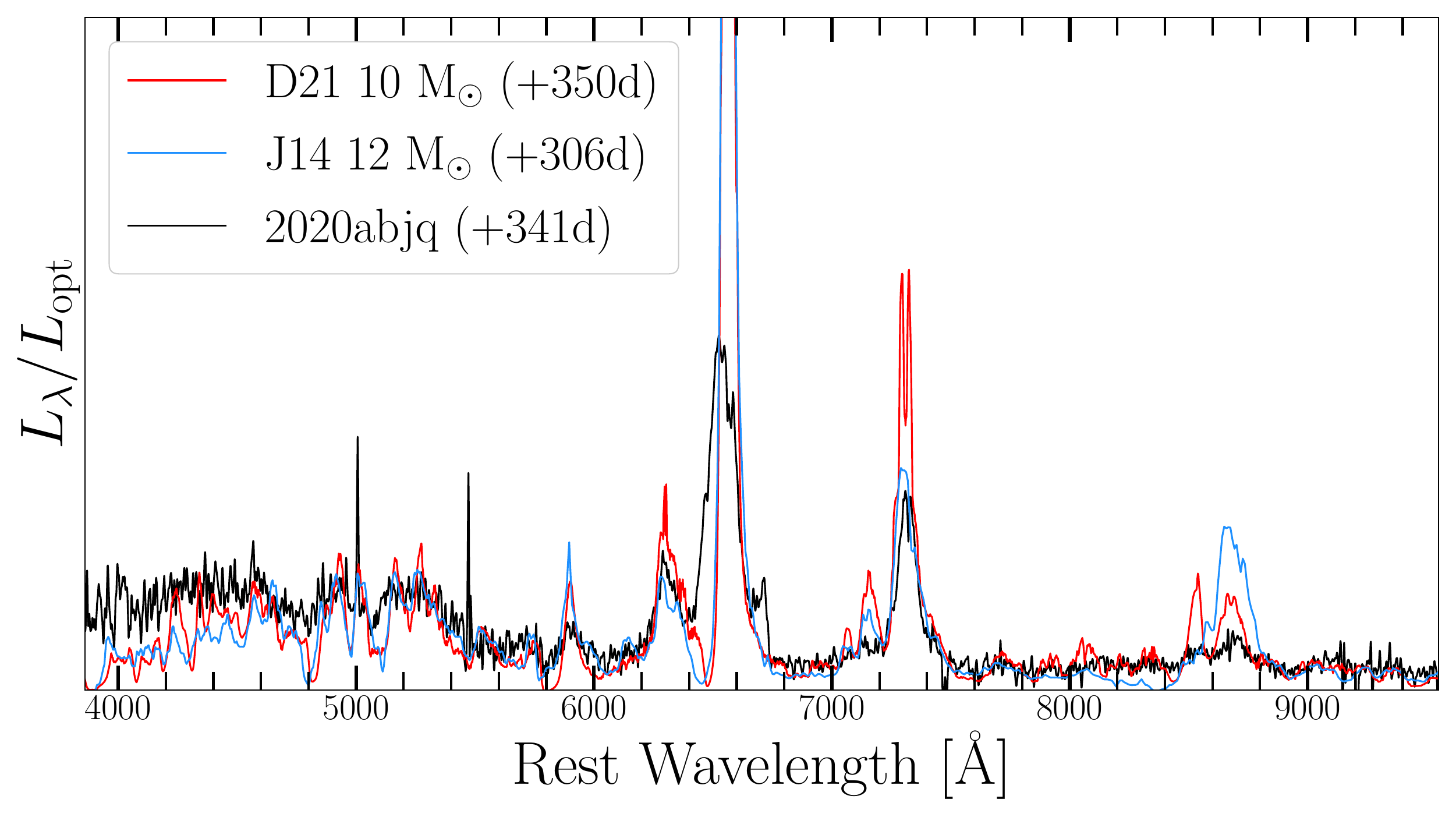}}
\subfigure{\includegraphics[width=0.44\textwidth]{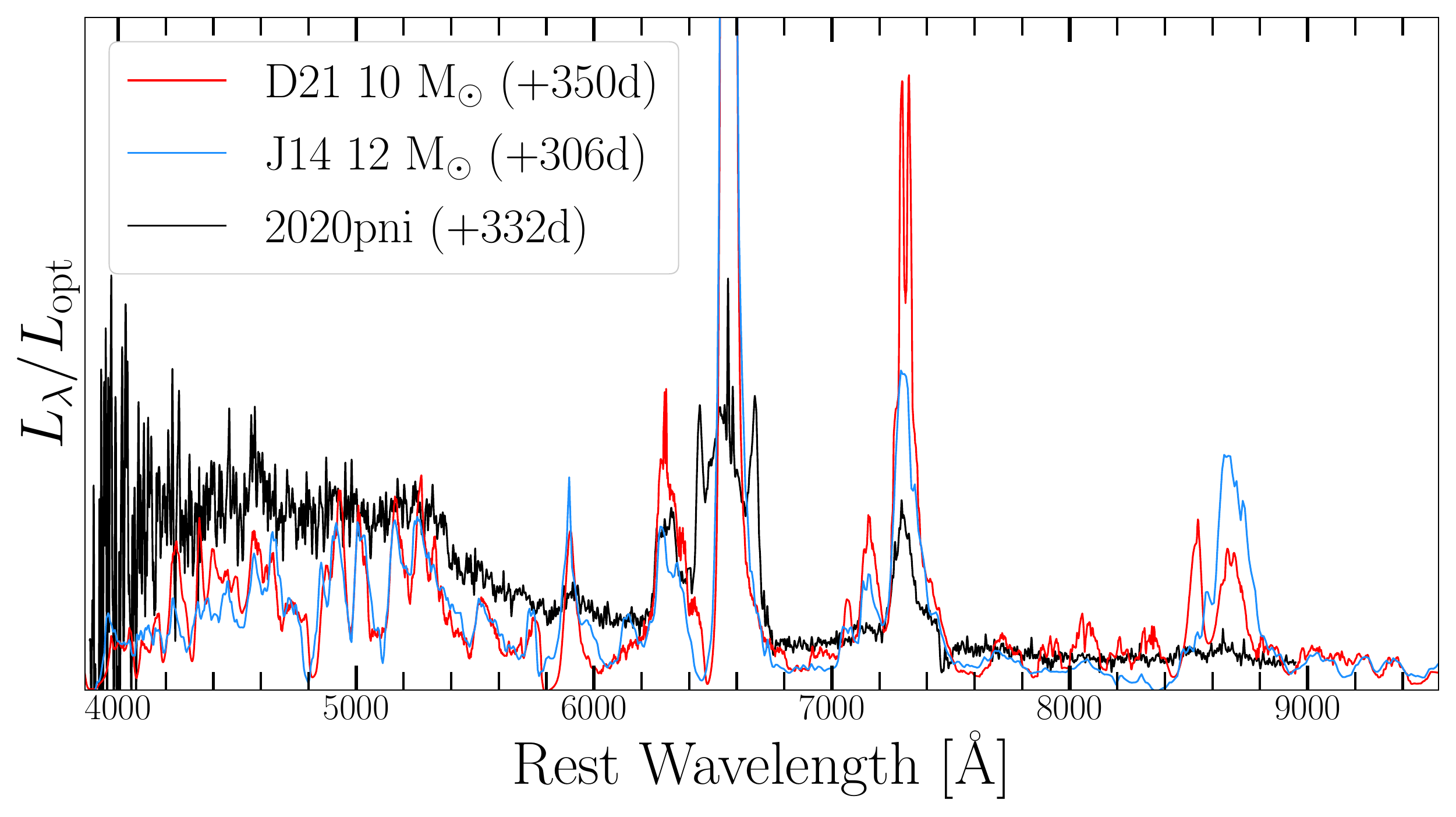}}\\
\subfigure{\includegraphics[width=0.44\textwidth]{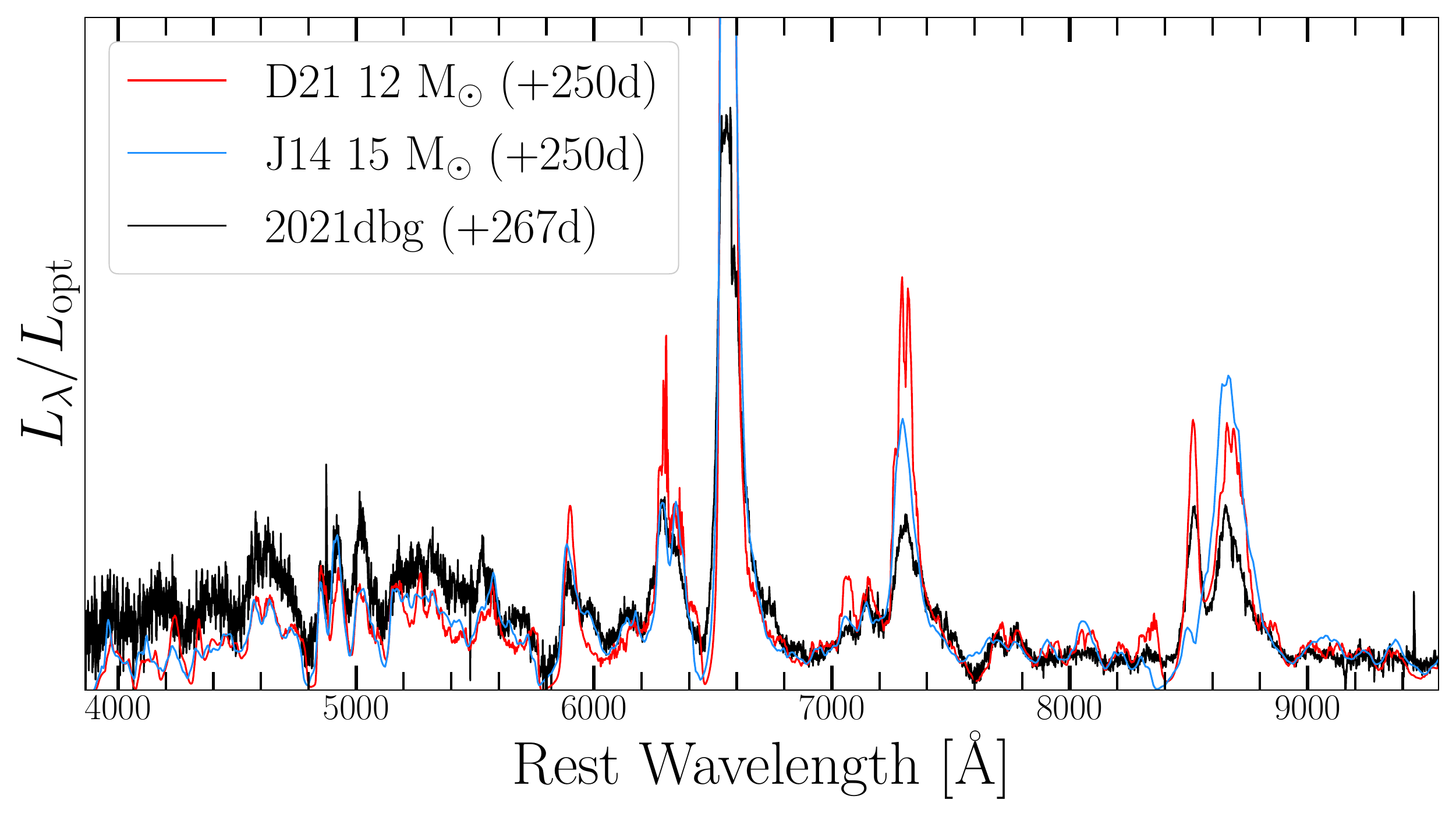}}
\subfigure{\includegraphics[width=0.44\textwidth]{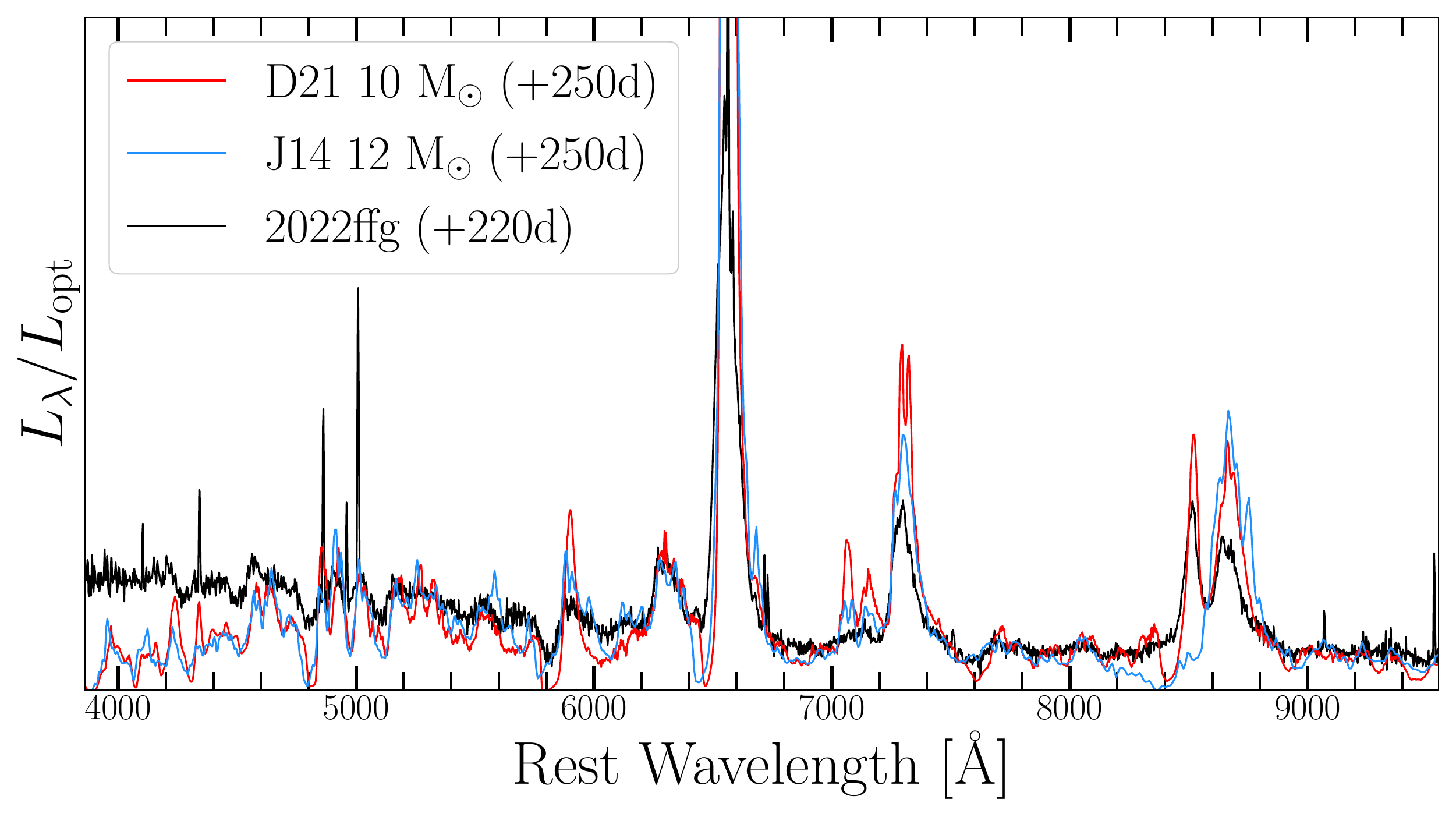}}\\
\subfigure{\includegraphics[width=0.44\textwidth]{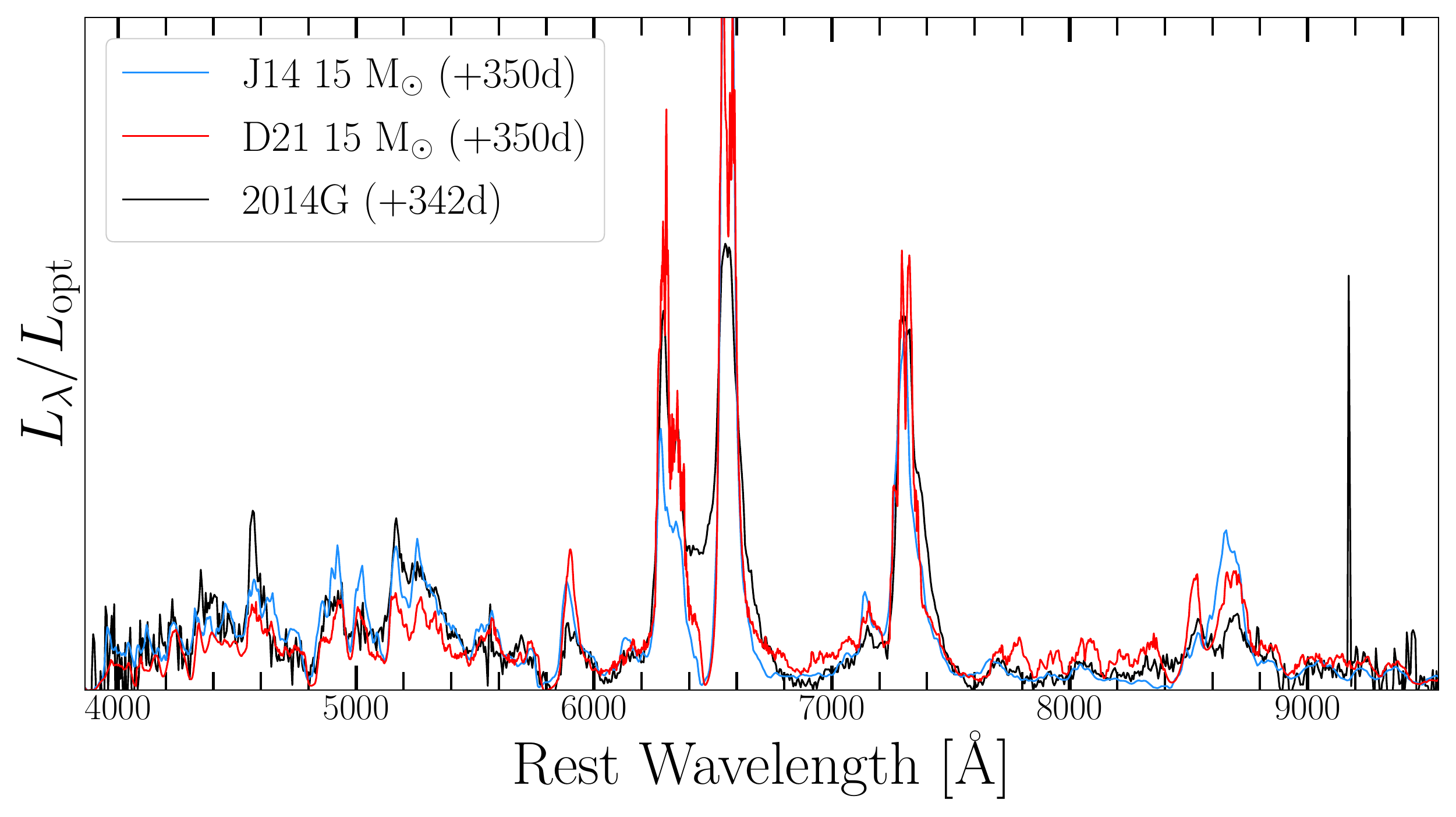}}
\subfigure{\includegraphics[width=0.44\textwidth]{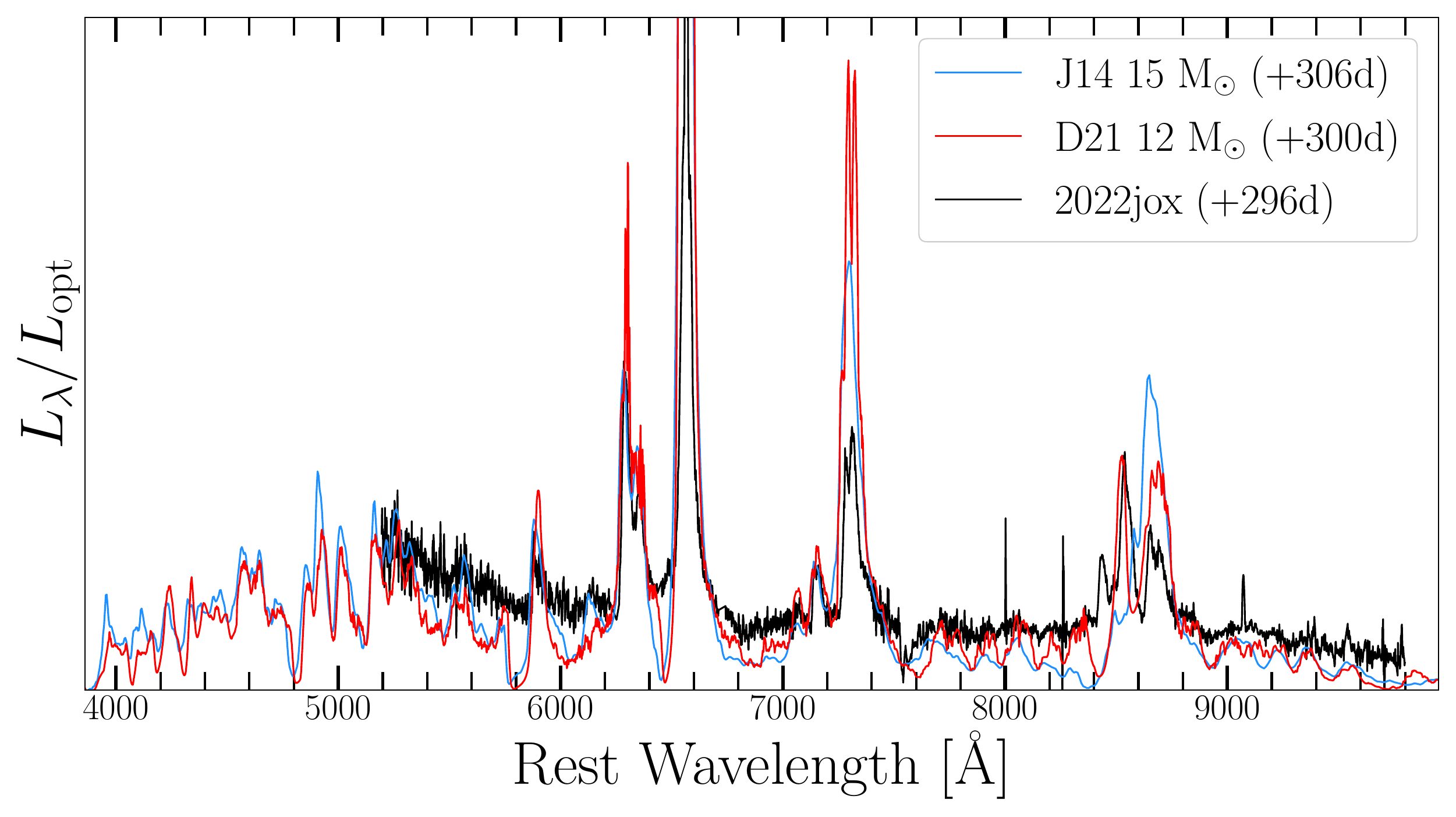}}\\
\subfigure{\includegraphics[width=0.44\textwidth]{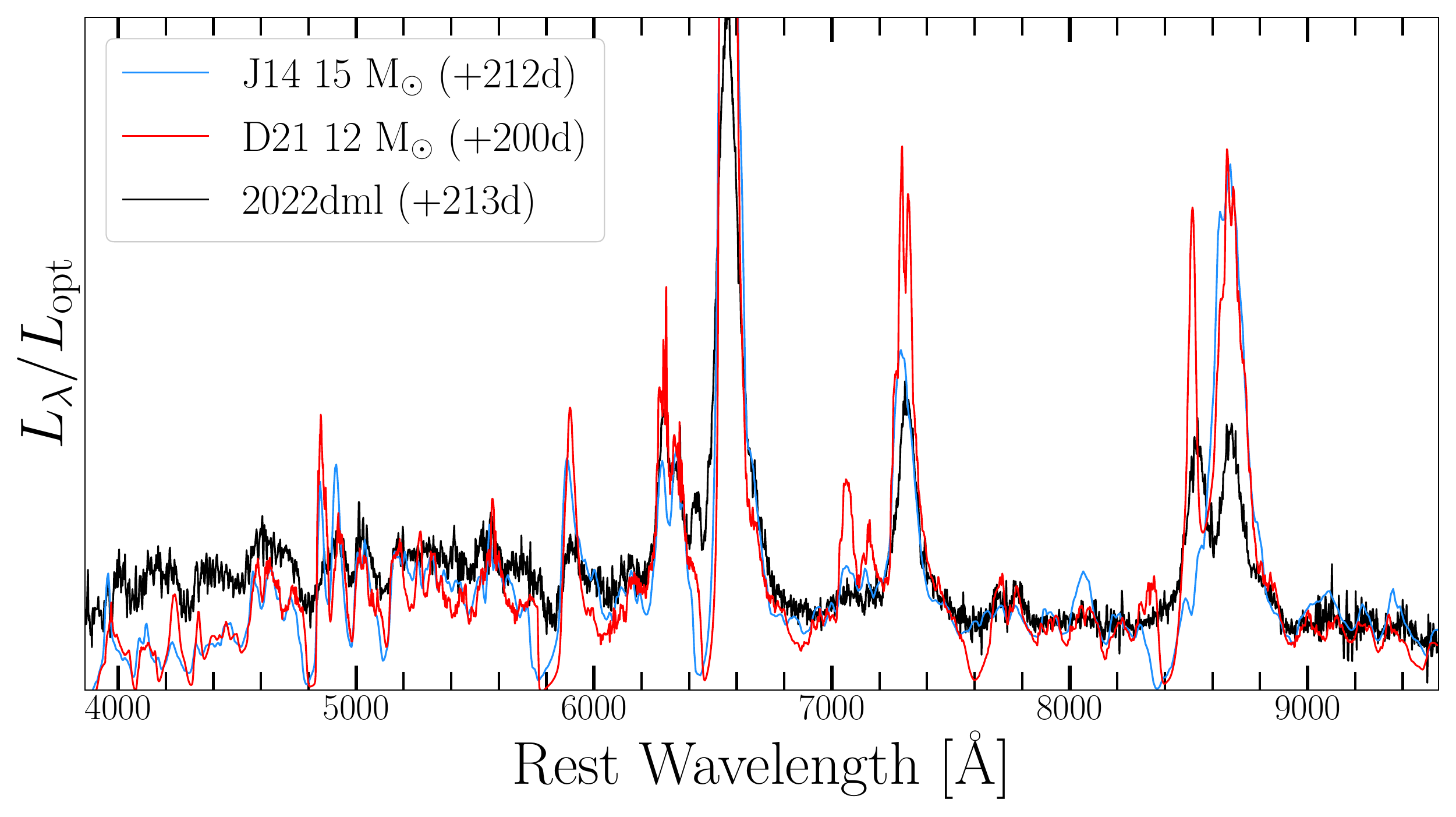}}
\subfigure{\includegraphics[width=0.44\textwidth]{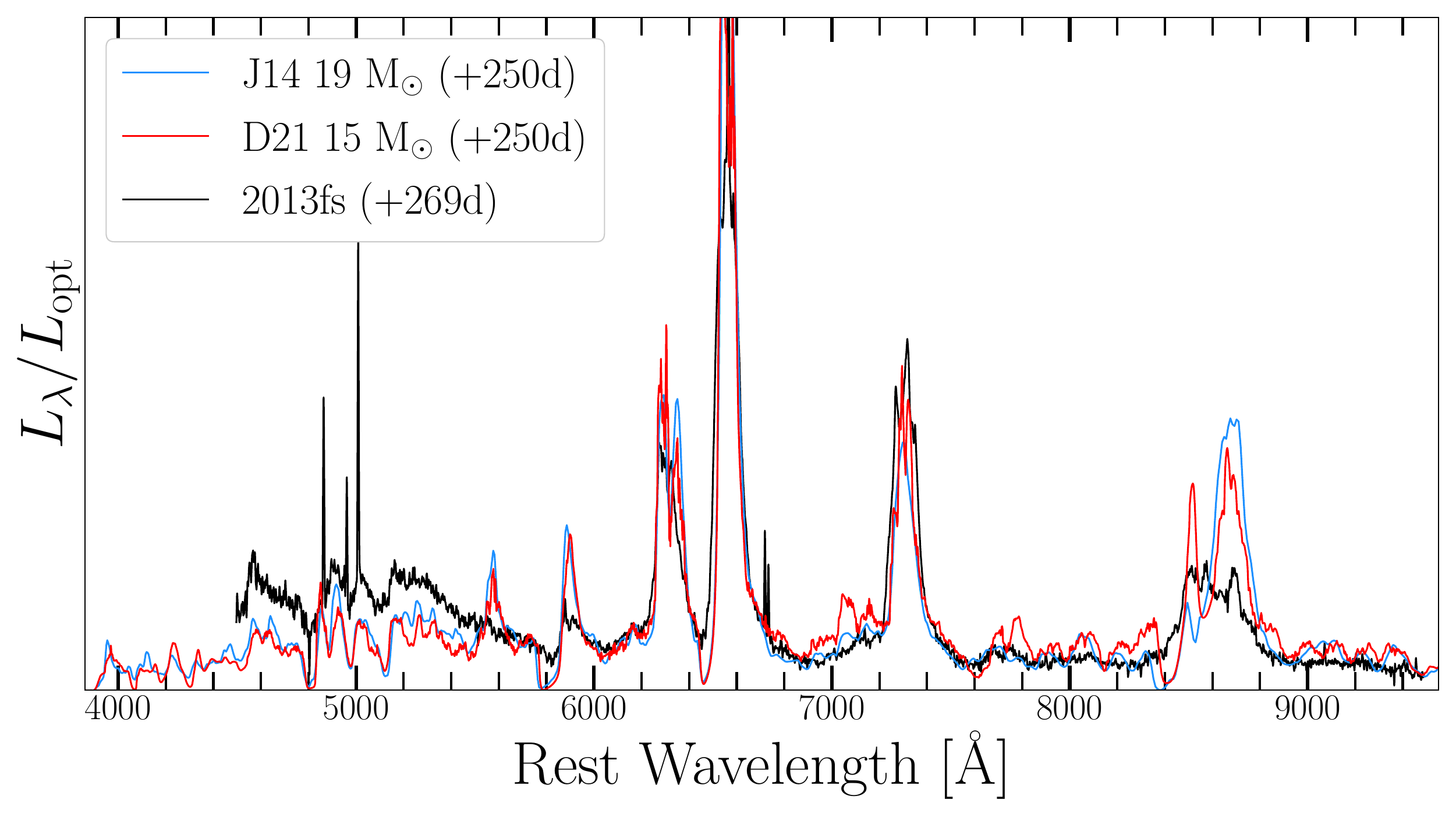}}
\caption{  Nebular spectra of sample objects (black) compared to best-matched model spectra from \cite{Dessart21} (red) and \cite{Jerkstrand14} (blue).  
\label{fig:neb_models_gold} }
\end{figure*}

\begin{figure*}[t!]
\centering
\subfigure{\includegraphics[width=0.44\textwidth]{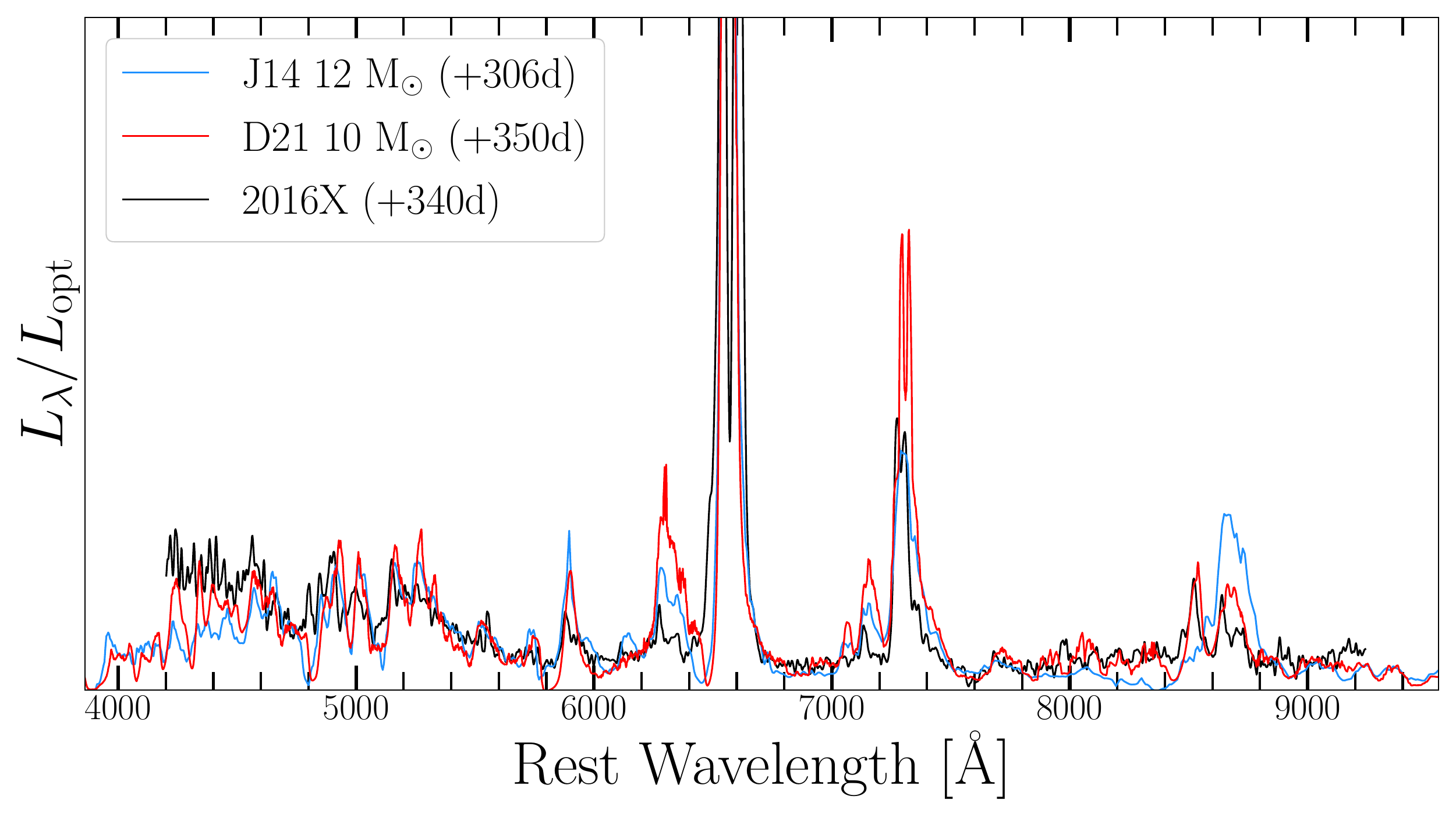}}
\subfigure{\includegraphics[width=0.44\textwidth]{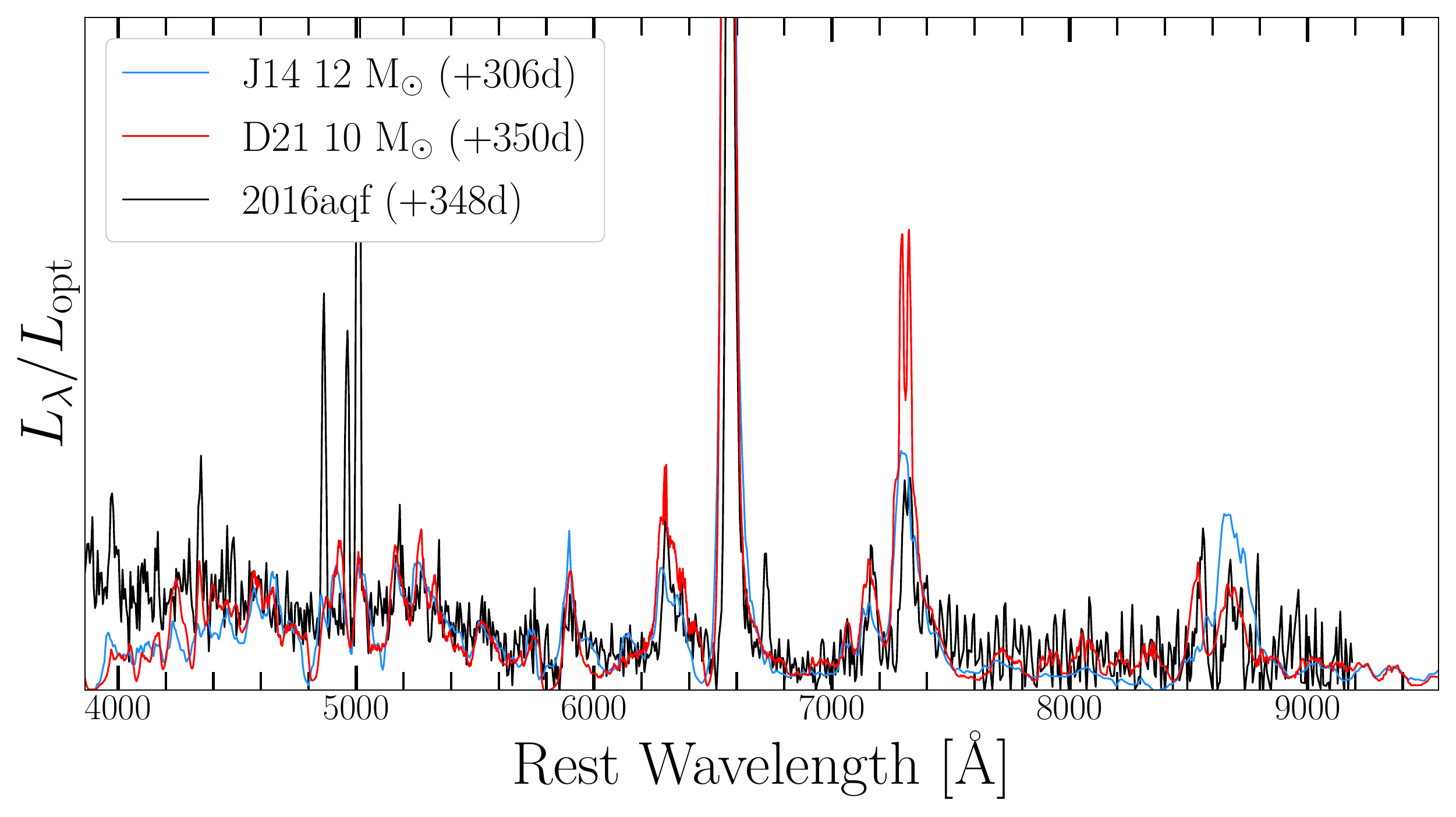}}\\
\subfigure{\includegraphics[width=0.44\textwidth]{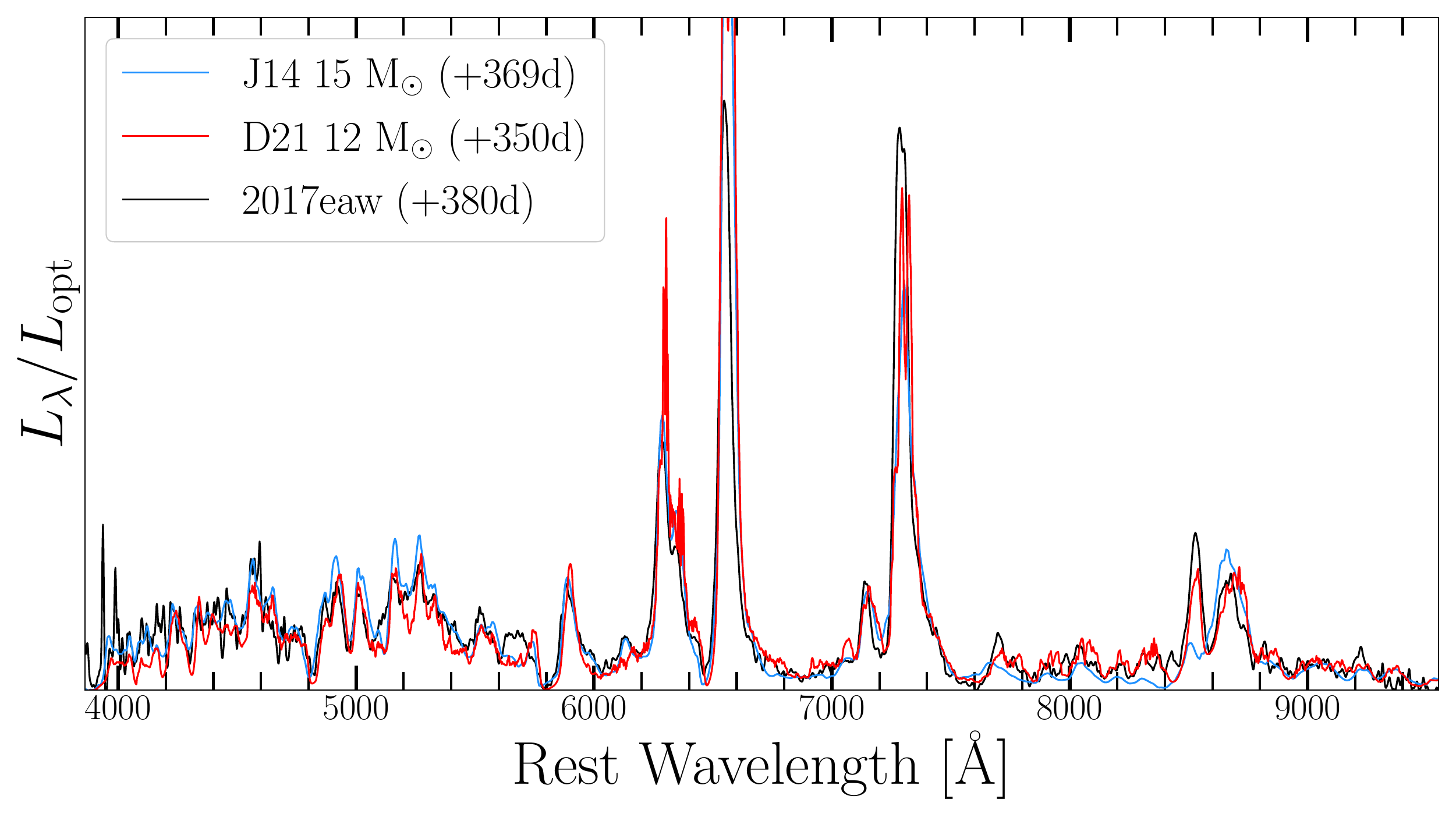}}
\subfigure{\includegraphics[width=0.44\textwidth]{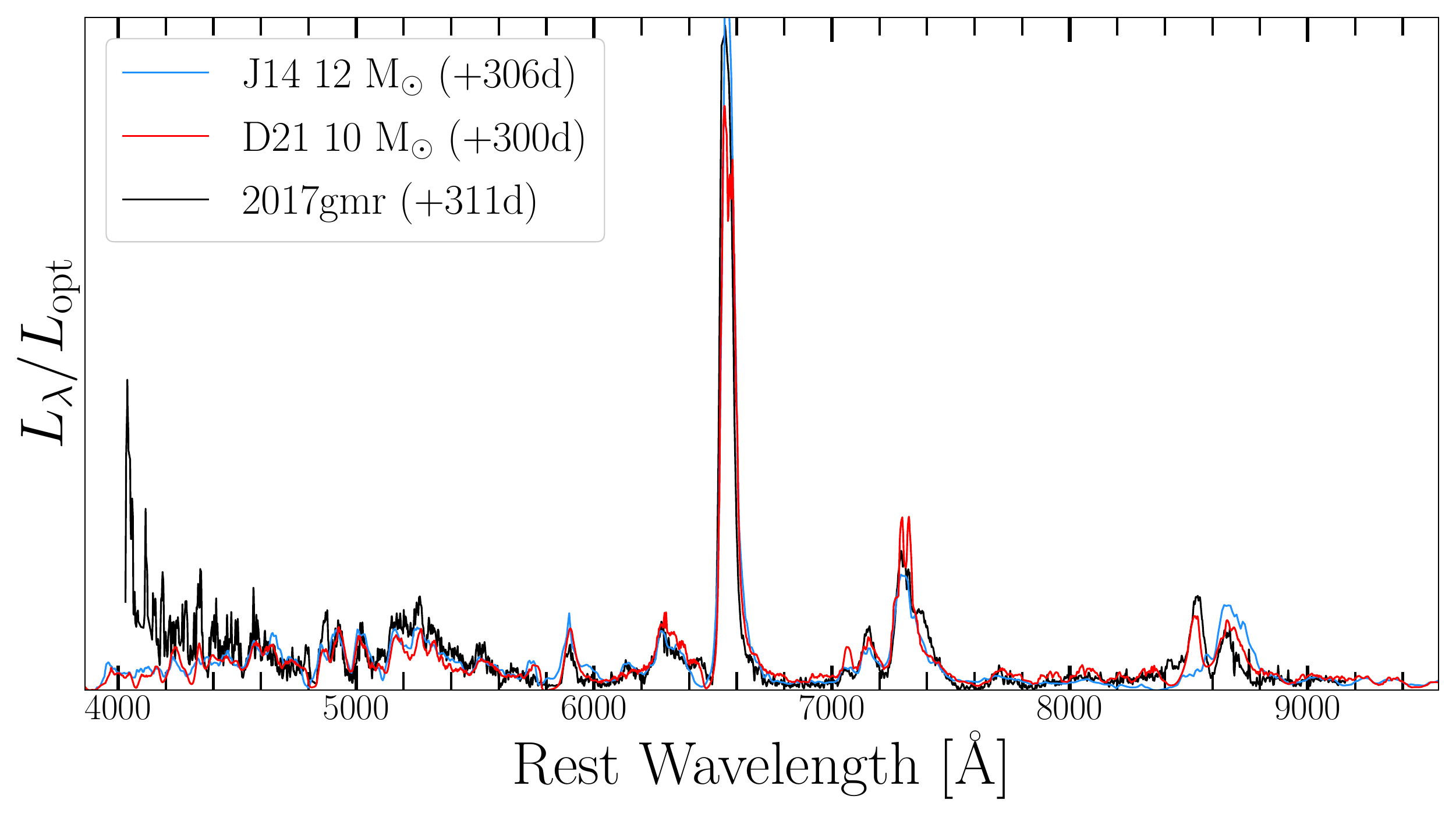}}\\
\subfigure{\includegraphics[width=0.44\textwidth]{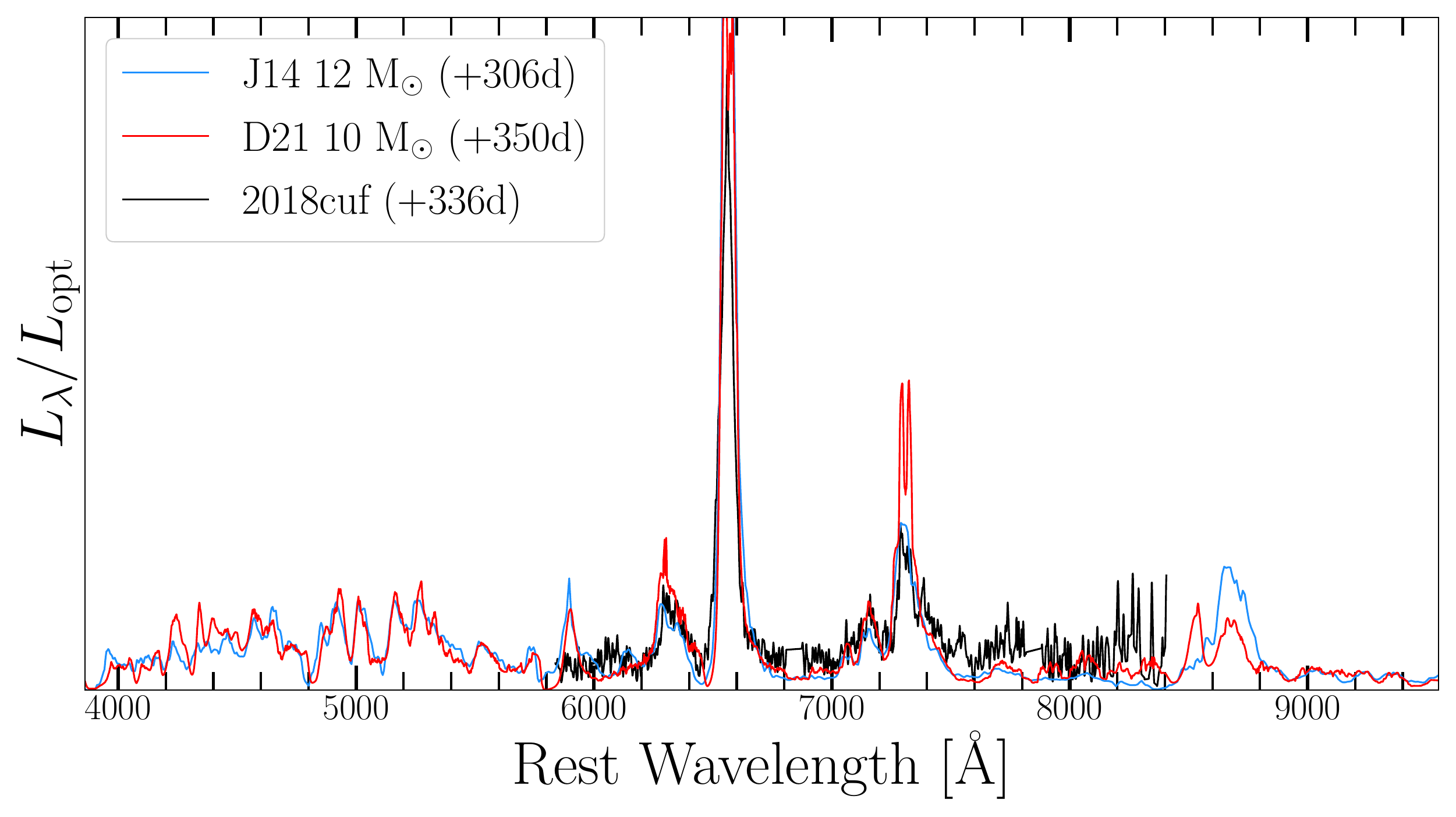}}
\subfigure{\includegraphics[width=0.44\textwidth]{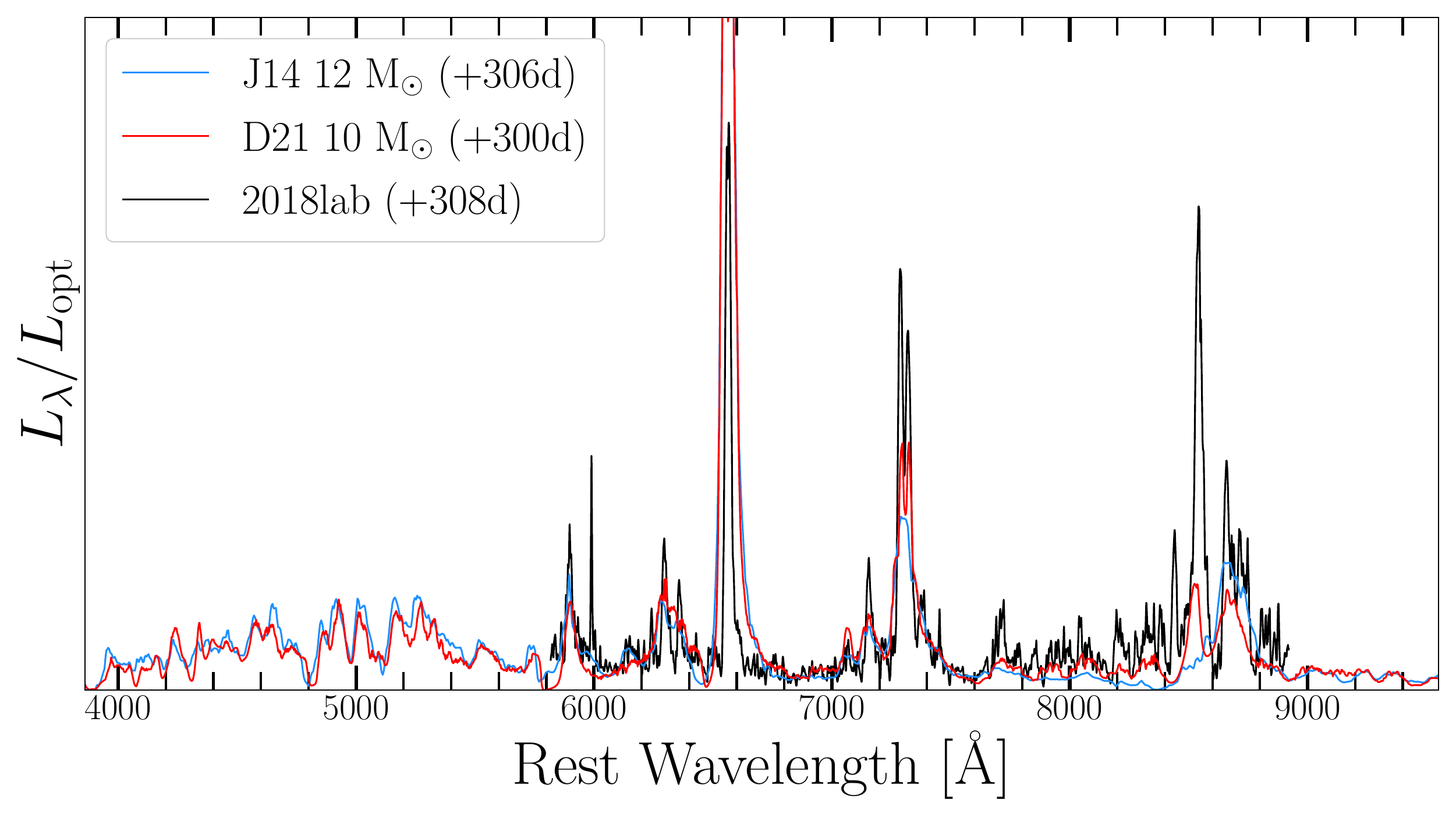}}\\
\subfigure{\includegraphics[width=0.44\textwidth]{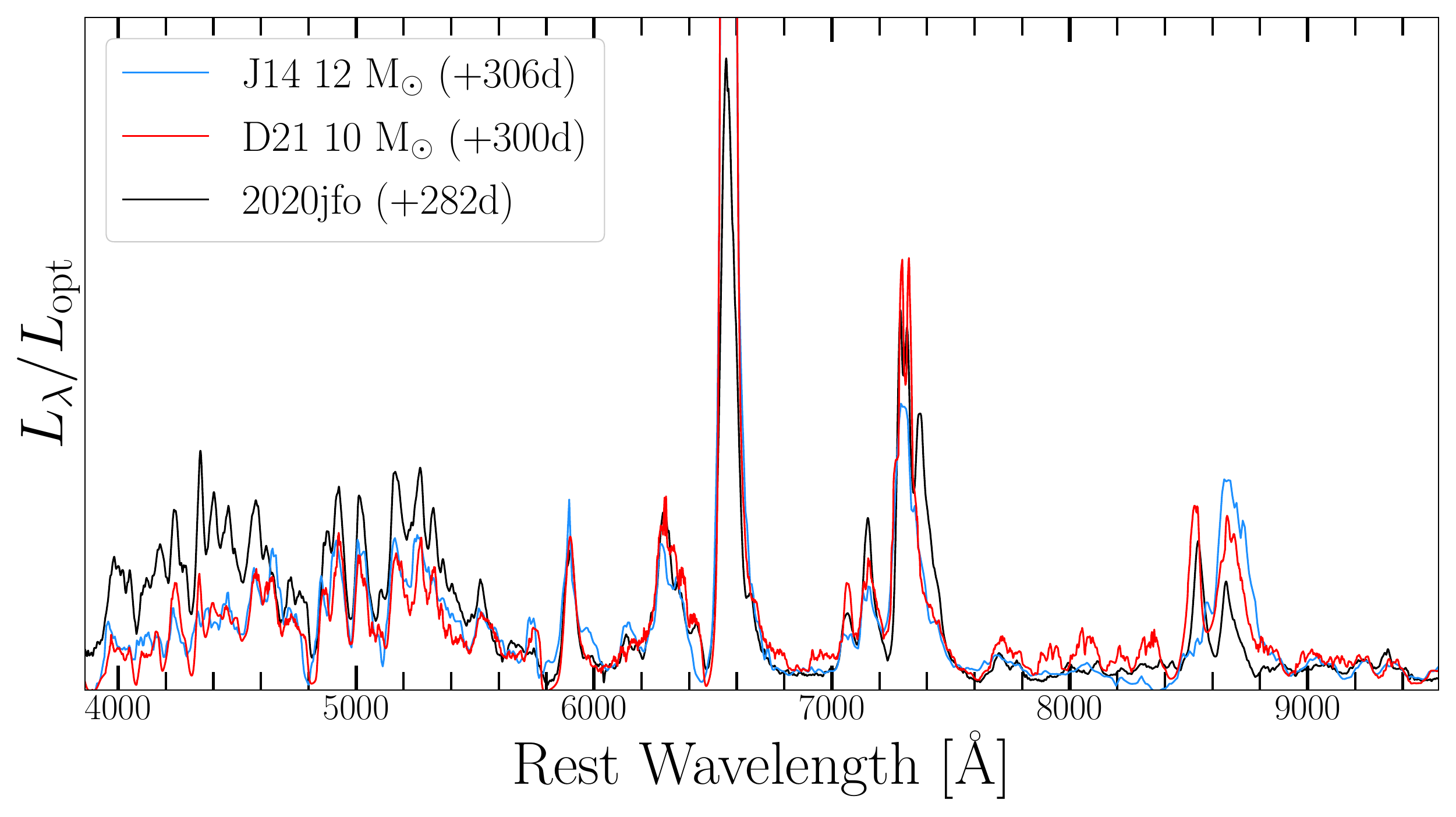}}
\subfigure{\includegraphics[width=0.44\textwidth]{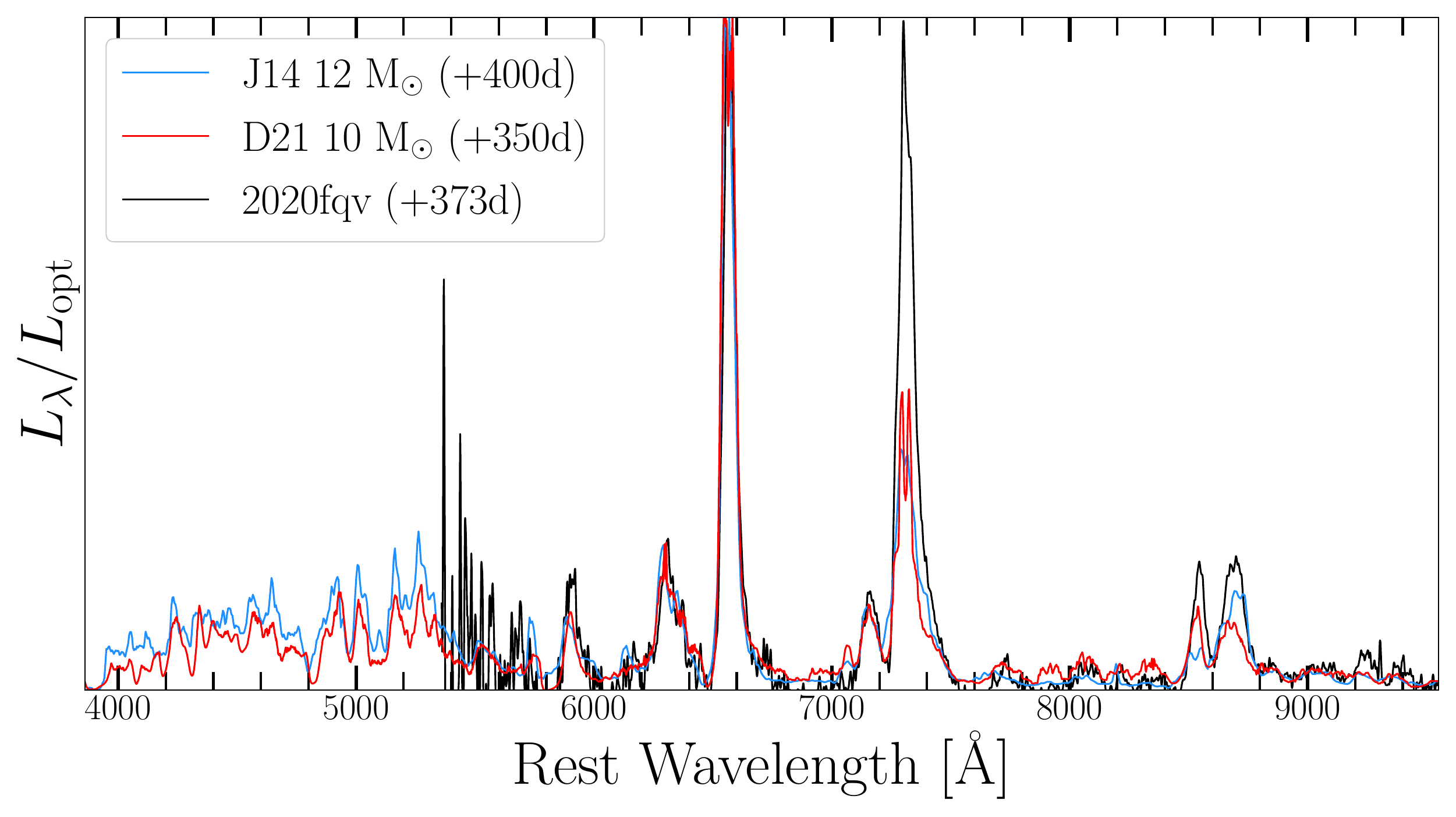}}\\
\subfigure{\includegraphics[width=0.44\textwidth]{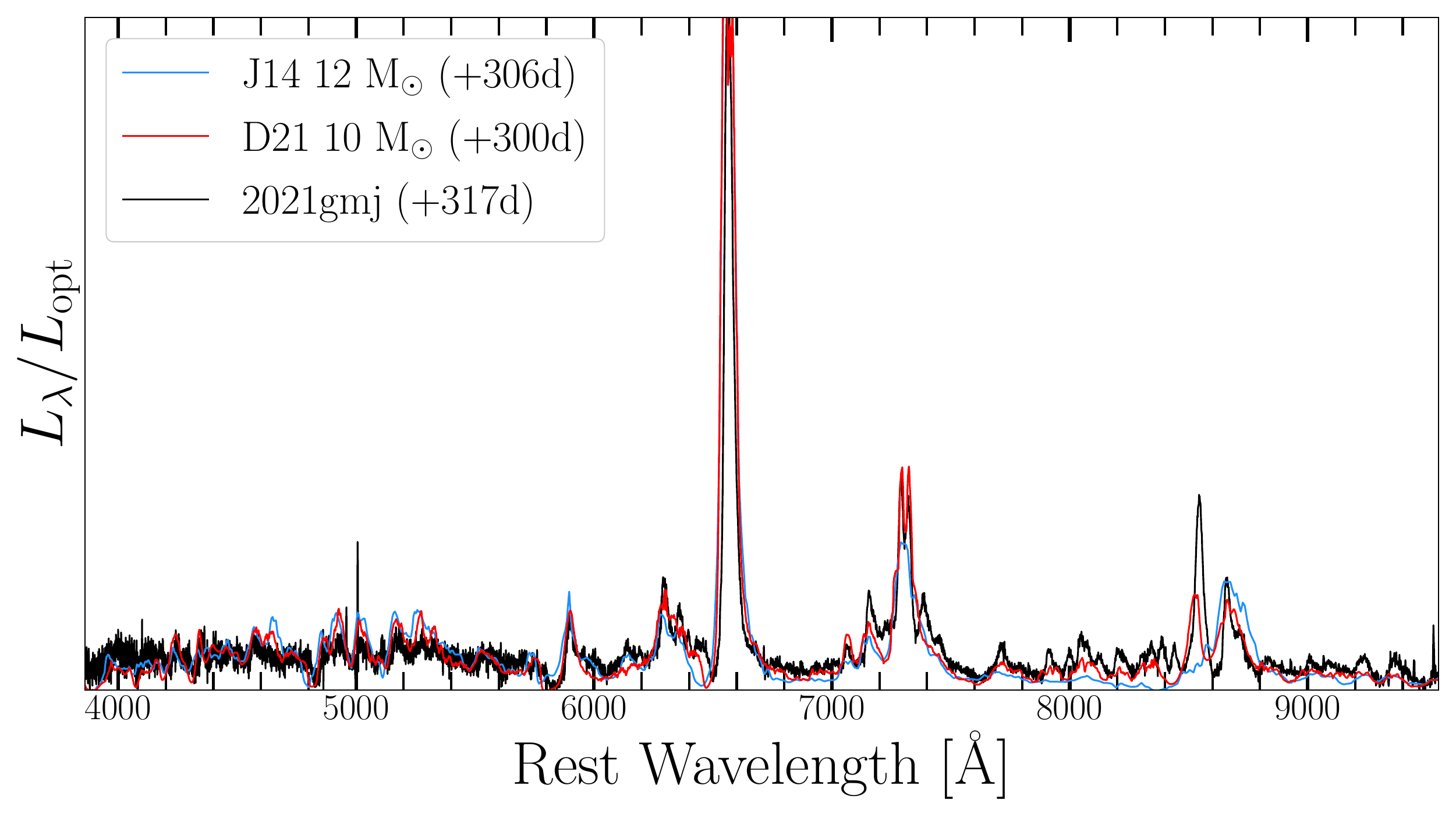}}
\subfigure{\includegraphics[width=0.44\textwidth]{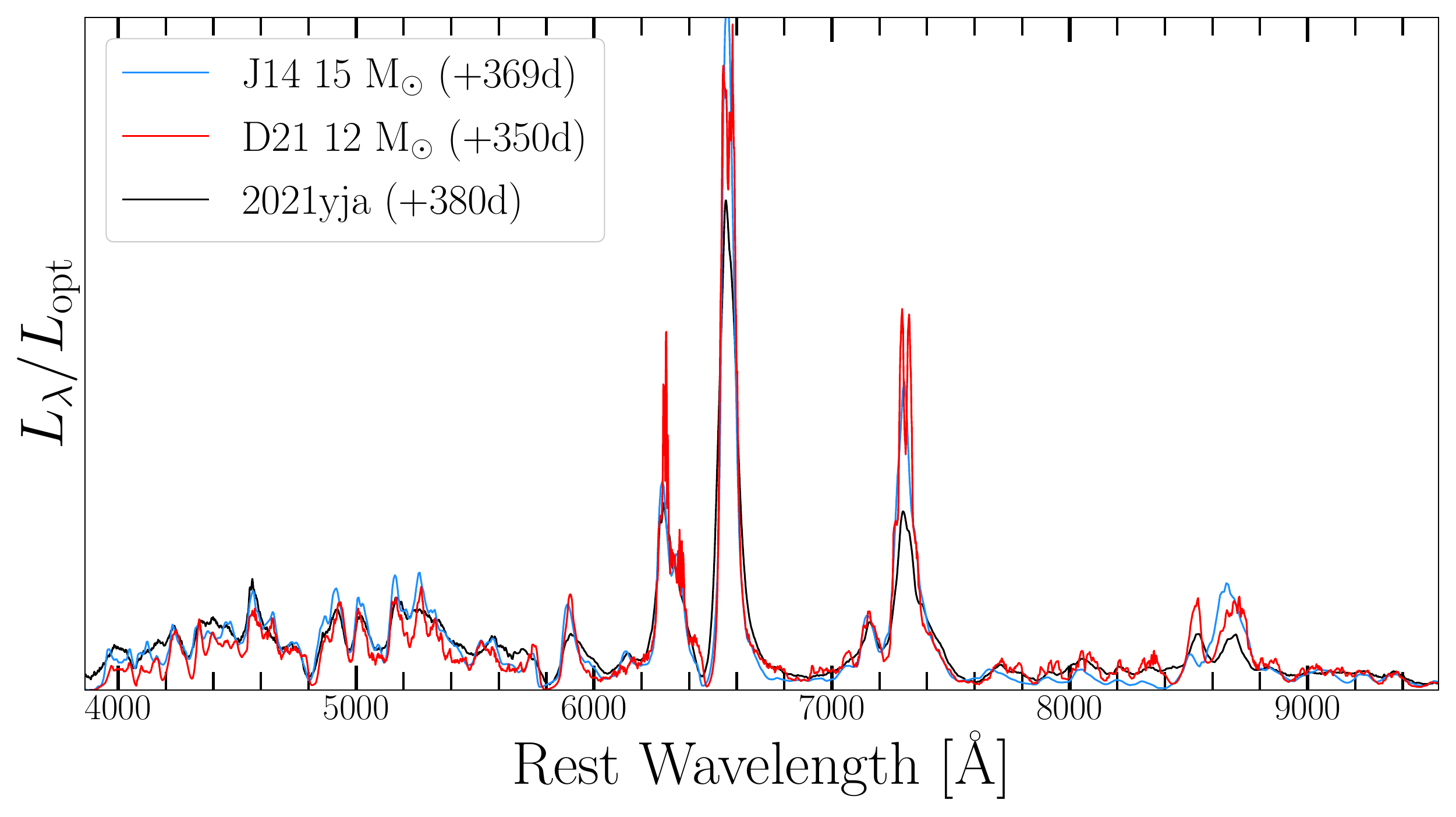}}\\
\caption{ Nebular spectra of sample objects (black) compared to best-matched model spectra from \cite{Dessart21} (red) and \cite{Jerkstrand14} (blue).   
\label{fig:neb_models_comp} }
\end{figure*}

Here we present previously unpublished nebular spectra for the gold/silver-sample objects in Table \ref{tab:spec_all}. Light-curve properties for gold/silver- and comparison-sample objects are given in Tables \ref{tab:sample_phot_gold}--\ref{tab:sample_phot_comp}. In Figure \ref{fig:EBV}, we display host-galaxy reddening compared to plateau luminosity at $\delta t = 50$~days and the estimated $^{56}$Ni mass for gold/silver- and comparison-sample objects. Figures \ref{fig:neb_models_gold} and \ref{fig:neb_models_comp} display best-matched nebular SN~II model spectra from \cite{Jerkstrand14}, \cite{Jerkstrand18}, and \cite{Dessart21} for all gold/silver- and comparison-sample objects with constrained $^{56}$Ni mass measurements.

\begin{deluxetable*}{cccccccc}
\tablecaption{Optical/Near-Infrared Spectroscopy \label{tab:spec_all}}
\tablecolumns{8}
\tablewidth{0.45\textwidth}
\tablehead{\colhead{SN Name} &
\colhead{UT Date} & \colhead{MJD} &
\colhead{Phase\tablenotemark{a}} &
\colhead{Telescope} & \colhead{Instrument} & \colhead{Wavelength Range} & \colhead{Data Source}\\
\colhead{} & \colhead{} & \colhead{} & \colhead{(days)} & \colhead{} & \colhead{} & \colhead{(\AA)}
}
\startdata
2020abjq & 2021-10-16T05:29:32 & 59503.2 & 319.8 & Gemini & GMOS-S & 5178-9885 & YSE \\
2020abjq & 2021-11-06T02:42:49 & 59857.1 & 340.7 & Keck & LRIS & 3162-10146 & YSE \\
2020pni & 2021-06-12T09:21:36 & 59377.4 & 331.6 & Keck & DEIMOS & 4018-9267 & TReX$^b$ \\
2021yja & 2022-09-22T13:31:10 & 59844.6 & 380.2 & Keck & LRIS & 3136-10192 & YSE \\
2021yja & 2023-11-17T12:00:00 & 60265.5 & 801.1 & Keck & LRIS & 5601-10250 & TReX \\
2021dbg & 2021-11-06T07:00:00 & 59524.3 & 267.2 & Keck & LRIS & 3162-10147 & YSE \\
2021dbg & 2022-01-31T07:12:00 & 59610.3 & 353.2 & Keck & LRIS & 3152-10276 & Filippenko \\
2021aaqn & 2022-09-22T13:16:19 & 59844.6 & 351.8 & Keck & LRIS & 3135-10195 & YSE \\
2022ibv & 2022-12-11T02:00:00 & 59924.1 & 233.6 & Gemini & GMOS-S & 5201-9885 & YSE \\
2022dml & 2022-09-25T05:48:29 & 59847.2 & 213.2 & Keck & LRIS & 3041-10283 & TReX \\
2022ffg & 2022-10-31T15:05:46 & 59883.6 & 219.8 & Keck & LRIS & 3040-10327 & TReX \\
2022ffg & 2023-02-28T02:00:00 & 60003.1 & 339.3 & Gemini & GMOS-S & 5251-9885 & YSE \\
2022ffg & 2023-11-11T12:00:00 & 60259.5 & 595.7 & Keck & LRIS & 5601-10199 & TReX \\
2022jox & 2023-02-28T02:00:00 & 60003.1 & 339.3 & Gemini & GMOS-S & 5701-9887 & YSE \\
\enddata
\tablenotetext{a}{Relative to first light.}
\tablenotetext{b}{TRansient EXtragalactic team at UC Berkeley (PIs Margutti and Chornock)}
\end{deluxetable*}

\begin{deluxetable*}{cccccccccc}[h!]
\tablecaption{Gold/Silver-Sample Photometric Properties \label{tab:sample_phot_gold}}
\tablecolumns{10}
\tablewidth{0pt}
\tablehead{\colhead{Name} & \colhead{log$_{10}(L_{50}$)} & \colhead{$M_{g,50}$} & \colhead{$M_{r,50}$} & \colhead{$S_{\rm UVOIR,50}$} & \colhead{$S_{g,50}$} & \colhead{$S_{g,50}$} & \colhead{$t_{\rm PT}$} & \colhead{$M(^{56}\rm Ni)$} & \colhead{$t_{\gamma}$}\\
\colhead{} & \colhead{(erg s$^{-1}$)} & \colhead{(mag)} & \colhead{(mag)} & \colhead{} & \colhead{} & \colhead{} & \colhead{(days)} & \colhead{($\Msun$)} & \colhead{(days)} }
\startdata
PTF11iqb$^1$ & 42.25$\pm$0.13 & -- & -17.69$\pm$0.15 & -1.5e-02$\pm$1.8e-03 & -- & 7.4e-03$\pm$3.4e-04 & 120.11$\pm$2.14 & $<$8.6e-02 &269.5$\pm$101.7 \\
2017ahn$^1$ & 41.87$\pm$0.40 & -16.26$\pm$0.43 & -16.86$\pm$0.43 & -1.8e-02$\pm$5.6e-03 & 4.7e-02$\pm$6.8e-04 & 3.0e-02$\pm$5.7e-04 & 60.51$\pm$2.22 & $<$3.8e-02 & 134.9$\pm$23.5 \\
2018zd$^1$ & 42.32$\pm$0.37 & -17.60$\pm$0.53 & -17.86$\pm$0.53 & -1.0e-02$\pm$5.4e-03 & 2.0e-02$\pm$9.2e-04 & 9.7e-03$\pm$9.2e-04 & 120.76$\pm$6.27 & 3.1e-02$\pm$4.0e-04 &481.1$\pm$25.2 \\
2019ust$^1$ & 42.60$\pm$0.20 & -17.63$\pm$0.16 & -18.01$\pm$0.15 & -1.1e-02$\pm$2.8e-03 & 3.7e-02$\pm$9.0e-04 & 1.9e-02$\pm$2.7e-04 & $>$93.32 & -- &-- \\
2019qch$^1$ & 42.63$\pm$0.13 & -17.57$\pm$0.10 & -17.80$\pm$0.09 & -1.1e-02$\pm$1.8e-03 & 1.9e-02$\pm$6.8e-04 & 7.8e-03$\pm$2.7e-04 & $>$73.11 & -- &-- \\
2020abjq$^1$ & 42.30$\pm$0.06 & -17.36$\pm$0.15 & -17.72$\pm$0.28 & -1.0e-02$\pm$8.8e-04 & 3.5e-02$\pm$3.0e-04 & 1.1e-02$\pm$3.5e-03 & $>$71.66 & $<$1.8e-01 &202.1$\pm$24.9 \\
2020pni$^1$ & 42.21$\pm$0.18 & -17.19$\pm$0.10 & -17.65$\pm$0.07 & -1.2e-02$\pm$2.7e-03 & 2.7e-02$\pm$1.4e-03 & 1.8e-02$\pm$7.8e-04 & $>$49.45 & $<$6.5e-02 &486.1$\pm$229.8 \\
2020tlf$^1$ & 42.45$\pm$0.05 & -17.26$\pm$0.08 & -17.47$\pm$0.08 & -7.5e-03$\pm$6.8e-04 & 2.3e-02$\pm$2.1e-04 & 1.3e-02$\pm$2.1e-04 & 122.66$\pm$1.52 & 2.0e-02$\pm$5.8e-04 &346.5$\pm$59.0 \\
2020abtf$^1$ & 42.19$\pm$0.05 & -16.44$\pm$0.16 & -16.80$\pm$0.16 & -3.7e-03$\pm$6.2e-04 & 1.8e-02$\pm$3.5e-04 & 8.7e-03$\pm$2.6e-04 & 125.00$\pm$5.00 & 2.1e-02$\pm$5.4e-04 &-- \\
2020sic$^1$ & 41.89$\pm$0.06 & -- & -17.68$\pm$0.19 & -2.9e-03$\pm$9.5e-04 & -- & 1.0e-02$\pm$1.6e-03 & -- & -- &-- \\
2021aek$^1$ & 42.54$\pm$0.12 & -17.73$\pm$0.25 & -17.87$\pm$0.23 & -1.3e-02$\pm$1.7e-03 & 2.1e-02$\pm$2.8e-03 & 5.4e-03$\pm$2.2e-03 & $>$136.03 & -- &-- \\
2021afkk$^1$ & 42.22$\pm$0.05 & -17.12$\pm$0.16 & -17.65$\pm$0.16 & -1.7e-02$\pm$7.4e-04 & 3.9e-02$\pm$6.0e-04 & 2.2e-02$\pm$5.3e-04 & 59.83$\pm$1.01 & 4.3e-02$\pm$8.5e-04 &-- \\
2021dbg$^1$ & 42.43$\pm$0.14 & -17.84$\pm$0.16 & -18.14$\pm$0.19 & -7.9e-03$\pm$2.0e-03 & 2.2e-02$\pm$6.6e-04 & 7.0e-03$\pm$1.7e-03 & $>$144.42 & 1.7e-01$\pm$4.8e-02 & -- \\
2021mqh$^1$ & 42.03$\pm$0.14 & -16.39$\pm$0.23 & -17.05$\pm$0.35 & -1.1e-02$\pm$1.8e-03 & 3.5e-02$\pm$2.5e-03 & 2.0e-02$\pm$4.9e-03 & $>$69.70 & 5.7e-02$\pm$4.3e-03 &-- \\
2021tyw$^1$ & 42.37$\pm$0.05 & -17.52$\pm$0.20 & -17.58$\pm$0.20 & -6.8e-03$\pm$6.7e-04 & 9.3e-03$\pm$4.6e-04 & 2.1e-03$\pm$5.0e-04 & 142.00$\pm$5.00 & 2.6e-02$\pm$3.4e-03 &-- \\
2021zj$^1$ & 42.61$\pm$0.14 & -17.48$\pm$0.17 & -17.89$\pm$0.17 & -4.6e-03$\pm$2.1e-03 & 2.4e-03$\pm$1.2e-03 & 2.7e-03$\pm$1.2e-03 & 137.38$\pm$2.61 & $<$6.1e-01 &-- \\
2021wvd$^1$ & -- & -- & -- & -- & -- & -- & $>$42.95 & -- &-- \\
2022ffg$^1$ & 42.59$\pm$0.05 & -17.86$\pm$0.15 & -17.97$\pm$0.16 & -1.5e-02$\pm$6.4e-04 & 3.5e-02$\pm$1.4e-04 & 1.9e-02$\pm$5.7e-04 & $>$92.88 & 6.0e-02$\pm$9.9e-04 & -- \\
2022pgf$^1$ & 42.82$\pm$0.03 & -18.14$\pm$0.44 & -18.36$\pm$0.43 & -1.6e-02$\pm$4.2e-04 & 4.3e-02$\pm$1.2e-03 & 3.2e-02$\pm$2.4e-04 & 150.00$\pm$5.00 & 9.0e-02$\pm$9.4e-03 &-- \\
2022prv$^1$ & 43.30$\pm$0.07 & -17.86$\pm$0.15 & -18.03$\pm$0.15 & -8.1e-03$\pm$1.1e-03 & 2.2e-02$\pm$2.4e-04 & 7.1e-03$\pm$3.6e-04 & $>$72.39 & -- &-- \\
PTF10gva$^2$ & 42.32$\pm$0.14 & -- & -17.92$\pm$0.17 & -9.5e-03$\pm$2.2e-03 & -- & 2.0e-02$\pm$1.2e-03 & $>$49.98 & -- &-- \\
PTF10abyy$^2$ & 42.17$\pm$0.03 & -- & -18.27$\pm$0.17 & -8.6e-03$\pm$4.1e-04 & -- & 1.2e-02$\pm$6.5e-04 & 65.02$\pm$0.53 & -- &-- \\
2014G$^2$ & 42.17$\pm$0.09 & -16.96$\pm$0.80 & -17.56$\pm$0.80 & -1.2e-02$\pm$1.2e-03 & 3.7e-02$\pm$8.9e-04 & 2.3e-02$\pm$7.2e-04 & 84.19$\pm$1.93 & 6.5e-02$\pm$4.4e-03 &182.2$\pm$12.7 \\
2015bf$^2$ & 42.33$\pm$0.39 & -- & -17.47$\pm$0.46 & -7.6e-03$\pm$5.4e-03 & -- & 1.1e-02$\pm$6.2e-03 & $>$56.67 & -- &-- \\
2016blz$^2$ & 42.08$\pm$0.13 & -- & -- & -8.1e-03$\pm$1.9e-03 & -- & -- & $>$149.54 & -- &-- \\
2018dfc$^2$ & 42.00$\pm$0.27 & -17.07$\pm$0.19 & -17.84$\pm$0.18 & -1.1e-02$\pm$3.8e-03 & 4.1e-02$\pm$1.6e-03 & 2.3e-02$\pm$1.4e-03 & $>$87.44 & -- &-- \\
2021can$^2$ & 42.20$\pm$0.14 & -16.42$\pm$0.08 & -17.23$\pm$0.12 & -1.4e-02$\pm$2.0e-03 & 5.2e-02$\pm$9.1e-04 & 2.4e-02$\pm$1.4e-03 & 61.09$\pm$1.73 & $<$1.0e-01 &177.2$\pm$27.9 \\
2021ont$^2$ & 42.54$\pm$0.12 & -17.58$\pm$0.19 & -17.95$\pm$0.18 & -1.0e-02$\pm$1.8e-03 & 4.1e-02$\pm$1.6e-03 & 2.6e-02$\pm$1.4e-03 & 58.95$\pm$2.69 & $<$4.2e-01 & -- \\
2021qvr$^2$ & 42.27$\pm$0.13 & -16.87$\pm$0.26 & -17.56$\pm$0.49 & -1.5e-02$\pm$1.8e-03 & 3.7e-02$\pm$8.2e-04 & 2.1e-02$\pm$5.9e-03 & 90.16$\pm$2.76 & $<$1.7e-02 & -- \\
2022dml$^2$ & 42.28$\pm$0.06 & -17.46$\pm$0.08 & -17.85$\pm$0.08 & -8.2e-03$\pm$8.2e-04 & 1.9e-02$\pm$5.2e-04 & 7.6e-03$\pm$4.8e-04 & 92.85$\pm$1.71 & $<$1.8e-01 &203.1$\pm$13.1 \\
2022jox$^2$ & 41.82$\pm$0.14 & -16.40$\pm$0.43 & -16.72$\pm$0.43 & -3.8e-03$\pm$1.9e-03 & 1.7e-02$\pm$4.6e-04 & 8.2e-03$\pm$4.1e-04 & $>$65.81 & 6.6e-02$\pm$1.2e-03 &478.3$\pm$40.9 \\
2013fs$^3$ & 41.93$\pm$0.04 & -16.50$\pm$0.38 & -17.07$\pm$0.36 & -5.7e-03$\pm$5.3e-04 & 1.4e-02$\pm$2.8e-03 & 6.6e-03$\pm$2.4e-03 & 84.37$\pm$2.34 & -- & -- \\
2018fif$^3$ & 41.78$\pm$0.08 & -16.22$\pm$0.16 & -17.05$\pm$0.15 & -5.5e-03$\pm$9.4e-04 & 1.9e-02$\pm$4.7e-04 & 1.1e-02$\pm$9.0e-05 & 115.00$\pm$5.00 & 8.4e-02$\pm$2.0e-02 & -- \\
2020lfn$^3$ & 42.36$\pm$0.17 & -17.67$\pm$0.23 & -18.28$\pm$0.14 & -8.9e-03$\pm$2.2e-03 & 3.5e-02$\pm$3.1e-03 & 1.1e-02$\pm$1.8e-03 & 75.24$\pm$2.40 & 1.1e-01$\pm$1.7e-02 & -- \\
2020nif$^3$ & 42.74$\pm$0.06 & -18.68$\pm$0.43 & -18.42$\pm$0.43 & -1.0e-02$\pm$9.4e-04 & 2.2e-02$\pm$9.2e-04 & 2.2e-02$\pm$9.2e-04 & $>$62.91 & 9.0e-02$\pm$3.1e-03 & 397.7$\pm$40.3 \\
2020xua$^3$ & 42.07$\pm$0.10 & -16.90$\pm$0.16 & -16.92$\pm$0.19 & -4.8e-03$\pm$1.4e-03 & 1.6e-02$\pm$8.2e-04 & 7.8e-03$\pm$1.6e-03 & $>$92.91 & -- & -- \\
2021aaqn$^3$ & 42.16$\pm$0.13 & -16.50$\pm$0.45 & -17.27$\pm$0.20 & -7.3e-03$\pm$2.0e-03 & 3.5e-02$\pm$7.0e-03 & 1.4e-02$\pm$2.1e-03 & 79.08$\pm$7.20 & -- & -- \\
2021jtt$^3$ & 42.82$\pm$0.04 & -18.00$\pm$0.26 & -18.21$\pm$0.21 & -1.0e-02$\pm$5.9e-04 & 2.7e-02$\pm$2.6e-03 & 1.6e-02$\pm$1.1e-03 & $>$60.44 & $<$6.6e-01 & -- \\
2022ibv$^3$ & -- & -16.24$\pm$0.16 & -16.85$\pm$0.16 & -- & 2.6e-02$\pm$5.8e-04 & 8.1e-03$\pm$3.8e-04 & $>$49.62 & -- & -- \\
\enddata
\end{deluxetable*}

\begin{deluxetable*}{cccccccccc}[h!]
\tablecaption{Comparison-Sample Photometric Properties \label{tab:sample_phot_comp}}
\tablecolumns{10}
\tablewidth{0pt}
\tablehead{\colhead{Name} & \colhead{log$_{10}(L_{50}$)} & \colhead{$M_{g,50}$} & \colhead{$M_{r,50}$} & \colhead{$S_{\rm UVOIR,50}$} & \colhead{$S_{g,50}$} & \colhead{$S_{g,50}$} & \colhead{$t_{\rm PT}$} & \colhead{$M(^{56}\rm Ni)$} & \colhead{$t_{\gamma}$}\\
\colhead{} & \colhead{(erg s$^{-1}$)} & \colhead{(mag)} & \colhead{(mag)} & \colhead{} & \colhead{} & \colhead{} & \colhead{(days)} & \colhead{($\Msun$)} & \colhead{(days)} }
\startdata
2013ft & 41.06$\pm$0.06 &-- & -14.59$\pm$0.48 & -3.4e-03$\pm$8.3e-04 & -- & -2.7e-03$\pm$2.3e-03 & $>$127.34 & -- &-- \\
2013am & 41.49$\pm$0.05 &-- & -15.98$\pm$0.41 & -1.9e-03$\pm$6.6e-04 & -- & 1.7e-03$\pm$4.2e-04 & 107.33$\pm$2.23 & 1.6e-02$\pm$3.3e-04 & -- \\
2013ab & 41.95$\pm$0.16 & -16.71$\pm$0.12 & -17.13$\pm$0.12 & -5.6e-03$\pm$2.2e-03 & 1.2e-02$\pm$1.1e-03 & 7.9e-03$\pm$1.3e-03 & 101.98$\pm$1.03 & 7.4e-02$\pm$8.1e-04 & -- \\
2016X & 41.68$\pm$0.05 & -15.80$\pm$0.43 & -16.42$\pm$0.43 & -4.7e-03$\pm$6.9e-04 & 2.1e-02$\pm$2.2e-04 & 3.0e-03$\pm$2.4e-04 & 95.42$\pm$0.57 & 2.7e-02$\pm$5.1e-04 & -- \\
2016aqf & 41.02$\pm$0.05 & -14.40$\pm$0.33 & -14.86$\pm$0.34 & -1.3e-03$\pm$7.2e-04 & 5.8e-03$\pm$2.2e-03 & 7.8e-04$\pm$2.5e-03 & $>$127.34 & 1.0e-02$\pm$6.1e-04 &312.5$\pm$24.6 \\
2017eaw & 42.01$\pm$0.05 & -16.94$\pm$0.08 & -17.21$\pm$0.08 & -2.9e-03$\pm$6.2e-04 & 5.0e-03$\pm$8.9e-04 & 5.1e-04$\pm$7.9e-04 & 116.82$\pm$0.22 & 6.5e-02$\pm$8.3e-04 & 408.3$\pm$6.0 \\
2017gmr & 42.44$\pm$0.05 & -17.71$\pm$0.16 & -18.15$\pm$0.16 & -5.1e-03$\pm$6.5e-04 & 1.8e-02$\pm$5.1e-04 & 4.4e-03$\pm$3.6e-04 & 94.49$\pm$0.55 & 2.2e-01$\pm$7.3e-03 & 253.0$\pm$24.8 \\
2018lab & 41.36$\pm$0.04 & -15.11$\pm$0.12 & -15.63$\pm$0.11 & -2.9e-03$\pm$5.5e-04 & 1.5e-02$\pm$1.6e-03 & 1.7e-03$\pm$1.3e-03 & $>$110.0 & 1.7e-02$\pm$7.6e-04 & -- \\
2018kpo & 41.73$\pm$0.21 & -16.39$\pm$0.16 & -16.87$\pm$0.16 & -6.2e-03$\pm$3.5e-03 & 1.1e-02$\pm$6.5e-04 & 3.2e-03$\pm$7.5e-04 & $>$63.21 & 5.1e-02$\pm$8.9e-03 & 346.0$\pm$87.6 \\
2018cuf & 41.99$\pm$0.02 & -16.72$\pm$0.30 & -17.13$\pm$0.30 & -4.7e-03$\pm$2.9e-04 & 1.2e-02$\pm$2.2e-04 & 4.2e-03$\pm$1.7e-04 & 111.39$\pm$0.75 & 6.5e-02$\pm$7.3e-04 & -- \\
2019edo & 41.73$\pm$0.06 & -16.02$\pm$0.43 & -16.62$\pm$0.43 & -3.1e-03$\pm$8.2e-04 & 1.0e-02$\pm$2.1e-04 & 3.5e-03$\pm$1.8e-04 & 88.47$\pm$1.85 & 4.2e-02$\pm$4.3e-03 &373.3$\pm$94.5 \\
2019nvm & 42.16$\pm$0.05 & -17.37$\pm$0.15 & -17.71$\pm$0.15 & -6.3e-03$\pm$6.9e-04 & 1.5e-02$\pm$2.0e-04 & 9.8e-03$\pm$1.4e-04 & $>$91.38 & 1.2e-01$\pm$5.3e-04 & -- \\
2019pjs & 41.55$\pm$0.05 & -15.70$\pm$0.17 & -16.15$\pm$0.16 & -2.8e-03$\pm$6.4e-04 & 6.1e-03$\pm$9.6e-04 & 3.5e-03$\pm$6.1e-04 & $>$75.40 & 2.7e-02$\pm$1.4e-03 &325.9$\pm$42.5 \\
2019enr &-- &-- &-- & -- & -- & -- & $>$32.65 & -- &-- \\
2020ekk & 41.86$\pm$0.08 & -16.41$\pm$0.22 & -16.95$\pm$0.26 & -6.1e-03$\pm$1.1e-03 & 2.1e-02$\pm$2.3e-03 & 1.1e-02$\pm$3.1e-03 & 105.42$\pm$1.21 & 2.0e-02$\pm$1.5e-03 &236.8$\pm$42.9 \\
2020jfo & 41.72$\pm$0.03 & -15.96$\pm$0.20 & -16.36$\pm$0.19 & -7.8e-03$\pm$5.1e-04 & 2.5e-02$\pm$2.9e-04 & 1.3e-02$\pm$1.1e-04 & 66.24$\pm$0.36 & 3.4e-02$\pm$5.1e-03 &194.9$\pm$19.8 \\
2020fqv & 42.17$\pm$0.03 & -16.98$\pm$0.45 & -17.53$\pm$0.45 & -3.6e-03$\pm$3.6e-04 & 2.1e-02$\pm$5.4e-04 & 6.0e-03$\pm$1.8e-04 & 113.26$\pm$1.10 & 6.0e-02$\pm$2.2e-03 &480.2$\pm$54.0 \\
2020mjm & 41.76$\pm$0.11 & -16.13$\pm$0.43 & -16.56$\pm$0.43 & -3.2e-03$\pm$1.5e-03 & 2.3e-03$\pm$2.8e-04 & -6.7e-04$\pm$2.5e-04 & $>$85.45 & 4.1e-02$\pm$5.9e-03 &208.4$\pm$20.9 \\
2020dpw & 41.44$\pm$0.06 & -15.06$\pm$0.43 & -15.80$\pm$0.43 & -2.9e-03$\pm$8.2e-04 & 1.2e-02$\pm$2.9e-04 & 1.7e-03$\pm$1.6e-04 & 114.07$\pm$3.49 & 2.4e-02$\pm$9.9e-05 & -- \\
2020acbm & 41.91$\pm$0.24 & -16.73$\pm$0.16 & -17.38$\pm$0.15 & -5.5e-03$\pm$3.0e-03 & 2.0e-02$\pm$9.0e-04 & 9.8e-03$\pm$4.8e-04 & $>$78.74 & 1.2e-01$\pm$6.7e-03 & -- \\
2021vaz & 42.07$\pm$0.07 & -16.83$\pm$0.15 & -17.26$\pm$0.16 & -5.2e-03$\pm$9.6e-04 & 1.3e-02$\pm$3.2e-04 & 1.0e-02$\pm$7.1e-04 & 104.90$\pm$1.20 & 7.2e-02$\pm$2.5e-03 & -- \\
2021ass & 41.45$\pm$0.04 & -15.37$\pm$0.37 & -15.98$\pm$0.22 & -3.1e-03$\pm$5.6e-04 & 2.1e-03$\pm$5.0e-03 & 3.2e-03$\pm$1.2e-03 & $>$65.45 & 2.8e-02$\pm$2.5e-04 &413.5$\pm$13.1 \\
2021gmj & 41.87$\pm$0.01 & -15.45$\pm$0.16 & -15.95$\pm$0.16 & -2.7e-04$\pm$1.2e-04 & 4.8e-03$\pm$3.9e-04 & 8.8e-04$\pm$2.2e-04 & 111.97$\pm$1.94 & 1.9e-02$\pm$3.3e-04 &465.5$\pm$49.6 \\
2021rhk &-- &-- &-- & -- & -- & -- & $>$23.93 & -- &-- \\
2021uoy & 41.99$\pm$0.25 & -16.63$\pm$0.30 & -17.37$\pm$0.18 & -7.3e-03$\pm$3.4e-03 & 2.1e-02$\pm$3.6e-03 & 1.4e-02$\pm$1.3e-03 & $>$98.81 & -- &-- \\
2021yja & 42.32$\pm$0.03 & -17.48$\pm$0.46 & -17.78$\pm$0.46 & -3.8e-03$\pm$4.2e-04 & 9.5e-03$\pm$3.8e-04 & 4.2e-03$\pm$2.7e-04 & 124.65$\pm$0.37 & 1.8e-01$\pm$3.0e-03 &404.6$\pm$15.6 \\
2021adly &-- &-- &-- & -- & -- & -- & $>$119.80 & -- &-- \\
2021apg & 41.89$\pm$0.07 & -15.92$\pm$0.17 & -16.75$\pm$0.19 & -1.8e-03$\pm$1.0e-03 & 1.5e-02$\pm$1.0e-03 & 3.8e-03$\pm$1.8e-03 & 106.61$\pm$3.61 & -- &-- \\
2021gvm & 42.12$\pm$0.06 & -16.70$\pm$0.11 & -17.71$\pm$0.09 & -9.1e-03$\pm$8.2e-04 & 3.0e-02$\pm$1.1e-03 & 1.2e-02$\pm$4.9e-04 & 96.04$\pm$0.61 & -- &-- \\
2021ucg & 42.05$\pm$0.04 & -16.88$\pm$0.16 & -17.26$\pm$0.16 & -2.8e-03$\pm$6.2e-04 & 3.6e-03$\pm$5.2e-04 & 2.9e-03$\pm$3.4e-04 & 123.53$\pm$1.11 & 1.2e-01$\pm$1.9e-03 & -- \\
2022inn &-- & -14.09$\pm$0.39 & -14.70$\pm$0.22 & -- & 8.6e-03$\pm$5.8e-03 & -2.6e-03$\pm$2.9e-03 & $>$48.97 & -- &-- \\
2022fuc & 41.34$\pm$0.07 & -15.09$\pm$0.43 & -15.63$\pm$0.43 & -4.2e-03$\pm$9.8e-04 & 9.2e-03$\pm$6.3e-04 & 5.2e-03$\pm$3.2e-04 & 107.48$\pm$10.26 & -- &-- \\
2022jzc & 40.70$\pm$0.03 & -13.46$\pm$0.44 & -14.12$\pm$0.43 & -7.8e-04$\pm$4.4e-04 & 4.1e-03$\pm$9.1e-04 & -5.9e-03$\pm$6.9e-04 & $>$59.61 & 2.3e-03$\pm$8.7e-05 & -- \\
2022ovb & 42.09$\pm$0.07 & -16.98$\pm$0.16 & -17.64$\pm$0.15 & -3.3e-03$\pm$9.3e-04 & 1.3e-02$\pm$3.8e-04 & 5.9e-03$\pm$2.8e-04 & 101.82$\pm$6.35 & -- &-- \\
2022frq & 42.00$\pm$0.04 & -16.83$\pm$0.18 & -17.22$\pm$0.16 & -2.8e-03$\pm$5.3e-04 & 9.5e-03$\pm$1.4e-03 & 3.4e-04$\pm$6.2e-04 & 120.00$\pm$5.00 & -- &-- \\
\enddata
\end{deluxetable*}

\begin{figure*}[t!]
\centering
\subfigure{\includegraphics[width=0.44\textwidth]{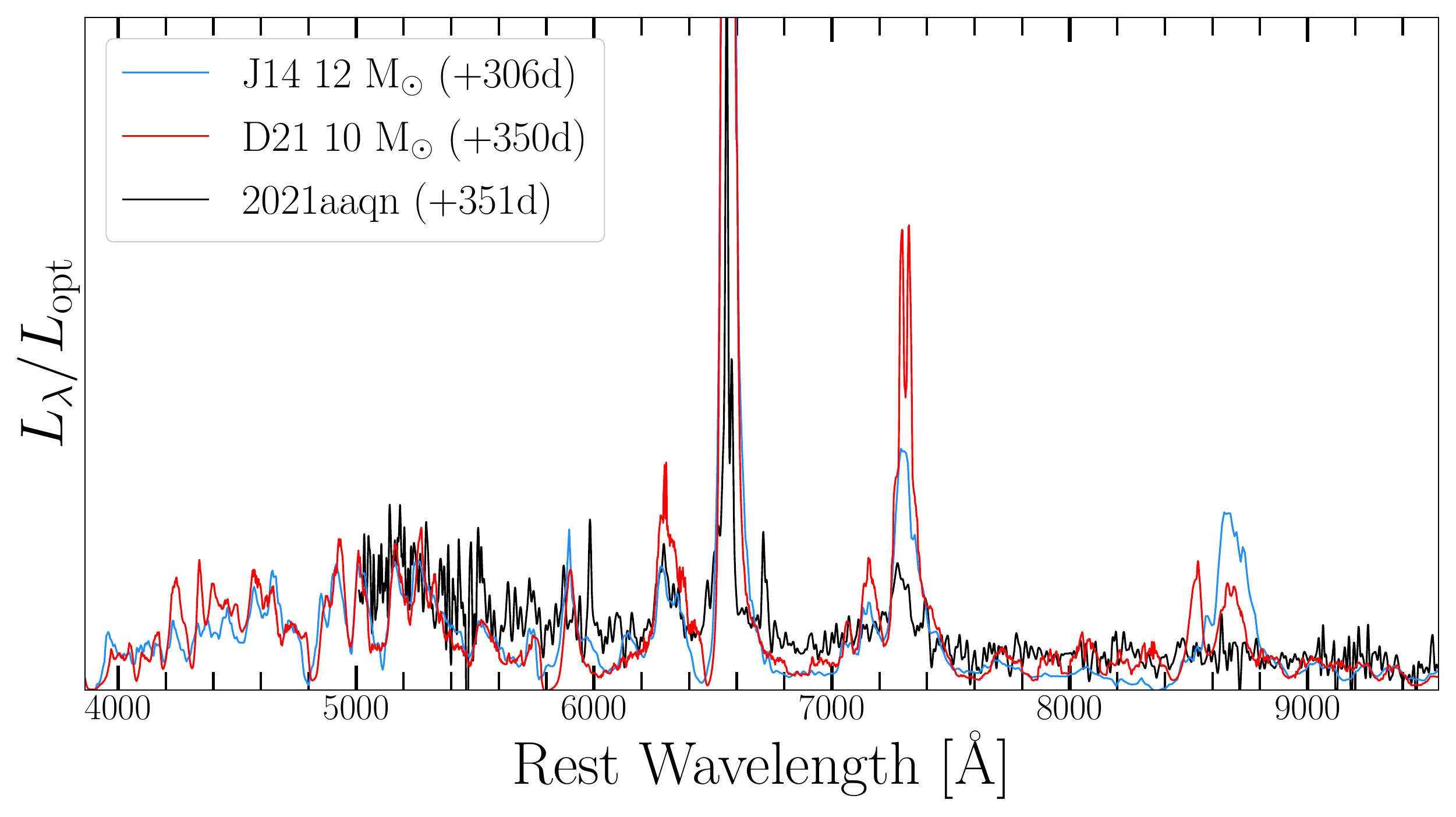}}
\subfigure{\includegraphics[width=0.44\textwidth]{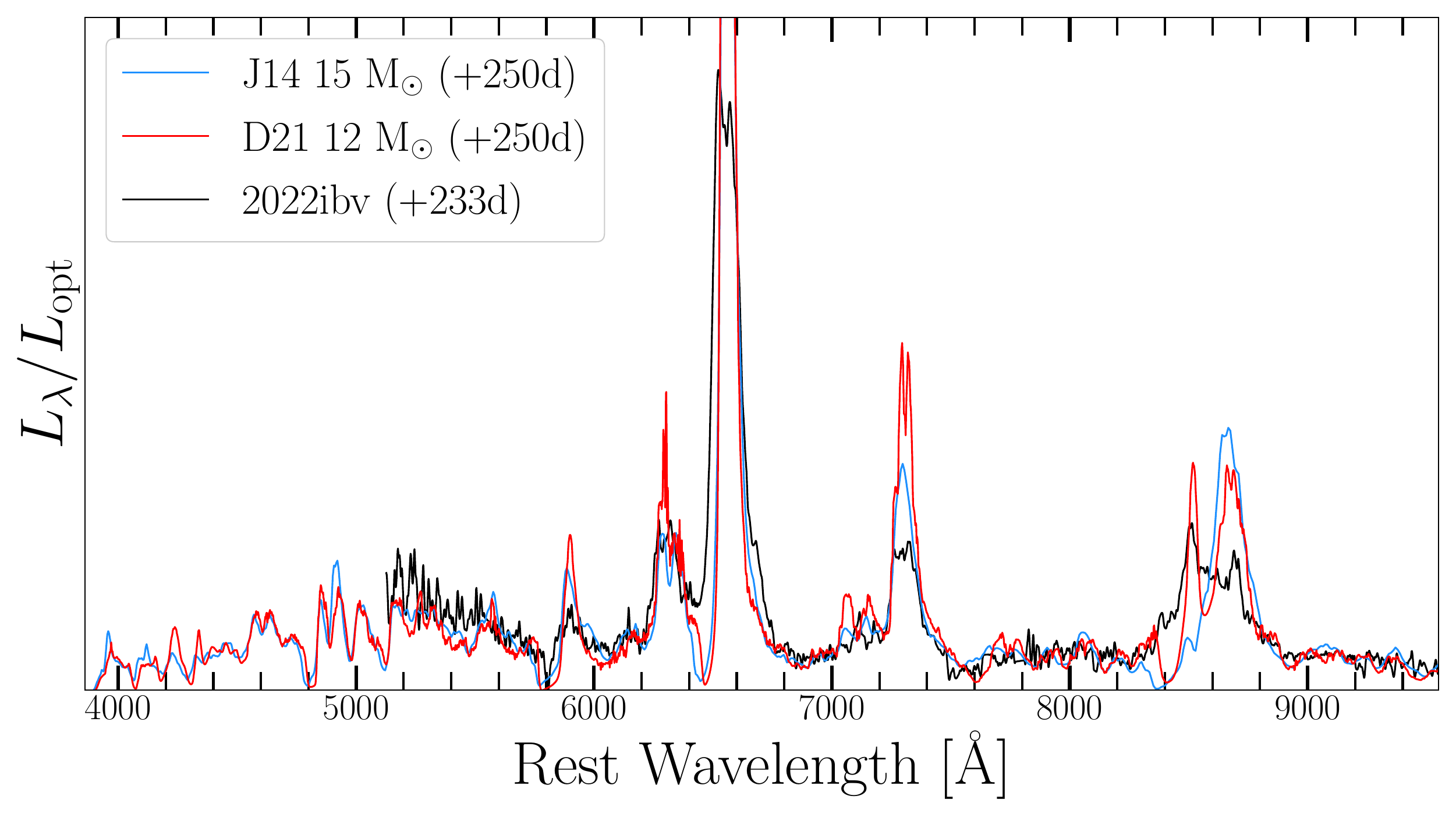}}\\
\subfigure{\includegraphics[width=0.44\textwidth]{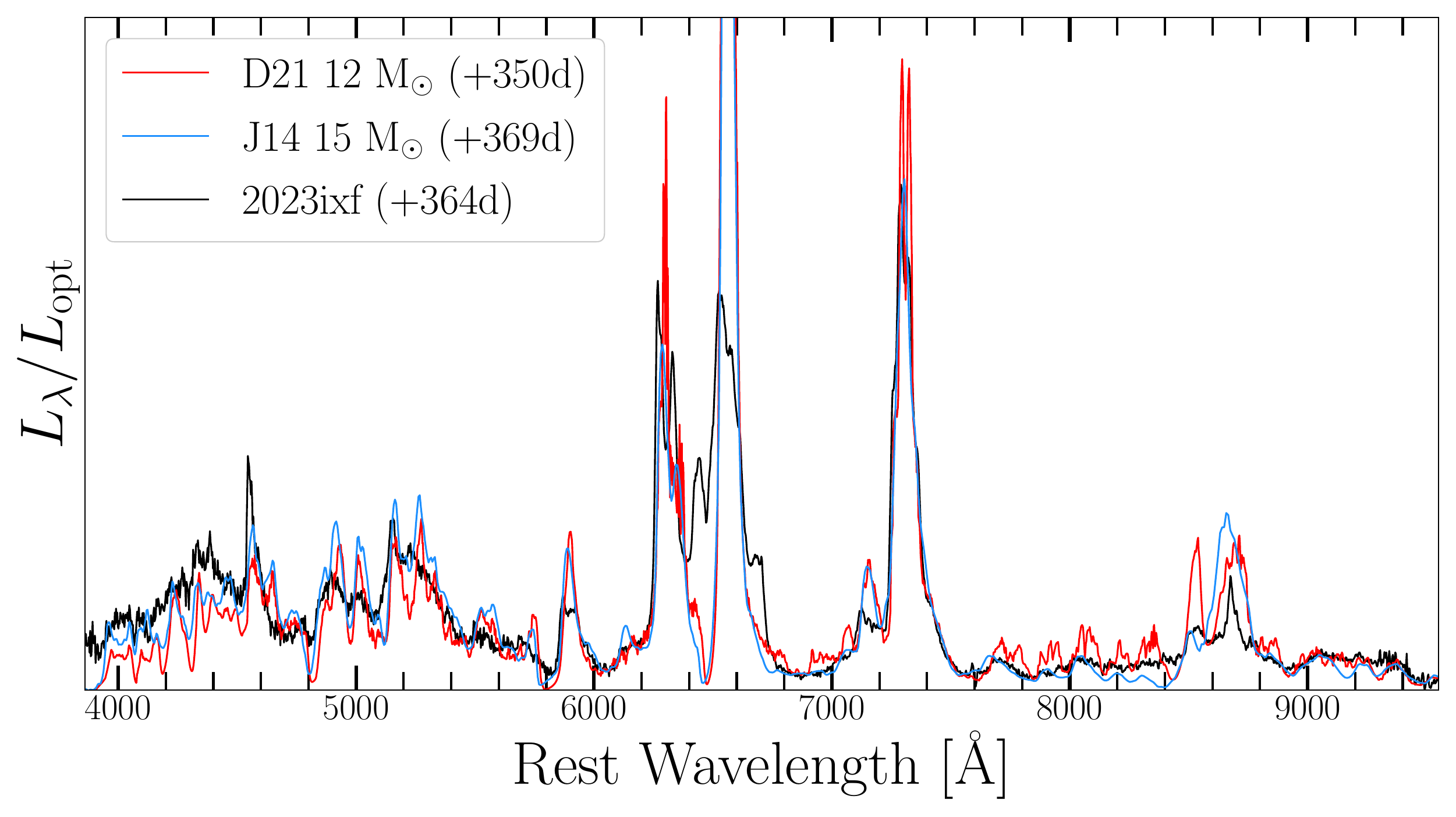}}
\subfigure{\includegraphics[width=0.44\textwidth]{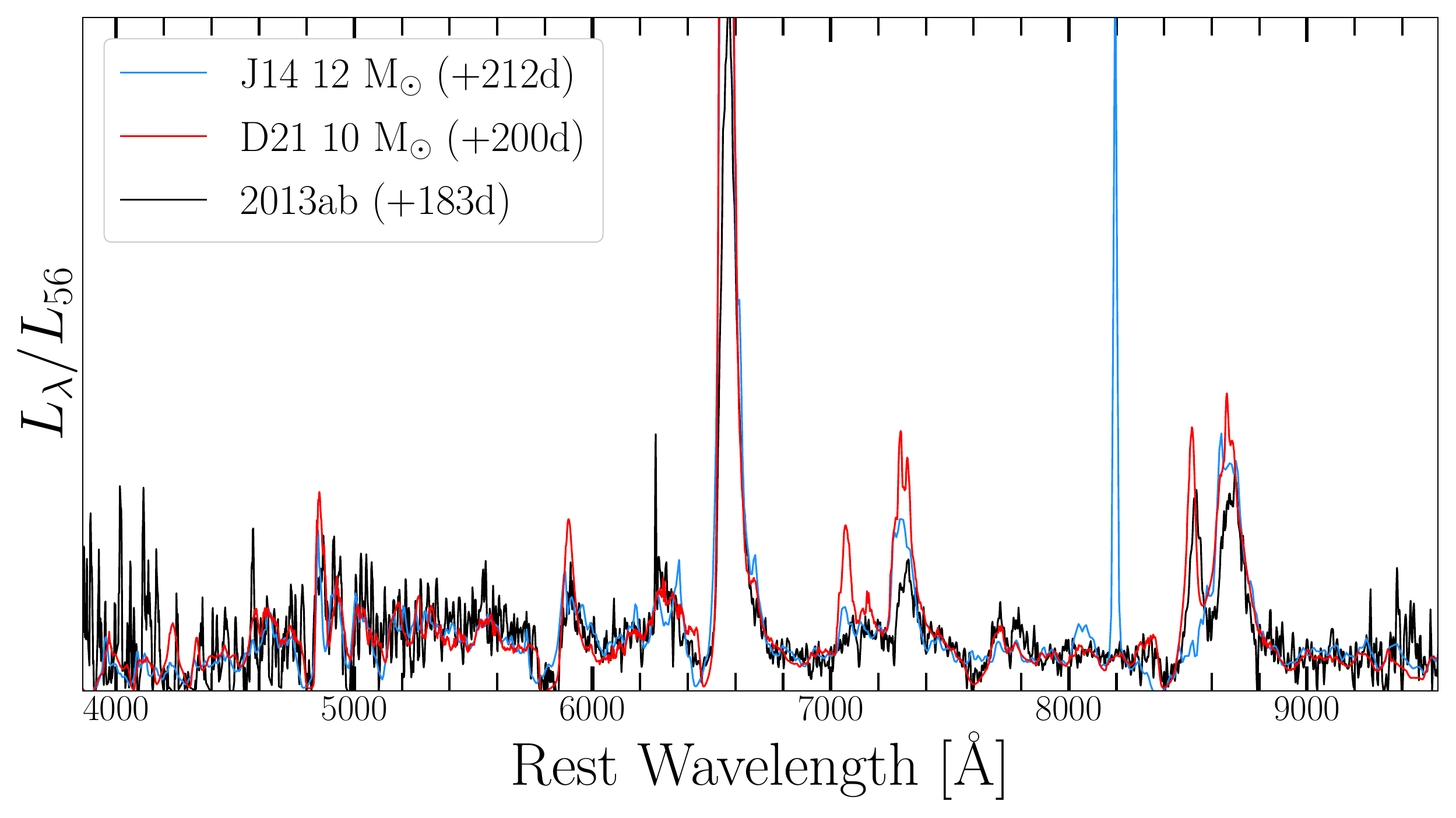}}
\caption{ Nebular spectra of sample objects (black) compared to best-matched model spectra from \cite{Dessart21} (red) and \cite{Jerkstrand14} (blue).   
\label{fig:neb_models_gold2} }
\end{figure*}

\end{document}